\newif\iffigs\figstrue
\documentclass[a4paper,11pt]{article}
\usepackage{latexsym,amssymb,lscape,graphics}
\usepackage{graphicx}        
\usepackage{longtable}
\usepackage{multirow}
\usepackage{color}
\usepackage[vsentermath]{youngtab}
\usepackage{slashed,epsfig}
\usepackage{amsfonts}

\newtheorem{definizione}{Definition}[section]

\newcommand{\bd}{\begin{definizione}}
\newcommand{\ed}{\end{definizione}}

\def\IC{\relax\,\hbox{$\inbar\kern-.3em{\rm C}$}}
\def\IG{\relax\,\hbox{$\inbar\kern-.3em{\rm G}$}}
\def\IB{\relax{\rm I\kern-.18em B}}
\def\ID{\relax{\rm I\kern-.18em D}}
\def\IL{\relax{\rm I\kern-.18em L}}
\def\IF{\relax{\rm I\kern-.18em F}}
\def\IH{\relax{\rm I\kern-.18em H}}
\def\II{\relax{\rm I\kern-.17em I}}
\def\IN{\relax{\rm I\kern-.18em N}}
\def\IP{\relax{\rm I\kern-.18em P}}
\def\IQ{\relax\,\hbox{$\inbar\kern-.3em{\rm Q}$}}
\def\bfzero{\relax\,\hbox{$\inbar\kern-.3em{\rm 0}$}}
\def\IK{\relax{\rm I\kern-.18em K}}
\def\IG{\relax\,\hbox{$\inbar\kern-.3em{\rm G}$}}
 \font\cmss=cmss10 \font\cmsss=cmss10 at 7pt
\def\IR{\relax{\rm I\kern-.18em R}}
\def\ZZ{\relax\ifmmode\mathchoice
{\hbox{\cmss Z\kern-.4em Z}}{\hbox{\cmss Z\kern-.4em Z}}
{\lower.9pt\hbox{\cmsss Z\kern-.4em Z}} {\lower1.2pt\hbox{\cmsss
Z\kern-.4em Z}}\else{\cmss Z\kern-.4em Z}\fi}
\def\bfone{\relax{\rm 1\kern-.35em 1}}

\def\Solv{\mathop{\rm Solv}\nolimits}

\def\inbar{\vrule height1.5ex width.4pt depth0pt}
\def\bfzero{\relax{\rm I\kern-.18em 0}}
\def\bfone{\relax{\rm 1\kern-.35em 1}}

\DeclareFontFamily{U}{rsf}{} \DeclareFontShape{U}{rsf}{m}{n}{
  <5> <6> rsfs5 <7> <8> <9> rsfs7 <10-> rsfs10}{}
\DeclareMathAlphabet\Scr{U}{rsf}{m}{n}

\newcommand{\ft}[2]{{\textstyle\frac{#1}{#2}}}

\def\1bar{1\hskip -.275cm -}
\def\2bar{2\hskip -.275cm -}
\def\3bar{3\hskip -.275cm -}

\newsavebox{\uuunit}
\sbox{\uuunit}
                 {\setlength{\unitlength}{0.825em}
                      \begin{picture}(0.6,0.7)
                                      \thinlines
                                      \put(0,0){\line(1,0){0.5}}
                                      \put(0.15,0){\line(0,1){0.7}}
                                      \put(0.35,0){\line(0,1){0.8}}
                                     \multiput(0.3,0.8)(-0.04,-0.02){10}{\rule{0.5pt}{0.5pt}}
                      \end {picture}}

\makeatletter \@addtoreset{equation}{section} \makeatother

\def\bfone{\relax{\rm 1\kern-.35em 1}}

\def\bfone{\relax{\rm 1\kern-.35em 1}}
\font\cmss=cmss10 \font\cmsss=cmss10 at 7pt

\newcommand{\so}{\mathfrak{so}}
\newcommand{\su}{\mathfrak{su}}

\newcommand{\uu}{\mathfrak{u}}
\newcommand{\sym}{\mathfrak{sp}}
\newcommand{\slal}{\mathfrak{sl}}

\begin{document}
\begin{titlepage}
\vskip 0.2cm
\begin{center}
{\Large {\bf Extremal Multicenter Black Holes:\\[0.5 cm]
Nilpotent Orbits  and Tits Satake Universality Classes}}\\[1cm]
{ \large Pietro Fr\'e$^{a}$\footnote{Prof. Fr\'e is presently fulfilling the duties of Scientific Counselor of the Italian Embassy in the Russian Federation, Denezhnij pereulok, 5, 121002 Moscow, Russia.}, Alexander S. Sorin$^{b}$ }
{}~\\
{}~\\
\quad \\
{{\em $^{a}$  Dipartimento di Fisica, Universit\'a di Torino,}}
\\
{{\em $\&$ INFN - Sezione di Torino}}\\
{\em via P. Giuria 1, I-10125 Torino, Italy}~\quad\\
{\tt fre@to.infn.it}
{}~\\
\quad \\
{{\em $^{b}$ Bogoliubov Laboratory of Theoretical Physics,}}\\
{{\em Joint Institute for Nuclear Research,}}\\
{\em 141980 Dubna, Moscow Region, Russia}~\quad\\
{\tt sorin@theor.jinr.ru}
\quad \\
\end{center}
~{}
\begin{abstract}
Four dimensional supergravity theories whose scalar manifold is a symmetric coset manifold $\mathrm{U_{D=4}/H_c}$ are arranged into a finite list of Tits Satake universality classes. Stationary  solutions of these theories, spherically symmetric or not,  are identified with those of an euclidian three-dimensional $\sigma$-model, whose target manifold is a Lorentzian coset $\mathrm{U_{D=3}/H^\star}$ and the extremal ones are associated with $\mathrm{H^\star}$ nilpotent orbits in the $\mathrm{K^\star}$ representation emerging from the orthogonal decomposition of the algebra $\mathbb{U}_{D=3}$ with respect to $\mathrm{H^\star}$. It is shown that the classification of such orbits can always be reduced to the Tits-Satake projection and it is a class property of the  Tits Satake universality classes. The construction procedure of  Bossard et al of extremal multicenter solutions by means of a triangular hierarchy of integrable equations is completed and converted into a closed algorithm by means of a general formula that provides the transition from the symmetric to the solvable gauge. The question of the relation between $\mathrm{H^\star}$ orbits and charge orbits $\mathbf{W}$ of the corresponding black holes is addressed and also reduced to the corresponding question within the Tits Satake projection. It is conjectured that on the vanishing locus of the Taub-NUT current the relation between $\mathrm{H}^\star$-orbit and $\mathbf{W}$-orbit is rigid and one-to-one. All black holes emerging from multicenter solutions associated with a given $\mathrm{H}^\star$ orbit have the same $\mathbf{W}$-type.  For the $S^3$ model we provide a complete survey of its multicenter solutions associated with all of the previously classified nilpotent orbits of $\slal(2)\times \slal(2)$ within $\mathfrak{g}_{2,2}$. We find a new intrinsic classification of the $W$-orbits of this model that might provide a paradigm for the analogous classification in all the other Tits Satake universality classes.
\end{abstract}
\end{titlepage}
\tableofcontents
\newpage
\section{Introduction}
The classical solutions of supergravity models that, in the gravitational sector,  include  metrics of black hole type, have attracted a lot of interest in the course of the last fifteen-seventeen years, with various peaks of attention and research activity when some new conceptual input was introduced, which has occurred repeatedly.
\par
The milestones in this new age of black hole physics have been:
\begin{description}
  \item[1)] The discovery of the attraction mechanism in the evolution of scalar fields \cite{ferrarakallosh}.
  \item[2)] The statistical interpretation of black hole entropy in terms of string microstates \cite{stromingerBH,Gibbons:1996af}.
  \item[3)] The ample development of black hole study \cite{CYandBH,criticalrefs,pioline,firstorderSUSY,iomatteo,olderBHliterature,ortino,orbifoldlimits} within the framework of special K\"ahler geometry, \cite{SKG,Andrianopoli:1996cm,specHomgeo} using in particular the first order equations that follow from the preservation of supersymmetry. This meant an in depth study of BPS black holes and of the relation of symplectic invariants of the duality groups with the black hole entropy \cite{dualitiessg}.
  \item[4)] The construction of black hole microstates in suitable conformal field theories \cite{microstateliterature}.
  \item[5)] The discovery of \textit{fake superpotentials} and of new attraction flows for non supersymmetric, non BPS black holes \cite{Lopes Cardoso:2007ky,Gaiotto:2007ag,fakeprepo,fakeprepo2,fakeprepoHJ}.
  \item[6)] The introduction of a three-dimensional $\sigma$-model approach for the derivation of black hole solutions \cite{Bergshoeff:2008be} and the association of these latter with nilpotent orbits of the $U$-duality group in the extremal case \cite{Fre:2009dg,Chemissany:2009af,Chemissany:2010zp}.
   \item[7)] The complete integrability of the supergravity equations in the spherical symmetric case and their association with a Lax pair formulation, in the case where the scalar manifold is a symmetric homogenous space $\mathrm{U/H}$ \cite{Fre:2009dg,Chemissany:2009af,Chemissany:2010zp,noiultimo,marioetal,noig22,titsblackholes}.
\end{description}
What mainly concerns the present paper are the developments related with points 6) and 7) of the above list.
\par
Concerning spherical symmetric supergravity black holes in the last couple of years there were significant parallel developments based on:
\begin{description}
  \item[A] The purely $D=4$ approach centered on the geodesic potential for radial flows, its reexpression in terms  of a \textit{fake superpotential} and the analysis of critical points of the latter, corresponding to  black hole horizons \cite{criticalrefs,itanteA,itanteB}. In this approach the classification of black holes is reduced to the classification of orbits of the symplectic $\mathbf{W}$-representation, namely the representation of the $D=4$ duality group $\mathrm{U_{D=4}}$, encoding the electromagnetic charges of the supergravity model.
  \item[B] The $D=3$ approach centered on the reduction of supergravity field equations to those of a $D=3$ $\sigma$-model with a Lorentzian target manifold which, for scalar manifolds that are symmetric coset spaces $\mathrm{U_{D=4}/H_c}$, is a also a symmetric coset $\mathrm{U_{D=3}/H^\star}$, the isotropy group $\mathrm{H^\star}$ being a non-compact  real section of the complexification of the maximal compact subgroup of $\mathrm{U_{D=3}}$. In this approach, which leads to a Lax pair formalism and to the development of explicit analytic formulae for the general integral \cite{noiultimo,Fre:2009dg,Chemissany:2010zp}, the classification of black holes is reduced to the classification of orbits of the non compact stability group $\mathrm{H^\star}$ in the $\mathbb{K}^\star$ representation that corresponds to the orthogonal decomposition of the $D=3$ Lie algebra: $\mathrm{\mathbb{U}_{D=3}}\, = \, \mathbb{H}^\star \, \oplus \, \mathbb{K}^\star$.
\end{description}
The main result of approach A) has been the classification of critical points in terms of certain special geometry invariants that are able to distinguish large and small black holes, BPS and non-BPS ones. Obviously the structure of these invariants  depend on the electromagnetic charges which determine the geodesic potential and hence on the orbit of the $\mathbf{W}$ representation yet such a link is not so clear in the $D=4$ approach.
On the other hand, also in the $D=3$ approach the relation between the $\mathrm{H}^\star$ orbits and the $\mathbf{W}$ orbits of the resulting black hole charges was not systematically investigated so far.
\par
In view of these facts the basic goal of the present paper is precisely a systematic investigation of the relation between these two approaches to the classification of black hole solutions.
\par
From the point of view of spherical symmetric solutions it was established  that non-extremal black holes correspond to orbits of regular, diagonalizable Lax operators in the $\mathbb{K}^\star$ representation of $\mathrm{H}^\star$, while extremal black holes are associated with nilpotent orbits in the same representation. For this reason a considerable amount of work was recently devoted to the study and classification of nilpotent orbits of the subgroup $\mathrm{H}^\star \subset \mathrm{U}$  extending and completing work done by mathematicians for the nilpotent orbits of the full group $ \mathrm{U}$. In the case of the so named $S^3$-model, leading to ${\mathbb{U}_{D=3}}\,= \, \mathfrak{g}_{2,2}$, partial results were obtained in \cite{bruxelles} confirmed and extended in \cite{noig22}.
\par
In \cite{titsblackholes} the present authors, in collaboration with M. Trigiante,  succeeded in elaborating a complete algorithm for the construction and classification of $\mathrm{H}^\star$ nilpotent orbits in the $\mathbb{K}^\star$ representation, applicable to all relevant supergravity models based on scalar manifolds that are symmetric coset spaces. The algorithm is based on the method of the standard triples $\{h,X,Y\}$, namely of the embedding of $\slal(2,\mathbb{R})$ algebras where $X,Y \in \mathbb{K}^\star$ and $h \in \mathbb{H}^\star$, with the additional essential use of the Weyl group $\mathcal{W}$ and of its subgroup $\mathcal{W}_H$ which preserves the splitting $\mathrm{\mathbb{U}_{D=3}}\, = \, \mathbb{H}^\star \, \oplus \, \mathbb{K}^\star$. The final labeling of the orbits is given by three set of eigenvalues named the $\alpha$, $\beta$ and $\gamma$ labels (see  \cite{titsblackholes}, \cite{nilorbits} for details).
\par
In the same paper where it was introduced, the constructive orbit classification algorithm was applied to the $\mathfrak{g}_{2,2}$ case, reobtaining the $7$ orbits found in previous calculations. It was also applied to the non maximally split case of the algebras $\so(4,4+2s)$ where, independently from the value of $s$, we were able to single out  a fixed pattern of $37$ nilpotent orbits. Indeed in \cite{titsblackholes} we showed that the structure of nilpotent orbits is a property of the Tits-Satake subalgebra of any algebra, giving a precise mathematical and physical relevance to the concept of \textit{Tits Satake universality classes} introduced several years ago in \cite{titsusataku}.
\par
In the present paper we address the systematic issue of organizing all supergravity models based on symmetric spaces into a finite list of \textit{Tits Satake universality classes} (see tables \ref{homomodels},\ref{exohomomodels}) and by means of an in depth group-theoretical analysis we show that for all elements of a given class there are always enough and appropriate parameters in the adjoint representation of $\mathbb{H}^\star$ to rotate a generic element of $\mathbb{K}^\star$ into its subspace $\mathbb{K}^\star_{\mathrm{TS}}$ obtained by means of the Tits Satake projection.
We show that the same is true for the $\mathbf{W}$-representation of charges: in the adjoint representation of $\mathrm{U_{D=4}}$ there are always enough and appropriate parameters to rotate any element of $\mathbf{W}$ into its Tits Satake projection $\mathbf{W}_{\mathrm{TS}}$. Therefore we show constructively that both at the $D=4$ and $D=3$ level the classification of black hole solutions can be restricted to the classification of orbit structures for a finite list of maximally split Tits Satake algebras.
\par
It is also appropriate to stress that the construction of nilpotent orbits for these universality classes has already consistently progressed, in view of the results of \cite{titsblackholes} and of \cite{marioetalf4} where the orbit pattern for the $\mathfrak{f}_{4,4}$ universality class was derived.
\par
Having established this crucial point the comparison between the $D=4$ and $D=3$ approaches is reduced to investigate the relation between the corresponding $\mathbf{W}_{\mathrm{TS}}$ and $\mathbb{K}^\star_{\mathrm{TS}}$ orbits. Although simplified to its essence, the problem remains, and in the spherical symmetric approach we obtain only some partial answers. We can summarize the issue in the following question:
\textit{The electromagnetic charges that can be obtained from Lax operators belonging to the same $\mathrm{H^\star}$ orbit do always fall in the same $\mathbf{W}^\star$ orbit of $\mathrm{U_{D=4}}$?}. The answer is far from being positive. Indeed one immediately gets counterexamples by showing that Lax operators belonging to the same $\mathrm{H^\star}$ orbit can yield both charges with vanishing and charges with non vanishing quartic invariant $\mathfrak{J}_4$. However there is a caveat. Typically we are interested in asymptotically flat  black holes so that we have to exclude a non vanishing Taub-NUT charge. The Taub-NUT charge is associated with the highest root of the algebra and the reduction of a dynamical system which kills the highest root corresponds to a consistent truncation. So we can consider the vanishing Taub-NUT locus in every $\mathrm{H}^\star$-orbit and it appears that all Lax operators belonging to such a locus produce charge vectors within the same $\mathbf{W}$-orbit. Since the vanishing Taub-NUT locus is not a coset manifold rather an algebraic surface within a coset manifold, we were not able to provide a formal proof of this fact, yet we can put it in the form of a \textbf{conjecture} since no counter example has been found.
\par
This is what we can say if we stick to spherically symmetric
solutions, yet there are entirely new perspectives that open up if
we take into account the brilliant strategy introduced in
\cite{Bossard:2009my,Bossard:2009at,Bossard:2009bw,Bossard:2009mz,Bossard:2011kz,Bossard:2012ge}
how to associate multicenter non spherically symmetric solutions
to each $\mathrm{H}^\star$ nilpotent orbit  of $\mathbb{K}^\star$.
\par
The main catch of this strategy relies on the use of a
\textit{symmetric}, rather than \textit{solvable gauge}, to
represent  the field equations of the three-dimensional
$\sigma$-model (equivalent to supergravity for stationary
solutions), that has the coset $\mathrm{U_{D=3}/H^{\star}}$ as
target space. The next crucial ingredient in this menu is the use
of the central element $h$ of each standard triple $\{h,X,Y\}$ as
a grading operator that determines a nilpotent subalgebra
$\mathbb{N}\subset \mathbb{U}_{\mathrm{D=3}}$, spanned by its
eigen-operators of positive grading. Denoting $\mathbb{N}\bigcap
\mathbb{K}^\star$ the intersection of such an algebra with the
$\mathbb{K}^\star$ subspace and $\mathfrak{h}(x)$ a map
$\mathcal{M}_3 \to \mathbb{N}\bigcap \mathbb{K}^\star$ from the
three-dimensional base space into such intersection, the symmetric
coset representative is just $\mathcal{Y}(x) \, = \,
\exp[\mathfrak{h}(x)]$ and obeys three-dimensional field equations
forming a solvable system. The components of $\mathfrak{h}(x)$
associated with lower gradings are just harmonic functions, while
those associated with higher gradings obey Laplace equation with a
source term provided by a functional of the lower grading
components. In this way one can construct non spherical symmetric
solutions of the $\sigma$-model representing  multicenter black
holes and also, as we will show in the sequel, Kerr like
solutions. A crucial technical problem that, in papers
\cite{Bossard:2009my,Bossard:2009at,Bossard:2009bw,Bossard:2009mz,Bossard:2011kz,Bossard:2012ge}
was only touched upon, or solved with ad hoc procedures for
specific cases,  concerns the final \textit{oxidation procedure}.
The correspondence between the fields of supergravity and the
fields of the three-dimensional $\sigma$-model is precise and
algorithmic in the solvable gauge realization of the coset
representative, not in the symmetric gauge. Hence, in order to
read off the supergravity solution one has to make such a gauge
transformation which, at first sight, appears a matter of art and
ingenuity, case by case. Fortunately this is not true, since the
problem was already solved in complete generality in
\cite{noiultimo,Fre:2009dg,Chemissany:2009af,Chemissany:2010zp},
by means of a formula (see eq.(\ref{solutioncosetrepr}) of the
present paper) originally found in the context of spherical
symmetric solutions, yet of much further reach. Indeed coupling
eq.(\ref{solutioncosetrepr})  with the strategy of papers
\cite{Bossard:2009my,Bossard:2009at,Bossard:2009bw,Bossard:2009mz,Bossard:2011kz,Bossard:2012ge},
the general construction of multicenter black hole solutions
associated with nilpotent orbits becomes truly algorithmic and
even implementable on a computer. We consider this one of the main
results of the present paper.
\par
We applied the algorithm to the case of the $S^3$ model, exploring the general form of multicenter solutions and we came to the following conclusion. The solutions associated with one nilpotent orbit have the property that, for each pole shared  by all the harmonic functions  springing from the corresponding nilpotent algebra, we have a black hole whose charges and other properties are those displayed by the spherical symmetric solution generated by the Lax operator in that orbit. However it may happen that a pole is located only in one subset of the harmonic building blocks, while other poles are located  in different subsets. In that case the  black holes springing from each pole have the charges and properties of  the spherical black hole pertaining to various  smaller orbits associated with each subgroups. This mechanism unveils that larger orbit black holes can be viewed as composite objects built from the coalescence of  smaller ones when their centers come together, namely when the corresponding poles overlap.
\par
Our analysis also shows that the solutions associated with very large orbits, typically those for which the corresponding Lax operator has nilpotency degree larger than three, are generically singular not only in the spherical symmetric case but also in the multicenter case, since at each common pole of the involved harmonic functions we just retrieve the same conditions pertaining to the spherical symmetric solution. This explicit result is somewhat different from the claims made at various places of ref.s \cite{Bossard:2009at,Bossard:2009bw,Bossard:2009my,Bossard:2009mz,Bossard:2011kz,Bossard:2012ge} that multicenter solutions with generic positions of the poles can be regular for higher orbits. In any case we would like to stress that the solutions associate with higher orbits can be quite complex and deserve a special detailed study. Several surprises might still lie ahead.
\par
In the course of our analysis we have found a number of other results that up to our knowledge were not known in the literature. We plan to summarize and high lite our findings in our conclusions, sect.\ref{concludone}
\par
The structure of our paper is displayed in the contents. We emphasize that it is organized in three parts. In the first part
sect.s \ref{genefram}-\ref{construo} we review the $\sigma$-model approach to stationary supergravity solutions, the construction of multicenter solutions attached to nilpotent orbits and we spell out the complete oxidation algorithm.
\par
In the second part, sect.s \ref{examplu1}-\ref{examplu2} we apply the general scheme to the case of the $S^3$ model and we explore all the solutions attached to each of the classified nilpotent orbits.
\par
Part three of the paper, sect.s \ref{sugrarelevant}-\ref{KWseczia} is devoted to organize all the available supergravity models based on symmetric scalar manifolds into Tits Satake universality classes, to show how within each class we can always reduce the analysis to the Tits Satake subalgebra both at the $D=4$ and $D=3$ level and finally to analyse the general structure of the $\mathbb{H}^\star$ subalgebra and the $\mathbb{K}^\star$ representation in the maximally split casse (Tits Satake projection).
\section{The general framework}
\label{genefram}
The addressed issue is that of stationary solutions of $D=4$ ungauged supergravity models. For such field theories that include, the metric, several abelian gauge fields and several scalar fields, we have a general form of the bosonic lagrangian, which is the following one:
\begin{eqnarray}
\mathcal{L}^{(4)} &=& \sqrt{|\mbox{det}\, g|}\left[\frac{R[g]}{2} - \frac{1}{4}
\partial_{ {\mu}}\phi^a\partial^{ {\mu}}\phi^b h_{ab}(\phi) \,
+ \,
\mbox{Im}\mathcal{N}_{\Lambda\Sigma}\, F_{ {\mu} {\nu}}^\Lambda
F^{\Sigma| {\mu} {\nu}}\right] \nonumber\\
&&+
\frac{1}{2}\mbox{Re}\mathcal{N}_{\Lambda\Sigma}\, F_{ {\mu} {\nu}}^\Lambda
F^{\Sigma}_{ {\rho} {\sigma}}\epsilon^{ {\mu} {\nu} {\rho} {\sigma}}\,,
\label{d4generlag}
\end{eqnarray}
where $F_{ {\mu} {\nu}}^\Lambda\equiv (\partial_{ {\mu}}A^\Lambda_{ {\nu}}-\partial_{ {\nu}}A^\Lambda_{ {\mu}})/2$.
In eq.(\ref{d4generlag}) $\phi^a$ denotes the whole set of $n_{\mathrm{s}}$ scalar fields
parameterizing the scalar manifold $ \mathcal{M}_{scalar}^{D=4}$
which, for supersymmetry $\mathcal{N} > 2$, is necessarily a coset manifold:
\begin{equation}
  \mathcal{M}_{scalar}^{D=4} \, =
  \,\frac{\mathrm{U_{D=4}}}{\mathrm{H}_c}
\label{cosettoquando}
\end{equation}
For $\mathcal{N}=2$ eq.(\ref{cosettoquando}) is not obligatory but it is
possible: a well determined class
 of symmetric homogeneous manifolds that are special K\"ahler manifolds
\cite{specHomgeo} falls into the set up of the present general discussion.
The theory includes also $n_{\mathrm{v}}$ vector fields $A_{{\mu}}^\Lambda$
for which
\begin{equation}
  \mathcal{F}^{\pm| \Lambda}_{ {\mu} {\nu}} \equiv \ft 12
  \left[{F}^{\Lambda}_{ {\mu} {\nu}} \mp \, {\rm i} \, \frac{\sqrt{|\mbox{det}\, g|}}{2}
  \epsilon_{ {\mu} {\nu} {\rho} {\sigma}} \, F^{ {\rho} {\sigma}} \right]
\label{Fpiumeno}
\end{equation}
denote the self-dual (respectively antiself-dual) parts of the field-strengths.
As displayed in
eq.(\ref{d4generlag}) they are non minimally coupled to the scalars via the symmetric complex matrix
\begin{equation}
  \mathcal{N}_{\Lambda\Sigma}(\phi)
  ={\rm i}\, \mbox{Im}\mathcal{N}_{\Lambda\Sigma}+ \mbox{Re}\mathcal{N}_{\Lambda\Sigma}
\label{scriptaenna}
\end{equation}
which transforms projectively under $\mathrm{U_{D=4}}$.
Indeed the field strengths ${\mathcal{F}}^{\pm|
\Lambda}_{\mu\nu}$ plus their magnetic duals:
\begin{equation}\label{defione}
    {G}_
{\Lambda|\mu\nu} \, \equiv \, \ft 12 \, \epsilon_{\mu\nu}^{\phantom{\mu\nu}\rho\sigma} \, \frac{\delta {\mathcal{L}}^{(4)}}{\delta{{F}_{\rho\sigma}^{\Lambda}}}
\end{equation}
 fill up a $2\,
n_\mathrm{v}$--dimensional symplectic representation of $\mathrm{\mathbb{U}_{D=4}}$
which we call by the name of $\mathbf{W}$.
\par
We rephrase the above statements
by asserting that there is always a symplectic embedding of the duality group
$\mathrm{U}_{D=4}$,
\begin{equation}
  \mathrm{U}_{D=4} \mapsto \mathrm{Sp(2n_\mathrm{v}, \mathbb{R})} \quad ; \quad
 n_{\mathrm{v}} \, \equiv \, \mbox{$\#$ of vector fields}
\label{sympembed}
\end{equation}
so that for each element $\xi \in \mathrm{U}_{D=4}$ we have its
representation by means of a  suitable real symplectic matrix:
\begin{equation}
  \xi \mapsto \Lambda_\xi \equiv \left( \begin{array}{cc}
     A_\xi & B_\xi \\
     C_\xi & D_\xi \
  \end{array} \right)
\label{embeddusmatra}
\end{equation}
satisfying the defining relation:
\begin{equation}
  \Lambda_\xi ^T \, \left( \begin{array}{cc}
     \mathbf{0}_{n \times n}  & { \mathbf{1}}_{n \times n} \\
     -{ \mathbf{1}}_{n \times n}  & \mathbf{0}_{n \times n}  \
  \end{array} \right) \, \Lambda_\xi = \left( \begin{array}{cc}
     \mathbf{0}_{n \times n}  & { \mathbf{1}}_{n \times n} \\
     -{ \mathbf{1}}_{n \times n}  & \mathbf{0}_{n \times n}  \
  \end{array} \right)
\label{definingsympe}
\end{equation}
Under an element of the duality group the field strengths transform
as follows:
\begin{equation}
  \left(\begin{array}{c}
     \mathcal{F}^+ \\
     \mathcal{G}^+ \
  \end{array} \right)  ^\prime \, = \,\left( \begin{array}{cc}
     A_\xi & B_\xi \\
     C_\xi & D_\xi \
  \end{array} \right) \,  \left(\begin{array}{c}
     \mathcal{F}^+ \\
     \mathcal{G}^+ \
  \end{array} \right) \quad ; \quad \left(\begin{array}{c}
     \mathcal{F}^- \\
     \mathcal{G}^- \
  \end{array} \right)  ^\prime \, = \,\left( \begin{array}{cc}
     A_\xi & B_\xi \\
     C_\xi & D_\xi \
  \end{array} \right) \,  \left(\begin{array}{c}
     \mathcal{F}^- \\
     \mathcal{G}^- \
  \end{array} \right)
\label{lucoidale1}
\end{equation}
where, by their own definitions:
\begin{equation}
    \mathcal{G}^+ = \mathcal{N} \, \mathcal{F}^+ \quad ; \quad \mathcal{G}^- = \overline{\mathcal{N}} \,
    \mathcal{F}^-
\label{lucoidale2}
\end{equation}
and the complex symmetric matrix $\mathcal{N}$ should transform as follows:
\begin{equation}
  \mathcal{N}^{\,\prime} = \left(  C_\xi + D_\xi \, \mathcal{N}\right) \, \left( A_\xi + B_\xi \,\mathcal{N}\right)
  ^{-1}
\label{Ntransfa}
\end{equation}
Choose a parametrization of the coset $\mathbb{L}(\phi) \in \mathrm{U_{D=4}}$, which assigns a definite group element to every coset point identified by the scalar fields. Through the symplectic embedding (\ref{embeddusmatra}) this produces a definite $\phi$-dependent symplectic matrix
\begin{equation}
   \left( \begin{array}{cc}
     A(\phi) & B(\phi) \\
     C(\phi) & D(\phi) \
  \end{array} \right)
\label{embeddusmatra2}
\end{equation}
in the $W$-representation of $\mathrm{U_{D=4}}$. In terms of its blocks the kinetic matrix $\mathcal{N}(\phi)$ is explicitly given by the Gaillard-Zumino formula:
\begin{eqnarray}
\mathcal{N}(\phi)& = &
\left [C(\phi) - i D(\phi)\right ] \left [ A(\phi) - i B(\phi)\right]^{-1}\,,\label{masterformula}
\end{eqnarray}
\subsection{The $\sigma$-model approach to extremal black holes}
A very powerful token in deriving stationary and in particular extremal black hole solutions of $D=4$ supergravity that depend only on three space-like coordinates $x_1,x_2,x_3$, is provided by the time-reduction of  the four-dimensional field equations  to those of an effective three-dimensional $\sigma$-model.
Let us shortly review this  procedure.
\par
In all $\mathcal{N}=2$ cases the number of vector fields in the theory is $n_\mathrm{v} = n+1$ where $n$ is the complex dimension of the scalar manifold ($n_s =2 \, n$), while in the case of other theories the relation between $n_\mathrm{v}$ and $n_s$ is different. Notwithstanding this difference, we can always introduce a $2 \, n_\mathrm{v} \, \times \, 2 \, n_\mathrm{v}$ field dependent matrix $\mathcal{M}_4$ defined as follows:
 \begin{eqnarray}
\mathcal{M}_4 & = &
\left(\begin{array}{c|c}
 { \mathrm{Im}}\mathcal{N}^{-1} & { \mathrm{Im}}\mathcal{N}^{-1}\,{\mathrm{Re}}\mathcal{N} \\
\hline
{\mathrm{Re}}\mathcal{N}\,{ \mathrm{Im}}\,\mathcal{N}^{-1} & {\mathrm{Im}}\mathcal{N}\,
+\, {\mathrm{Re}}\mathcal{N} \, { \mathrm{Im}}\mathcal{N}^{-1}\, {\mathrm{Re}}\mathcal{N} \
\end{array}\right) \label{quaternionic}\\
\mathcal{M}_4^{-1} & = &
\left(\begin{array}{c|c}
{\mathrm{Im}}\mathcal{N}\,
+\, {\mathrm{Re}}\mathcal{N} \, { \mathrm{Im}}\mathcal{N}^{-1}\, {\mathrm{Re}}\mathcal{N} & \, -{\mathrm{Re}}\mathcal{N}\,{ \mathrm{Im}}\,\mathcal{N}^{-1}\\
\hline
-\, { \mathrm{Im}}\mathcal{N}^{-1}\,{\mathrm{Re}}\mathcal{N}  & { \mathrm{Im}}\mathcal{N}^{-1} \
\end{array}\right) \label{inversem4}
\end{eqnarray}
and we can introduce the following set of $2+n_s + 2n_\mathrm{v}$ fields  depending on  the three parameters $x_i$:
\begin{center}
\begin{tabular}{|l|cc||cc|}
  \hline
  \null & Generic &\null & $\mathcal{N}=2$ & \null\\
  \cline{1-2}\cline{4-5}
  \hline
  warp factor & $U(x)$ & $1$ & 1 & \null \\
  Taub-NUT field & $a(x)$ & 1 & 1 & \null \\
   D=4 scalars& $\phi^a(x)$ & $n_s$ &  $ 2 n$ &\null \\
  Scalars from vectors & $Z^M(x) \, = \, \left(Z^\Lambda(x)\, , \, Z_\Sigma(x) \right)$ & $2 n_\mathrm{v}$ &$ 2 n + 2$ & \null \\
  \hline
  \textbf{Total} & \null & $2+ n_s + 2 n_\mathrm{v}$ & $ 4 n + 4$ & \null \\
  \hline
\end{tabular}
\end{center}
the fields $\{U,a,\phi,Z\}$ are interpreted as the coordinates of a new $(2+ n_s + 2 n_\mathrm{v})$-dimensional  manifold
${\mathcal{Q}}$,
whose metric we declare to be the following:
\begin{eqnarray}
ds^2_{\mathcal{Q}} &=& \frac{1}{4} \, \left [ d{U}^2+\,h_{rs}\,d{\phi}^r\,d{\phi}^s
+ e^{-2\,U}\,(d{a}+{\bf Z}^T\mathbb{C}d{{\bf
Z}})^2\,+\,2 \, e^{-U}\,d{{\bf
Z}}^T\,\mathcal{M}_4\,d{{\bf Z}}\right ]
\label{geodaction}
\end{eqnarray}
having denoted by $\mathbb{C}$ the constant symplectic invariant metric in $2 n_\mathrm{v}$ dimensions that underlies the construction of the matrix $\mathcal{N}_{\Lambda\Sigma}$. The metric (\ref{geodaction}) has the following indefinite signature
\begin{equation}
\mbox{sign}\left[ds^2_{\mathcal{Q}}\right] \, = \, \left(\underbrace{+,\dots,+}_{2+\mathrm{n_s}},\underbrace{-,\dots ,-}_{2\mathrm{n_v}+2}\right)
\end{equation}
since the matrix $\mathcal{M}_4 $ is negative definite.
\par
Moreover one very important point to be stressed is that the metric (\ref{geodaction}) admits a typically large group of isometries. Certainly it admits all the isometries of the original scalar manifold $\mathcal{M}_{scalar}$ enlarged with additional ones related to the new fields that have been introduced $\{U,a,Z^M\}$. In the case when the $D=4$ scalar manifold is a homogeneous symmetric space:
 \begin{equation}\label{cosettoD4}
    \mathcal{M}_{scalar} \, = \, \frac{\mathrm{U_{D=4}}}{\mathrm{H_c}}
 \end{equation}
 one can show \cite{Breitenlohner:1987dg,cmappa,Bergshoeff:2008be,Fre':2005si}, that the manifold $\mathcal{Q}$ with the metric (\ref{geodaction}) is a new homogeneous symmetric space
 \begin{equation}\label{qcosetto}
    \mathcal{Q} \, = \, \frac{\mathrm{U}_{D=3}}{\mathrm{H}^\star}
 \end{equation}
 whose structure is universal and can be described in general terms.
\subsubsection{General structure of the $\mathbb{U}_{D=3}$ Lie algebra}
 The Lie algebra $\mathbb{U}_{D=3}$ of the numerator group always contains, as
subalgebra, the duality algebra $\mathbb{U}_{D=4}$ of the parent
supergravity theory in $D=4$ and a universal  ${\slal(2,\mathbb{R})_E}$
algebra which is associated with the gravitational degrees of freedom $\{U,a\}$
 Furthermore, with respect to this subalgebra, $\mathbb{U}_{D=3}$ admits the following universal decomposition,
holding for all $\mathcal{N}$-extended supergravities:
\begin{equation}
\mbox{adj}(\mathbb{U}_{D=3}) =
\mbox{adj}(\mathbb{U}_{D=4})\oplus\mbox{adj}({\slal(2,\mathbb{R})_E})\oplus
W_{(2,\mathbf{W})}
\label{gendecompo}
\end{equation}
where $\mathbf{W}$ is the {\bf symplectic} representation of
$\mathbb{U}_{D=4}$ to which the electric and magnetic field
strengths are assigned. Indeed the scalar fields associated with
the generators of $W_{(2,\mathbf{W})}$ are just those coming from
the vectors in $D=4$. Denoting the generators of
$\mathbb{U}_{D=4}$ by $T^a$, the generators of
${\slal(2,\mathbb{R})_E}$ by $\mathrm{L^x}$ and denoting by
$W^{iM}$ the generators in $W_{(2,\mathbf{W})}$, the commutation
relations that correspond to the decomposition (\ref{gendecompo})
have the following general form:
\begin{eqnarray}
\nonumber && [T^a,T^b] = f^{ab}_{\phantom{ab}c} \, T^c  \\
\nonumber && [L^x,L^y] = f^{xy}_{\phantom{xy}z} \, L^z , \\
&&\nonumber [T^a,W^{iM}] = (\Lambda^a)^M_{\,\,\,N} \, W^{iN},
\\ \nonumber && [L^x, W^{iM}] = (\lambda^x)^i_{\,\, j}\, W^{jM}, \\
&&[W^{iM},W^{jN}] = \epsilon^{ij}\, (K_a)^{MN}\, T^a + \,
\mathbb{C}^{MN}\, k_x^{ij}\, L^x \label{genGD3pre}
\end{eqnarray}
where the $2 \times 2$ matrices $(\lambda^x)^i_j$, are the canonical generators of ${\slal(2,\mathbb{R})}$
in the fundamental, defining representation:
\begin{equation}
  \lambda^3 = \left(\begin{array}{cc}
     \ft 12 & 0 \\
     0 & -\ft 12 \
  \end{array} \right) \quad ; \quad \lambda^1 = \left(\begin{array}{cc}
     0 & \ft 12  \\
     \ft 12 & 0\
  \end{array} \right) \quad ; \quad \lambda^2 = \left(\begin{array}{cc}
     0 & \ft 12  \\
     -\ft 12 & 0\
  \end{array} \right)
\label{lambdax}
\end{equation}
while $\Lambda^a$ are the generators
of $\mathbb{U}_{D=4}$ in the symplectic representation $\mathbf{W}$. By
\begin{equation}
  \mathbb{C}^{MN} \equiv \left( \begin{array}{c|c}
     \mathbf{0}_{n_\mathrm{v}\times n_\mathrm{v}} & \mathbf{1}_{n_\mathrm{v}\times n_\mathrm{v}} \\
     \hline
     -\mathbf{1}_{n_\mathrm{v}\times n_\mathrm{v}} & \mathbf{0}_{n_\mathrm{v}\times n_\mathrm{v}} \
  \end{array}\right)
\label{omegamatra}
\end{equation}
we denote the antisymmetric symplectic metric in $2n_\mathrm{v}$ dimensions, $n_\mathrm{v}$
being the number of vector fields in $D=4$ as we have already stressed. The symplectic character
of the representation $\mathbf{W}$ is asserted by the identity:
\begin{equation}
  \Lambda^a\, \mathbb{C} + \mathbb{C}\, \left( \Lambda^a \right )^T = 0
\label{Lamsymp}
\end{equation}
The fundamental doublet representation of ${\slal(2,\mathbb{R})}$
is also symplectic and we have denoted by $\epsilon^{ij}= \left( \begin{array}{cc}
  0 & 1 \\
  -1 & 0
\end{array}\right) $ the
$2$-dimensional symplectic metric, so that:
\begin{equation}
    \lambda^x\, \epsilon + \epsilon\, \left( \lambda^x \right )^T = 0,
\label{lamsymp}
\end{equation}
In eq. (\ref{genGD3pre}) we have used the standard
convention according to which symplectic indices are raised
and lowered with the appropriate symplectic metric, while adjoint representation
indices are raised and lowered with the Cartan-Killing metric.
\subsubsection{General form of the three-dimensional $\sigma$-model}
Next we consider a gravity coupled three-dimensional euclidian $\sigma$-model, whose fields
$$\Phi^A(x) \, \equiv \, \{U(x),a(x),\phi(x),Z(x)\}$$ describe mappings:
\begin{equation}\label{corneato}
    \Phi \, : \quad \mathcal{M}_3 \, \rightarrow \, \mathcal{Q}
\end{equation}
from a three-dimensional manifold $\mathcal{M}_3$, whose metric we denote by $\gamma_{ij}(x)$, to the target space $\mathcal{Q}$.
The action of this $\sigma$-model is the following:
\begin{eqnarray}\label{beddamatre}
\mathcal{A}^{[3]} &= &  \int  \sqrt{\mbox{det}\gamma}\, \mathfrak{R}[\gamma]\,  d^3x\,+ \, \int  \sqrt{\mbox{det}\gamma}\,  \mathcal{L}^{(3)} \, d^3x \label{aczia3}\\
\mathcal{L}^{(3)} & = &   \left( \partial_i U \, \partial_j U+\,h_{rs}\,\partial_i{\phi}^r\,\partial_j{\phi}^s
\right. \nonumber\\
&& \left.+ e^{-2\,U}\,\left (\partial_i{a}+{\bf Z}^T\mathbb{C}\partial_i {{\bf
Z}}\right ) \, \left (\partial_j{a}+{\bf Z}^T\mathbb{C}\partial_j {{\bf
Z}}\right )\,+\,2 \, e^{-\,U}\,\partial_i{{\bf
Z}}^T\,\mathcal{M}_4\,\partial_j{{\bf Z}} \right ) \, \gamma^{ij}\label{trilagranka}
\end{eqnarray}
where $\mathfrak{R}[\gamma]$ denotes the scalar curvature of the metric $\gamma_{ij}$.
\par
The field equations of the $\sigma$-model are obtained by varying the action both in the metric $\gamma_{ij}$ and in the fields $\Phi^A(x)$. The Einstein equation reads as usual:
\begin{equation}\label{canfora}
    \mathfrak{R}_{ij}\, - \, \ft 12 \gamma_{ij} \, \mathfrak{R} \, = \, \mathfrak{T}_{ij}
\end{equation}
where:
\begin{equation}\label{tensorostressato}
    \mathfrak{T}_{ij} \, = \, \frac{\delta\mathcal{L}^{(3)}}{\delta \gamma^{ij}} \, - \, \gamma_{ij} \, \mathcal{L}^{(3)}
\end{equation}
is the stress energy tensor, while the matter field equations assume the standard form:
\begin{equation}\label{Fiequa}
    \frac{1}{\sqrt{\mbox{det}\gamma}}\, \gamma^{ij} \, \partial_i \, \left[\sqrt{\mbox{det}\gamma} \frac{\delta \mathcal{L}^{(3)}}{\delta \, \partial^j \Phi^A}\right] \, - \, \frac{\delta \mathcal{L}^{(3)}}{\delta \Phi^A} \, = \, 0
\end{equation}
As it is well known, in $D=3$ there is no propagating graviton and the Riemann tensor is completely determined by the Ricci tensor, namely, via Einstein equations, by the stress-energy tensor of the matter fields.
\par
Extremal solutions of the $\sigma$-model are those for which the three-dimensional metric can be consistently chosen flat:
\begin{equation}\label{rutillo}
    \gamma_{ij} \, = \, \delta_{ij}
\end{equation}
corresponding to a vanishing stress-energy tensor:
\begin{eqnarray}
    \left( \partial_i U \, \partial_j U+\,h_{rs}\,\partial_i{\phi}^r\,\partial_j{\phi}^s
+ e^{-2\,U}\,\left (\partial_i{a}+{\bf Z}^T\mathbb{C}\partial_i {{\bf
Z}}\right ) \, \left (\partial_j{a}+{\bf Z}^T\mathbb{C}\partial_j {{\bf
Z}}\right )\,+\,2 \, e^{-\,U}\,\partial_i{{\bf
Z}}^T\,\mathcal{M}_4\,\partial_j{{\bf Z}} \right )\, = \,0 \nonumber\\
\label{forbice}
\end{eqnarray}
We will see in the sequel how the nilpotent orbits of the group $\mathrm{H}^\star$ in the $\mathbb{K}^\star$ representation can be systematically associated with general extremal solutions of the field equations.
\subsection{Oxidation rules for extremal multicenter black holes}
\label{genoxo}
Let us now describe the oxidation rules, namely the procedure by means of which to every configuration of the three-dimensional fields $\Phi(x) \, = \,\{U(x),a(x),\phi(x),Z(x)\}$, satisfying the field equations (\ref{Fiequa}) and also the extremality condition (\ref{forbice}), we can associate a well defined configuration of the four-dimensional fields satisfying the field equations of supergravity that follow from the lagrangian (\ref{d4generlag}).
We might write such oxidation rules for general solutions of the $\sigma$-model, also non extremal, yet given the goal of the present paper we confine ourselves to spell out such rule in the extremal case, which is somewhat simpler since it avoids the extra complications related with the three-dimensional metric $\gamma_{ij}$.
\par
In order to write the $D=4$ fields, the first necessary item we have to determine is the Kaluza-Klein vector field $\mathbf{A}^{[KK]}\, = \, A^{[KK]}_i \, dx^i$. This latter is worked out  through the following dualization procedure:
\begin{eqnarray}
  \mathbf{F}^{[KK]} &=& d\mathbf{A}^{[KK]} \nonumber\\
  \mathbf{F}^{[KK]} &=& - \epsilon_{ijk}\, dx^i \wedge dx^j \, \left[ \exp[-2 U] \, \left( \partial^k a \, + \, Z \, \mathbb{C} \, \partial^k Z\right)\right] \label{rangallo}
\end{eqnarray}
Given the Kaluza-Klein vector we can write  the four-dimensional metric which is  the following:
\begin{equation}\label{metruzza}
    ds^2 \, = \, - \, \exp[U] \, \left( dt + \mathbf{A}^{[KK]}\right)^2 \, + \, \, \exp[- U] \, dx^i \otimes dx^j \, \delta_{ij}
\end{equation}
The vielbein description of the same metric is immediate. We just write:
\begin{eqnarray}
  ds^2 &=& - E^0 \otimes E^0 \, + \, E^i \otimes E^i \nonumber\\
  E^0 &=& \exp[\ft U 2] \left( dt + \mathbf{A}^{[KK]}\right) \nonumber\\
  E^i&=& \exp[-\ft U 2] \, dx^i \label{vielbino}
\end{eqnarray}
Next we can present the form of the electromagnetic field strengths:
\begin{eqnarray}
  \mathbf{F}^\Lambda &=& \mathbb{C}^{\Lambda M}\partial_i Z_M \, dx^i \, \wedge \, (dt +\mathbf{A}^{[KK]}) \nonumber  \\
  &\null& + \, \epsilon_{ijk} dx^i \, \wedge \, dx^j \, \left [ \exp[-U] \, \left(\mathrm{Im}\mathcal{N}^{-1}\right)^{\Lambda\Sigma}\left(\partial^k Z_\Sigma \, + \,\mathrm{Re}\mathcal{N}_{\Sigma\Gamma} \partial^k Z^\Gamma \right)\right]
\end{eqnarray}
Next we define the electromagnetic charges and the Taub-NUT charges for multicenter solutions. Considering the metric (\ref{metruzza}) the black hole centers are defined by the zeros of the warp-factor $\exp[U(\vec{x})]$.  In a composite $m$-black hole solution there are $m$ three-vectors $\vec{r}_{\alpha}$ ($\alpha=1,\dots,m$), such that:
\begin{equation}\label{fruttolo}
    \lim_{\vec{x} \rightarrow \vec{r}_{\alpha}}\, \exp[U(\vec{x})] \, = \, 0
\end{equation}
Each of these zeros defines a non trivial homology two-cycle $\mathbb{S}^2_{\alpha}$ of the 4-dimensional space-time which surrounds the singularity $\vec{r}_{\alpha}$. The electromagnetic charges of the individual holes are obtained by integrating the field strengths and their duals on such homology cycles.
\begin{equation}\label{chiffo}
    \left ( \begin{array}{c}
              p^\Lambda \\
              q_\Sigma
            \end{array}
    \right )_\alpha \, = \, \frac{1}{4\pi\sqrt{2}}\,\left ( \begin{array}{c}
                                 \int_{\mathbb{S}^2_{\alpha}}\, \mathbf{F}^\Lambda\\
                                 \int_{\mathbb{S}^2_{\alpha}} \, \mathbf{G}_\Sigma
                               \end{array}
    \right) \, \equiv \, \frac{1}{4\pi} \, \int_{\mathbb{S}^2_{\alpha}} \, j^{EM}
\end{equation}
Utilizing the form of the field strengths  we obtain the explicit formula:
\begin{equation}\label{cane}
\mathcal{Q}_\alpha \, \equiv \,      \left ( \begin{array}{c}
              p^\Lambda \\
              q_\Sigma
            \end{array}
    \right )_\alpha \, = \, \frac{1}{4\pi\sqrt{2}}\,\int_{\mathbb{S}^2_{\alpha}}\, \epsilon_{ijk} dx^i \wedge dx^j \,  \left [   \exp[-U] \, \mathcal{M}_{4} \, \partial^k Z \, + \, \exp[-2 U] \, \left( \partial^k a \, + \, Z \, \mathbb{C} \, \partial^k Z\right) \, \mathbb{C} \,Z \right]
\end{equation}
which provides $m$-sets of electromagnetic charges associated with the solution. Similarly we have $m$ Taub-NUT charges defined by:
\begin{equation}\label{finkle}
    \mathbf{n}_\alpha \, = \, - \, \frac{1}{4\pi} \, \int_{\mathbb{S}^2_{\alpha}} \, \epsilon_{ijk} dx^i \wedge dx^j \, \exp[-2 U] \, \left( \partial^k a \, + \, Z \, \mathbb{C} \, \partial^k Z\right) \, \equiv\, \frac{1}{4\pi} \, \int_{\mathbb{S}^2_{\alpha}} \, j^{TN}
\end{equation}
\subsubsection{Reduction to the spherical case}
The  spherical symmetric one-center solutions are retrieved from the general case by assuming that all the three-dimensional fields depend only on one radial coordinate:
\begin{equation}\label{tauvaria}
    \tau \, = \, - \, \frac{1}{r} \quad ; \quad r \, = \, \sqrt{x_1^2 + x_2^2 + x_3^2}
\end{equation}
On functions only of $\tau$ we have the identity:
\begin{equation}\label{giorgio}
    \partial_i f(\tau) \, = \, - x^i \, \tau^3 \, \frac{d}{d\tau}\, f(\tau)
\end{equation}
and introducing polar coordinates:
\begin{eqnarray}
  x_1 &=&\frac{1}{\tau} \, \cos \, \theta \nonumber \\
  x_2 &=& \frac{1}{\tau} \, \sin\, \theta \, \sin \varphi \nonumber  \\
   x_3 &=& \frac{1}{\tau} \, \sin\, \theta \, \cos \varphi
\end{eqnarray}
we obtain:
\begin{equation}\label{volumetto}
   \, \tau^3 \epsilon_{ijk} \, {x^i dx^j \wedge dx^k} \, = \, - \, 2 \, \sin \theta \, d\theta \wedge d\varphi
\end{equation}
By using these identities and restricting one's attention to the extremal case, the action of the $\sigma$-model (\ref{beddamatre}) reduces to:
\begin{eqnarray}\label{fortedeichiodi}
    \mathcal{A} & = & \int \, d\tau \, \mathcal{L} \nonumber\\
    \mathcal{L} & = & \dot{U}^2+\,h_{rs}\,\dot{\varphi}^r\,\dot{\varphi}^s
+ e^{-2\,U}\,(\dot{a}+{\bf Z}^T\mathbb{C}\dot{{\bf
Z}})^2\,+\,2 \, e^{-\,U}\,\dot{{\bf
Z}}^T\,\mathcal{M}_4\,\dot{{\bf Z}}
\end{eqnarray}
where the dot denotes derivatives with respect to the $\tau$ variable. The $\sigma$-model field equations take the standard form of the Euler Lagrangian equations:
\begin{eqnarray}\label{eulag}
     \frac{\mathrm{d}}{\mathrm{d}\tau} \frac{\mathrm{\mathrm{d}} \mathcal{L}}{\mathrm{d} \dot{\Phi}}&=& \, \frac{\mathrm{d} \mathcal{L}}{\mathrm{d} \Phi}\nonumber\\
\end{eqnarray}
and the extremality conditions (\ref{forbice}) reduces to:
\begin{equation}\label{beddopatre}
\mathcal{L} \, = \,    \dot{U}^2+\,h_{rs}\,\dot{\varphi}^r\,\dot{\varphi}^s
+ e^{-2\,U}\,(\dot{a}+{\bf Z}^T\mathbb{C}\dot{{\bf
Z}})^2\,+\,2 \, e^{-\,U}\,\dot{{\bf
Z}}^T\,\mathcal{M}_4\,\dot{{\bf Z}} \, = \, 0
\end{equation}
It appears from this that spherical extremal black holes are in one-to-one correspondence with light-like geodesics of the manifold $\mathcal{Q}$.
\paragraph{The reduced  oxidation rules}
In the spherical case the above discussed oxidation rules reduce as follows. For the metric we have
\begin{equation}\label{metricona}
    ds_{(4)}^2 \, = \, - \, e^{U(\tau)} \, \left(dt + \, 2\,\mathbf{n}\, \cos\theta \, d\varphi\right)^2 \, + \, e^{-\, U(\tau)}\left[ \frac{1}{\tau^4} \, d\tau^2 \, + \,
    \frac{1}{\tau^2}\,\left( d\theta^2 + \sin^2\theta \, d\phi^2\right)\right]
\end{equation}
where $\mathbf{n}$ denotes the Taub-NUT charge obtained from the form of the Kaluza-Klein field strength:
\begin{eqnarray}\label{fildaforza}
    \mathbf{F}^{KK} &  = & - 2\, \mathbf{n} \, \sin\theta \, d\theta \, \wedge \, d\varphi \nonumber\\
    \mathbf{n} & = & \left(\dot a \, + \, Z \, \mathbb{C} \, \dot{Z}\right)
\end{eqnarray}
\par
The electromagnetic field-strengths are instead the following ones:
\begin{equation}\label{finalfildone}
    F^\Lambda \, = \, 2 \, p^\Lambda \, \sin\theta \, d\theta \wedge d\varphi \, + \, \dot{Z}_\Lambda d\tau \wedge
    \left(dt + 2\mathbf{n} \, \cos\theta \, d\varphi\right)
\end{equation}
where the magnetic charges $p^\Lambda$ are extracted from the reduction of the general formula (\ref{cane}), namely:
\begin{equation}\label{cariche}
    \mathcal{Q}^M  \, = \, \left(\begin{array}{c}
                                                                p^\Lambda \\
                                                                q_\Sigma
                                                              \end{array}
    \right)\, = \, \sqrt{2} \,\left[ e^{-\,U} \, \mathcal{M}_4 \, \dot{Z} \, - \, \mathbf{n} \, \mathbb{C} \, Z\right]^M
\end{equation}
\subsection{A counter example: the extremal Kerr metric}
In this section, in order to better clarify the notion of extremality provided by conditions (\ref{rutillo}-\ref{forbice}) we consider the physically relevant counter-example  of the extremal Kerr metric. Such static solution of Einstein equations is certainly encoded in the $\sigma$-model approach yet it is not extremal in the sense of eq.s (\ref{rutillo}-\ref{forbice}) and therefore it is not related to any nilpotent orbit. Indeed the extremal Kerr metric is a solution of pure gravity and as such its $\sigma$-model representation lies in the euclidian submanifold:
\begin{equation}\label{gravitaziu}
    \frac{\mathrm{SL(2,\mathbb{R})}}{\mathrm{O(2)}}
\end{equation}
for which the coset tangent space $\mathbb{K}$ contains no nilpotent elements.
\par
Instead the so named BPS Kerr-Newman metric, which is not extremal in the sense of General Relativity and actually displays a naked singularity, is extremal in the sense of eq.s (\ref{rutillo}-\ref{forbice}) and can be retrieved in one of the nilpotent orbits of the $S^3$-model. We will show that explicitly in section \ref{kerronumano}.
\par
As a preparation to such discussions let us recall the general form of the Kerr-Newmann metric which we represent in polar coordinates as it follows:
\begin{eqnarray}
  ds_{KN}^2 &=& -V^0 \otimes V^0 + \sum_{i=1}^3 \, V^i \otimes V^i \label{KNmetra}\\
  V^0 &=& \frac{\delta (r)}{\sigma (r,\theta )}\, \left( {dt}\, - \, \alpha  \sin ^2\theta  \, {d \phi  }\right) \\
V^1&=& \frac{ \sigma (r,\theta )}{\delta (r)} \, dr\\
V^2 &=&  \sigma (r,\theta ) \, {d\theta }\\
V^3 &=& \frac{ \sin (\theta )}{\sigma (r,\theta )}\,\left(\left(r^2+\alpha ^2\right) \, {d\phi } \, -\,  \alpha \, {dt}\right) \\
\delta(r) &=& \sqrt{q^2+r^2+\alpha ^2-2 m r} \\
\sigma(r,\theta) &=& \sqrt{r^2+\alpha ^2 \cos ^2(\theta )}
\end{eqnarray}
Parameters of the Kerr-Newman solution are the mass $m$, the electric charge $q$ and the angular momentum $J=m \, \alpha$ of the Black Hole.
The two particular cases we shall consider in this paper correspond to:
\begin{description}
  \item[a)] The extremal Kerr solution: $q=0$ and $m=\alpha$.
  \item[b)] The BPS Kerr-Newman solution $q=m$, arbitrary $\alpha$.
\end{description}
Let us then focus now on the extremal Kerr solution. With the choice $m=\alpha$, $q=0$, the metric (\ref{KNmetra}) can be rewritten in the following form:
\begin{eqnarray}\label{gongolo}
    ds^2_{EK} \, = \, - \, \exp[U]\,  \left( dt + \mathbf{A}^{[KK]} \right )^2 \, + \, \exp[-U]\, \gamma_{ij} \, dy^i \otimes dy^j
\end{eqnarray}
where $y^i \, = \, \{r,\theta,\phi\}$ are the polar coordinates, the three dimensional metric $\gamma_{ij}$ is the following one:
\begin{equation}\label{gammametra}
    \gamma_{ij} \, = \, \left(
\begin{array}{lll}
 \frac{2 r^2-\alpha ^2+\alpha ^2 \cos (2 \theta )}{2 r^2} & 0 & 0 \\
 0 & r^2-\frac{\alpha ^2}{2}+\frac{1}{2} \alpha ^2 \cos (2 \theta ) & 0 \\
 0 & 0 & r^2 \sin ^2(\theta )
\end{array}
\right)
\end{equation}
the warp factor is:
\begin{equation}\label{ziatosa}
  U \, = \,  \log\left[ \frac{r^2-\alpha ^2 \sin ^2(\theta )}{(r+\alpha )^2+\alpha ^2 \cos ^2(\theta )}\right]
\end{equation}
and the Kaluza Klein vector has the following appearance:
\begin{equation}\label{ziotonto}
    \mathbf{A}^{[KK]} \, = \, \frac{2 \alpha ^2 (r+\alpha ) \sin ^2(\theta )}{r^2-\alpha ^2 \sin ^2(\theta )}\,  d\phi
\end{equation}
In presence of the metric $\gamma_{ij}$ the duality relation between the Kaluza Klein vector field and the $\sigma$-model scalar field $a$ reads as follows: \begin{equation}\label{formicoliere}
    \mathbf{F}_{ij}^{[KK]} \, \equiv \, \partial_{[i}\mathbf{A}^{[KK]}_{j]} \, = \, \exp[-2U] \, \sqrt{\mbox{det} \,\gamma} \, \epsilon_{ijk} \, \gamma^{k\ell} \, \partial_\ell \, a
\end{equation}
and it is solved by:
\begin{equation}\label{girotullo}
    a \, = \, -\frac{2 \alpha ^2 \cos (\theta )}{2 r^2+4 \alpha  r+3 \alpha ^2+\alpha ^2 \cos (2 \theta )}
\end{equation}
In this way, by means of inverse engineering we have showed how the extremal Kerr metric is retrieved in the $\sigma$-model approach. The crucial point is that the metric $\gamma_{ij}$ is not flat and hence such a configuration of the $U,a$ fields does not correspond to an extremal solution of the $\sigma$-model field equations. Indeed calculating the curvature two-form of the three-dimensional metric (\ref{gammametra}) we find
\begin{eqnarray}
  \mathfrak{R}^{12} &=& \frac{4 \alpha ^2 \left(2 r^2+\alpha ^2-\alpha ^2 \cos (2 \theta )\right) }{\left(2 r^2-\alpha ^2+\alpha ^2 \cos (2 \theta )\right)^3} \, e^1\wedge e^2\\
  \mathfrak{R}^{13}  &=& \frac{4 \alpha ^2 }{\left(2 r^2-\alpha ^2+\alpha ^2 \cos (2 \theta )\right)^2} \, e^1\wedge e^3\\
  \mathfrak{R}^{23}  &=& -\frac{4 \alpha ^2 }{\left(2 r^2-\alpha ^2+\alpha ^2 \cos (2 \theta )\right)^2} \, e^2\wedge e^3
\end{eqnarray}
where
\begin{eqnarray}
  e^1 &=& \frac{{dr} \, \sqrt{\frac{\cos (2 \theta ) \alpha ^2}{r^2}-\frac{\alpha ^2}{r^2}+2}}{\sqrt{2}} \\
  e^2 &=& {d\theta } \, \sqrt{r^2-\frac{\alpha ^2}{2}+\frac{1}{2} \alpha ^2 \cos (2 \theta )} \\
  e^3 &=& {d\phi } \, r \sin (\theta )
\end{eqnarray}
is the \textit{dreibein} corresponding to  (\ref{gammametra}).
\par
Hopefully this explicit calculation should have convinced the reader that the extremal Kerr solution and, by the same token, also the extremal Kerr-Newman solution are not extremal in the $\sigma$-model sense and are retrieved in regular rather than in nilpotent orbits of  $\mathrm{U}/\mathrm{H}^\star$.
\section{Construction of multicenter solutions associated with nilpotent orbits}
\label{construo}
For spherically symmetric black holes the construction of solutions is associated with nilpotent orbits in the following way. A representative of the $\mathrm{H}^\star$ orbit is a standard triple $\{h,X,Y\}$ and hence an embedding of an $\slal(2,\mathbb{R})$ Lie algebra:
 \begin{equation}\label{gornayalavanda}
    \left [ h, X\right] \, = \,2 \,  X \quad ; \quad  \left [ h, Y\right] \, = \, -\, 2\, Y \quad ; \quad \left [ X, Y\right] \, = \, 2 \, h
 \end{equation}
  into $\mathbb{U}_{\mathrm{D=3}}$ in such a way that $h\in \mathbb{H}^\star$ and $X,Y \in \mathbb{K}^\star$. The nilpotent operator $X$ is identified with the Lax operator $L_0$ at euclidian time $\tau =0$ and the corresponding solution depending on $\tau$ is constructed by using the algorithm described in \cite{Fre:2009dg,Chemissany:2010zp,noig22}. In the multicenter approach of \cite{Bossard:2009my,Bossard:2009at,Bossard:2009bw,Bossard:2009mz,Bossard:2011kz,Bossard:2012ge} one utilizes the standard triple to single out a nilpotent subalgebra $\mathbb{N}$, as follows.
  One diagonalizes the adjoint action of the central element $h$ of the triple on the Lie Algebra $\mathbb{U}_{\mathrm{D=3}}$:
  \begin{equation}\label{toffolone}
    \left [ h \, , \, C_\mu \right ] \, = \, \mu \, C_\mu
  \end{equation}
 The set of all eigen-operators $C_\mu$ corresponding  to positive gradings $\mu \, > \, 0$ spans a subalgebra $\mathbb{N} \, \subset \mathbb{U}_{\mathrm{D=3}}$ which is necessarily nilpotent
 \begin{equation}\label{crutilonico}
    \mathbb{N} \, = \, \mbox{span}\left [ C_2 \, , \, C_3 \, , \,\dots \, , \, C_{max} \right ]
 \end{equation}
 Such a nilpotent subalgebra has an intersection $\mathbb{N}\bigcap\mathbb{K}^\star$ with the space $\mathbb{K}^\star$ which  is not empty since at least the operator $C_2=X$ is present by definition of a standard triple. The next steps of the construction are as follows.
\subsection{The coset representative in the symmetric gauge}
Given a basis $A^i$ of the space $\mathbb{N}_\mathbb{K} \, \equiv \, \mathbb{N}\bigcap\mathbb{K}^\star$, whose dimension we denote:
\begin{equation}\label{dimensiaNK}
    \ell \, \equiv \, \mbox{dim} \, \mathbb{N}_\mathbb{K}
\end{equation}
and a basis $B^\alpha$ of the subalgebra $\mathbb{N}_\mathbb{H} \, \equiv \, \mathbb{N}\bigcap\mathbb{H}^\star$,
whose dimension we denote
\begin{equation}\label{dimensiaNH}
    \mathfrak{m} \, \equiv \, \mbox{dim} \, \mathbb{N}_\mathbb{H}
\end{equation}
we can construct a map:
\begin{equation}\label{mippo}
    {\mathfrak{H}} \quad : \quad \mathbb{R}^3 \, \rightarrow \,  \mathbb{N}_\mathbb{K}
\end{equation}
by writing:
\begin{equation}\label{moppo}
    \mathbb{N}_\mathbb{K} \, \ni \, {\mathfrak{H}}(\vec{x}) \, = \, \sum_{i=1}^\ell \,\mathfrak{h}_i(\vec{x}) \, A^i
\end{equation}
By construction, the point dependent Lie algebra element ${\mathfrak{H}}(\vec{x})$ is  nilpotent of a certain maximal degree $\mathrm{d_n}$, so that its exponential map to the nilpotent group $\mathrm{N} \subset \mathrm{U_{D=3}}$ truncates to a finite sum:
\begin{equation}\label{girarigira}
    \mathcal{Y}(x) \, = \, \exp \left[ {\mathfrak{H}}(\vec{x})  \right ] \, = \,  \mathbf{1} \, + \, \sum_{a=1}^{\mathrm{d_n}} \, \frac{1}{a!} \, {\mathfrak{H}}^a(\vec{x})
\end{equation}
The above constructed object realizes an explicit $\vec{x}$-dependent coset representative from which we can
construct the Maurer Cartan left-invariant one form:
\begin{equation}\label{omegone}
    \Sigma \, = \, \mathcal{Y}^{-1}\partial_i \mathcal{Y} \, dx^i
\end{equation}
Next let us decompose $\Sigma$  along the $\mathbb{K}^\star$ subspace and the $\mathbb{H}^\star$ subalgebra, respectively. This is done by setting:
\begin{equation}\label{zio}
   \mathbf{P} \, = \, \mbox{Tr}(\Sigma\, K^A) K_A \quad ; \quad \Omega \, = \, \mbox{Tr}(\Sigma\, H^m)  H_m
\end{equation}
where $K_A$ and $H_m$ denote a basis of generators for the two considered subspaces, $K^A$ and $H^m$ being their duals:
\begin{equation}\label{nascoli}
    \mbox{Tr}(K^A \, K_B) \, =\, \delta^A_B \quad ; \quad \mbox{Tr}(H^m \, H_n)  \, = \, \delta^m_n \quad ; \quad \mbox{Tr}(K^A \, H_n) \, = \, 0
\end{equation}
Denoting:
\begin{equation}\label{ziotta}
    {\null^\star}\mathbf{P}\, \equiv \, \ft 12 \, \epsilon_{ijk}\, \delta^{im} \, \mathbf{P}_m \, dx^j \wedge dx^k
   \end{equation}
the Hodge-dual of the coset vielbein
\begin{equation}\label{nipotina}
     \mathbf{P} \, = \,  \mathbf{P}_m \, dx^m
\end{equation}
the field equations of the three dimensional $\sigma$-model reduce to the following one:
\begin{equation}\label{equatata}
    d{\null^\star}\mathbf{P} \, = \, \Omega\, \wedge \, {\null^\star}\mathbf{P} \, - \, \, {\null^\star}\mathbf{P} \, \wedge \, \Omega\,
\end{equation}
Actually, since $\mathbb{N}\subset \mathbb{U}_{D=3}$ forms a nilpotent subalgebra the constructed object $\mathcal{Y}$ realizes a map from the three-dimensional space to the much smaller coset manifold:
\begin{equation}\label{arsenico}
    \mathcal{Y} \quad : \quad \mathbb{R}^3 \, \rightarrow \, \frac{\mathrm{N}}{\mathrm{N}_\mathrm{H}}
\end{equation}
and due to the polynomial form of the coset representative the final equations of motion obtain a  triangular solvable form that we  describe here below. Since the algebra $\mathbb{N}$ is nilpotent, its derivative series terminates, namely we have:
\begin{equation}\label{deriseria}
    \mathbb{N} \, \supset \, \mathcal{D}\mathbb{N} \, \supset \, \dots \, \supset \,\mathcal{D}^{n}\mathbb{N} \, \supset \, \mathcal{D}^{n+1}\mathbb{N} \, = \, \mathbf{0}
\end{equation}
where at each step $\mathcal{D}^{i}\mathbb{N}$ is a proper subspace of $\mathcal{D}^{i-1}\mathbb{N}$. Correspondingly let us define:
\begin{equation}\label{rullo}
    \mathcal{D}^{i}\mathbb{N}_{\mathbb{K}} \, = \, \mathcal{D}^{i}\mathbb{N} \bigcap \mathbb{K}^\star
\end{equation}
the intersections of the derivative subalgebras with the $\mathbb{K}^\star$ subspace and let us introduce the complementary orthogonal subspaces:
\begin{equation}\label{complespazi}
    \mathcal{D}^{i}\mathbb{N}_{\mathbb{K}} \, = \, \mathbb{N}_{K}^{(i)} \, \oplus \, \mathcal{D}^{i+1}\mathbb{N}_{\mathbb{K}}
\end{equation}
This yields an orthogonal graded decomposition of the space $\mathbb{N}_{\mathbb{K}}$ of the following form:
\begin{equation}\label{ortodecongo}
   \mathbb{N}_{\mathbb{K}} \, = \, \bigoplus_{a=0}^{n} \, \mathbb{N}_{\mathbb{K}}^{(a)}
\end{equation}
The space $\mathbb{N}_{\mathbb{K}}^{(0)}$ contains those generators that cannot be produced by any commutator within the algebra, $\mathbb{N}_{\mathbb{K}}^{(1)}$ contains those generators that are produced in simple commutators, $\mathbb{N}_{\mathbb{K}}^{(2)}$ contains those that are produced in double commutators and so on. Let us name
\begin{equation}\label{cirillini}
    \ell_a \, = \, \mbox{dim} \, \mathbb{N}_{\mathbb{K}}^{(a)} \quad ; \quad \sum_{a}^n \ell_a \, = \, \ell
\end{equation}
Correspondingly we can arrange the $\ell$ functions $\mathfrak{h}_i(\vec{x})$ according to the graded decomposition (\ref{ortodecongo}), by writing:
\begin{equation}\label{fischiaincurva}
    {\mathfrak{H}}(\vec{x}) \, = \, \sum_{\alpha=0}^n \, \underbrace{\sum_{i=1}^{\ell_\alpha} \,\mathfrak{h}^{(\alpha)}_i(\vec{x}) \, A^i_\alpha}_{\in \, \mathbb{N}_{\mathbb{K}}^{(\alpha)}}
\end{equation}
and equations (\ref{equatata}) take the following triangular form:
\begin{eqnarray}
  \nabla^2 \mathfrak{h}^{(0)}_i  &=& 0 \nonumber\\
   \nabla^2 \mathfrak{h}^{(1)}_i   &=& {\mathfrak{F}}^{(1)}_i \left(\mathfrak{h}^{(0)},\nabla\mathfrak{h}^{(0)}\right) \nonumber\\
  \nabla^2 \mathfrak{h}^{(2)}_i   &=& {\mathfrak{F}}^{(2)}_i \left(\mathfrak{h}^{(0)},\nabla\mathfrak{h}^{(0)}, \mathfrak{h}^{(1)},\nabla\mathfrak{h}^{(1)}\right) \nonumber\\
  \dots &=& \dots \nonumber\\
\nabla^2 \mathfrak{h}^{(n)}_i   &=& {\mathfrak{F}}^{(n)}_i \left( \mathfrak{h}^{(0)},\nabla\mathfrak{h}^{(0)}, \mathfrak{h}^{(1)},\nabla\mathfrak{h}^{(1)}, \dots ,\mathfrak{h}^{(n-1)},\nabla\mathfrak{h}^{(n-1)}\right), \label{failbrodo}
\end{eqnarray}
where $\nabla^2$ denotes the three-dimensional Laplacian and at each level $\alpha$, by ${\mathfrak{F}}^{(\alpha)}_i (\dots)$ we denote an $\so(3)$ invariant polynomial of all the functions $h^{\beta}$  up to level $\alpha-1$ and of their derivatives.
\par
Therefore the first $\ell_0$ functions $\mathfrak{h}^{(0)}_i$ are just harmonic functions, while the higher ones satisfy Laplace equation with a source that is provided by the previously determined functions.
\subsection{Transformation to the solvable gauge}
\label{zurlina}
\par
Given the symmetric coset representative $\mathcal{Y}(\vec{x})$, parameterized by functions $\mathfrak{h}_i^{(\alpha)}(\vec{x})$ which satisfy the field equations (\ref{failbrodo}), in order to retrieve the corresponding supergravity fields satisfying supergravity field equations, we need to solve a technical, yet quite crucial problem. We need to construct a new upper triangular coset representative:
\begin{equation}\label{graffo}
    \mathbb{L}(\mathcal{Y}) \, = \, \left(\begin{array}{ccccc}
                                  L_{1,1}(\mathcal{Y}) & L_{1,2}(\mathcal{Y}) & \cdots &L_{1,n-1}(\mathcal{Y})  & L_{1,n}(\mathcal{Y}) \\
                                  0& L_{2,2}(\mathcal{Y})& \cdots & L_{2,n-1}(\mathcal{Y})  & L_{2,n}(\mathcal{Y})  \\
                                  0 & 0 & L_{3,3}(\mathcal{Y}) & \cdots & L_{3,n}(\mathcal{Y})  \\
                                  \vdots & \ldots & 0 &\cdots & \vdots \\
                                  0 & 0& \cdots & 0 & L_{3,n}(\mathcal{Y})
                                \end{array}
     \right)
 \end{equation}
which depends algebraically on the matrix entries of $\mathcal{Y}$ and satisfies the following equivalence condition
\begin{equation}\label{cribbio}
     \mathbb{L}(\mathcal{Y}) \, \mathcal{Q}(\mathcal{Y}) \, = \, \mathcal{Y} \quad ; \quad \mathcal{Q}(\mathcal{Y}) \, \in \, \mathrm{H}^\star
 \end{equation}
 where, as specified above, $\mathcal{Q}(\mathcal{Y})$ is a suitable element of the subgroup $\mathrm{H}^\star$.
 It should be stressed that in the existing literature, this  transition from the symmetric to the solvable gauge, which is compulsory in order to make the construction of the black hole solutions explicit, has been advocated, yet it has been to ad hoc procedures to be invented case by case.
 \par
 Actually a universal and very elegant solution of such a problem exists and was found, from a different perspective, by the authors of the present paper in \cite{Fre:2009dg},\cite{noiultimo},\cite{Chemissany:2009af,Chemissany:2010zp},\cite{noig22}. Indeed
 defining the following determinants:
\begin{equation}\label{DktN}
    \mathfrak{D}_{i}(\mathcal{Y}) \, := \, \mbox{Det} \, \left ( \begin{array}{ccc}
    \mathcal{Y}_{1,1}  & \dots & \mathcal{Y}_{1,i} \\
    \vdots & \vdots & \vdots \\
    \mathcal{Y}_{i,1}  & \dots & \mathcal{Y}_{i,i}
    \end{array}\right)  \, , \quad
    \mathfrak{D}_{0}(\mathcal{Y}):=1 \,
    \end{equation}
the matrix elements of the inverse of the upper triangular coset representative satisfying both equations (\ref{graffo}) and (\ref{cribbio}) are given by the following expressions:
\begin{eqnarray}
\left(\mathbb{L}(\mathcal{Y})^{-1}\right)_{ij}
 &\equiv&\frac{1}{\sqrt{\mathfrak{D}_i(\mathcal{Y})\mathfrak{D}_{i-1}(\mathcal{Y})}}\mathrm{Det}\left(\begin{array}{cccc} \mathcal{Y}_{1,1} &\dots
&\mathcal{Y}_{1,i-1} &
 \mathcal{Y}  _{1,j}\\
\vdots&\vdots&\vdots&\vdots\\
\mathcal{Y}_{i,1} &\dots &
\mathcal{Y}_{i,i-1} &  \mathcal{Y}  _{i,j}\\
\end{array}\right)\label{solutioncosetrepr}
\end{eqnarray}
 Equation (\ref{solutioncosetrepr}) provides a universal non-trivial and very elegant  solution to the gauge-change problem  and makes the entire construction based on harmonic functions truly algorithmic from the start to the very end.
\subsection{Extraction of the three dimensional scalar fields}
The result of the procedure described in the previous section is a triangular coset representative $\mathbb{L}(\mathfrak{h}^{(\alpha)}_i)$ whose entries are polynomial and square root of polynomials in the functions $\mathfrak{h}^{(\alpha)}_i(x)$. The extraction of the scalar fields $\{U(x),a(x),Z(x),\phi(x)\}$ can now be performed according to the rules already presented in \cite{noig22}, which we recall here for completeness.
\par
The general form of the solvable coset representative in terms of the fields is the following one:
\begin{equation}\label{genaforma}
    \mathbb{L}(\Phi) \, = \, \exp\left[ - a \, L_+^E \right]\, \exp\left[ \sqrt{2}\, Z^M \,  \mathcal{W}_M \right] \,
    \mathbb{L}_4(\phi) \, \exp\left[ U \, L_0^E \right]
\end{equation}
where $L_0^E,L_\pm^E$ are the generators of the Ehlers group and $\mathcal{W}^M \equiv W^{1M}$ are the generators in the $W$-representation, according to the general structure (\ref{genGD3pre}) of the $\mathbb{U}_{D=3}$ Lie algebra; furthermore $\mathbb{L}_4(\phi)$ is the coset representative of the $D=4$ scalar coset manifold immersed in the $\mathrm{U_{D=3}}$ group.
From this structure, identifying $\mathbb{L}(\Phi) \, = \, \mathbb{L}(\mathfrak{h}^{(\alpha)}_i)$ we deduce the following iterative procedure for the extraction of the relevant fields:
\par
First of all we can determine the warp factor $U$ by means of the following simple formula:
\begin{equation}\label{warpfactor}
    U (\mathfrak{h})\, = \, \log \left[\ft 12 \, \mbox{Tr} \left (\mathbb{L}(\mathfrak{h}) \, L_+^E \, \mathbb{L}^{-1}(\mathfrak{h})\, L_-^E \right)\right]
\end{equation}
Secondly we obtain the fields $\phi_i$  as follows. Defining the functionals
\begin{eqnarray}
  \Xi_i(\mathfrak{h}) &=&  \mbox{Tr} \left (\mathbb{L}^{-1}(\mathfrak{h})\, T_i \,\mathbb{L}(\mathfrak{\tau})\, \right)
   \label{fiytau}
\end{eqnarray}
from the form of the coset representative (\ref{genaforma}) it follows that $\Xi_i$ depend only on the $D=4$ scalar fields and, according to the explicit form of the $D=4$ coset, one can work out the scalar fields $\phi_i$.
\par
The knowledge of $U,\phi_i$  allows to define:
\begin{equation}\label{omtau}
    \Omega (\mathfrak{h})\, = \, \mathbb{L}(\mathfrak{h})\, \exp \left[ - \, U \, L_0^E \right] \, \mathbb{L}_4(\phi)^{-1}\,
\end{equation}
from which we extract the $Z^M$ fields by means of the following formula:
\begin{equation}\label{ZMtau}
    Z^M(\mathfrak{h}) \, = \, \frac{1}{2\sqrt{2}} \, \mbox{Tr} \left[ \Omega(\mathfrak{h})\, \mathcal{W}_M^T\right]
\end{equation}
where $T$ means transposed. Finally the knowledge of $Z^M(\mathfrak{h})$ allows to extract the $a$ field by means of the following trace:
\begin{equation}\label{atau}
    a(\mathfrak{h}) \, = \,-\, \ft 12 \mbox{Tr} \left[\Omega (\mathfrak{h}) \, \exp\left[-\sqrt{2} \, Z^M(\mathfrak{h}) \, \mathcal{W}_M\right]  \, L_+^E\right]
\end{equation}
\subsection{General properties of the black hole solutions and structure of  their poles}
\label{genepoles}
Having discussed the structure of supergravity solutions in terms of black-boxes that are a set of harmonic functions and of their descendants generated through the solution of the hierarchical equations (\ref{failbrodo}), it is appropriate to study the general form of the geometries one obtains in this way and the properties of the available harmonic functions.
\par
First of all, naming:
\begin{equation}\label{warp}
    {\mathfrak{W}}\,=\,\exp[U(x)]
\end{equation}
the warp factor that defines the $4$-dimensional metric (\ref{metruzza}), we would like to investigate the general properties of the corresponding geometries. For the case where the Kaluza-Klein monopole is zero $\mathbf{A}^{[KK]}\, = \,0$ we can write the general form of the curvature two-form of such spaces and therefore the intrinsic form of the Riemann tensor. Using the vielbein formalism introduced in eq.(\ref{vielbino}) we obtain:
\begin{eqnarray}
  \mathfrak{R}^{0i} &=& - \, \mathfrak{W} \, \nabla^i \nabla_k  \mathfrak{W} \,\, E^0 \, \wedge \, E^k \, - \,2 \, \nabla^i \mathfrak{W}
  \nabla_k \mathfrak{W} \,\, E^0 \, \wedge \, E^k \nonumber\\
  \mathfrak{R}^{ij} &=& - \,2\, \mathfrak{W} \, \nabla^{[i} \nabla_k  \mathfrak{W} \,\, E^{j]} \, \wedge \, E^k \, + \,\left( \nabla \mathfrak{W}\cdot\nabla \mathfrak{W}\right) \,
  \nabla_k \mathfrak{W} \,\, E^i \, \wedge \, E^j \label{riemanianotto}
\end{eqnarray}
where the derivatives used in the above equations are defined as follows. Let the flat metric in three dimension be described by a euclidian \textit{dreibein} $e^i$ such that:
\begin{eqnarray}\label{gomorra}
    ds^2_{flat} & = & \sum_{i=1}^3 \, e^i \, \otimes \, e^i\nonumber\\
    E^i & = & \frac{1}{\mathfrak{W}}\, e^i
\end{eqnarray}
then the total differential of the warp factor expanded along $e^i$ yields the derivatives $\nabla_k  \mathfrak{W}$, namely:
\begin{equation}\label{grandato}
    d \, \mathfrak{W} \, = \, \nabla_k  \mathfrak{W}\, e^k
\end{equation}
Next let us consider the general form of harmonic functions. These latter form a linear space since
 any linear combination of harmonic functions is still harmonic. There are  three types of building blocks that we can use:
 \begin{description}
   \item[a] Real center pole:
   \begin{equation}\label{realpole}
    \mathcal{H}_\alpha(\vec{x}) \, = \, \frac{1}{|\vec{x} \, - \, \vec{x}_\alpha |}
   \end{equation}
   \item[b] Real part of an imaginary center pole:
   \begin{equation}\label{realimpole}
    \mathcal{R}_\alpha (\vec{x}) \, = \, \mbox{Re} \left[\frac{1}{|\vec{x} \, - \, {\rm i} \,\vec{x}_\alpha |}\right]
   \end{equation}
   \item[c] Imaginary part of an imaginary center pole:
   \begin{equation}\label{realimpole}
    \mathcal{J}_\alpha (\vec{x}) \, = \, \mbox{Im} \left[\frac{1}{|\vec{x} \, - \, {\rm i} \,\vec{x}_\alpha |}\right]
   \end{equation}
 \end{description}
Hence the most general harmonic function can be written as the following sum:
\begin{equation}\label{rolanda}
    \mathrm{Harm}(\vec{x}) \, = \, h_\infty \, + \, \sum_{\alpha} \, \frac{p_\alpha}{|\vec{x} \, - \, \vec{x}_\alpha |}
    + \, \sum_{\beta} \, q_\beta \,\mbox{Re} \left[\frac{1}{|\vec{x} \, - \, {\rm i} \,\vec{x}_\beta |}\right] \,
    + \, \sum_{\gamma} \, k_\gamma \,\mbox{Im} \left[\frac{1}{|\vec{x} \, - \, {\rm i} \,\vec{x}_\gamma |}\right]
\end{equation}
where the constant $h_\infty$ is the boundary value of the harmonic function at infinity far from all the poles.
In order to study the behavior of $\mathrm{Harm}(\vec{x})$ in the vicinity of a real pole $\left(|\vec{x} \, - \, \vec{x}_\alpha |<< 1\right)$ it is convenient to adopt local polar coordinates:
\begin{eqnarray}
  x^1- x_\alpha^1&=& r \, \cos \, \theta \nonumber \\
  x^2- x_\alpha^2&=& r \, \sin  \, \theta \, \sin \, \phi\nonumber \\
  x^3- x_\alpha^3&=& r \, \sin  \, \theta \, \cos \, \phi\nonumber \\
\end{eqnarray}
In this coordinates the harmonic function is approximated by:
\begin{equation}\label{crucchis}
    \mathrm{Harm}(\vec{x}) \, \simeq \, h_\alpha \, + \, \frac{p_\alpha}{r}
\end{equation}
where the effective constant $h_\alpha$ encodes the finite part of the function contributed by all the other poles.
In polar coordinates the Laplacian operator on functions of $r$ becomes:
\begin{equation}\label{laplacino}
    \Delta \, = \, \frac{d^2}{dr^2} \, + \,  \frac{2}{r} \, \frac{d}{dr}
\end{equation}
The general outcome of the construction procedure outlined in the previous section is that the warp factor is the square root
of a rational function of $n$ harmonic functions, where $n \, =\, \mbox{dim} \,\mathbb{N}_{\mathbb{K}}$
 \begin{equation}\label{francioso}
    \mathfrak{W} (\vec{x})\, = \, \sqrt{ \frac{ \mathbb{P}\,\left(\widehat{\mathrm{Harm}}_1(\vec{x}), \,\dots ,\,\widehat{\mathrm{Harm}}_n(\vec{x})\right)}{\mathbb{\mathbb{Q}}\,\left(\widehat{\mathrm{Harm}}_1(\vec{x}), \,\dots ,\,\widehat{\mathrm{Harm}}_n(\vec{x})\right)}}
 \end{equation}
 where $\mathbb{P}$ and $\mathbb{Q}$ are two polynomials. By $\widehat{\mathrm{Harm}}_1(\vec{x})$ we denote both harmonic functions and their descendants generated by the hierarchical system (\ref{failbrodo}).
 For a given multicenter solution it is convenient to enumerate all the poles displayed by one or the other of the harmonic functions and in the vicinity of each of those poles we will have:
 \begin{equation}\label{crucchis}
    \widehat{\mathrm{Harm}}_i(\vec{x}) \, \simeq \,  \, \frac{p_i}{r^{m_i}}
\end{equation}
where $p_i \ne 0$ if the considered pole belongs to the considered function and it is zero otherwise. Furthermore if $\mathrm{Harm}_i(\vec{x}) $ is one of the level one harmonic function the exponent $m_i=1$. Otherwise it is bigger, but in any case
$m_i\ge 1$. Taking this into account the effective behavior of the warp factor will always be of the following form:
\begin{equation}\label{pornot}
    \mathfrak{W} (\vec{x})\, \simeq \, r^{\ell_\alpha} \, \sqrt{c_\alpha}
\end{equation}
where $\ell$ is some integer or half integer power (positive or negative) and $c_\alpha$ is a constant. In order for the pole to be a regular point of the solution, two conditions have to be satisfied:
\begin{enumerate}
  \item The constant $c_\alpha >0 $ must be positive so that the warp factor is real.
  \item The power $\ell_\alpha \ge 1$ so that the Riemann tensor does not diverge at the pole.
\end{enumerate}
The second condition follows from the form (\ref{riemanianotto}) of the Riemann tensor which implies that all of its components behave as:
\begin{equation}\label{gerarco}
    \mathfrak{R}^{ab}_{\phantom{ab}cd} \, \simeq \, r^{2\ell_\alpha -2} \, \times \, \mbox{const}
\end{equation}
Near the pole the metric behaves as follows:
\begin{equation}\label{cucuccu}
    ds^2 \, \simeq \, - \,\sqrt{c_\alpha}\,  r^{\ell_\alpha} dt^2 \, + \, \frac{1}{\sqrt{c_\alpha}}\, \frac{1}{r^{\ell_\alpha}} \,
    \left[ dr^2 \, + \, r^2 \left( d\theta^2 \, + \, \sin^2\theta \, d\phi^2\right) \right]
\end{equation}
In order for the pole to be an event horizon of finite or of vanishing area, we must have $2-\ell_\alpha >0$, so that the volume of the two-sphere described by $\left( d\theta^2 \, + \, \sin^2\theta \, d\phi^2\right) $ does not diverge. Hence for regular black holes we have only three possibilities:
\begin{equation}\label{grugnando}
    \underbrace{\ell_\alpha \, = \, 2}_{\mbox{Large Black Holes}} \quad ; \quad \underbrace{\ell_\alpha \, = \, \ft 32}_{\mbox{Small Black Holes}} \quad ; \quad \underbrace{\ell_\alpha \, = \, 1}_{\mbox{Very Small Black Holes}}
\end{equation}
When we are in the case of Large Black Holes, the near horizon geometry is approximated by that:
\begin{equation}\label{crinollo}
    \mathrm{AdS_2} \, \times \, \mathbb{S}^2
\end{equation}
\par
The case of the harmonic functions with an imaginary center requires a different treatment. Their near singularity behavior is best analyzed by using spheroidal coordinates.
\par
These are easily introduced by setting:
\begin{eqnarray}
  x^1 &=& \sqrt{r^2 +\alpha^2} \, \sin\,\theta \, \sin\phi \nonumber\\
  x^2 &=& \sqrt{r^2 +\alpha^2} \, \sin\,\theta \, \cos\phi \nonumber\\
  x^3 &=& r \cos\,\theta \, \label{spheroidal}
\end{eqnarray}
where $r,\theta,\phi$ are the new coordinates and $\alpha$ is a deformation parameter which represents the position of the center in the complex plane. In terms of these coordinates the flat euclidian three-dimensional metric takes  the following form:
\begin{equation}\label{flateuclimet}
  ds^2_{\mathbb{E}^3} \, = \,  d\Omega_{spheroidal}^2 \, \equiv \, \frac{\left(r^2+\alpha ^2 \cos ^2\theta
   \right) {dr}^2}{r^2+\alpha
   ^2}+ \left(r^2+\alpha
   ^2\right) \sin ^2\theta \, {d\phi }^2+\left(r^2+\alpha ^2 \cos ^2\theta
   \right)\,{d\theta
   }^2
\end{equation}
and the two harmonic functions that correspond to the real and imaginary part of a complex harmonic function with center on the imaginary $z$-axis at $\alpha$-distance from zero are:
\begin{eqnarray}
\label{frantaccio1}
  \mathcal{P}_\alpha(r,\theta) &=& \frac{r}{r^2+\alpha ^2 \cos ^2\theta } \\
  \mathcal{R}_\alpha(r,\theta) &=& \frac{\alpha  \cos \theta }{r^2+\alpha ^2
   \cos ^2\theta }
\end{eqnarray}
and the Hodge duals of their gradients, in spheroidal coordinates have the following form:
\begin{eqnarray}
  \star \, \nabla \mathcal{P}_\alpha &=& \frac{\sin \theta}{\left(r^2+\alpha ^2 \cos ^2\theta\right)^2} \, \left[2 \alpha ^2\, r \cos \theta\sin \theta\, {dr}\wedge {d\phi } +\left(r^2+\alpha ^2\right) \left(r^2-\alpha ^2 \cos ^2\theta
   \right) {d\theta }\wedge {d\phi }\right] \\
 \star \nabla \mathcal{R}_\alpha &=& \frac{\alpha  \sin \theta  }{\left(r^2+\alpha ^2 \cos ^2\theta \right)^2} \, \left[\left(\alpha ^2 \cos ^2\theta -r^2\right) \sin \theta  {dr}\wedge {d \phi }+2 r \left(r^2+\alpha ^2\right) \cos \theta
   {d\theta }\wedge {d\phi}\right] \nonumber\\
   \label{frantaccio2}
\end{eqnarray}

These are the building blocks we can use to construct Kerr-Newman like solutions and we shall outline a pair of examples in the sequel.
\section{The example of the $S^3$ model: classification of the nilpotent orbits}
\label{examplu1}
 As an illustration of the general procedure we explore the case of the $S^3$ model, leading to the $\mathrm{G_{2,2}}$ group in $D=3$. The spherical symmetric black hole solutions of this model where already discussed in a full-fledged way in \cite{noig22}, and in \cite{titsblackholes} the detailed classification of the corresponding nilpotent orbits was derived. Here we reconsider the same case from the point  of view of the multicenter construction outlined in the previous section, relying on the results of \cite{titsblackholes}.
According to that paper, for the case of the coset manifold\footnote{For the rationale of our notation we refer the reader to the later section \ref{hstarrostrucco}.}:
\begin{equation}\label{definog22}
    \frac{\mathrm{U_{D=3}}}{\mathrm{H}^\star} \, = \, \frac{\mathrm{G_{(2,2)}}}{\widehat{\mathrm{SL(2,\mathbb{R})}}\times \mathrm{SL(2,\mathbb{R})_{\mathrm{h}^\star}}}
\end{equation}
we have just seven distinct nilpotent orbits of the $\mathrm{H}^\star \, = \, \widehat{\mathrm{SL(2,\mathbb{R})}}\times \mathrm{SL(2,\mathbb{R})_{\mathrm{h}^\star}}$ subgroup in the $\mathbb{K}^\star$ representation $\left(2,\ft 32\right)$, which are enumerated by the three set of labels $\alpha\beta\gamma$ and are denoted $\mathcal{O}^\alpha_{\beta\gamma}$ as described in table \ref{tablettina} \footnote{For the notation and organization of the $\alpha\beta\gamma$ labels we refer the reader to \cite{titsblackholes}.}. An explicit choice of a representative for each of the seven orbits  is provided below.
{\begin{equation}\label{111}
 \mathcal{O}^1_{11} \, = \,  \left(
\begin{array}{lllllll}
 \sqrt{\frac{3}{2}} & \frac{\sqrt{\frac{5}{2}}}{2} & \sqrt{\frac{3}{2}} & \frac{\sqrt{5}}{2} & 0 & \frac{\sqrt{\frac{5}{2}}}{2} & 0 \\
 \frac{\sqrt{\frac{5}{2}}}{2} & \sqrt{6} & -\frac{\sqrt{\frac{5}{2}}}{2} & -\sqrt{3} & -\frac{\sqrt{\frac{5}{2}}}{2} & 0 & \frac{\sqrt{\frac{5}{2}}}{2} \\
 -\sqrt{\frac{3}{2}} & \frac{\sqrt{\frac{5}{2}}}{2} & -\sqrt{\frac{3}{2}} & \frac{\sqrt{5}}{2} & 0 & \frac{\sqrt{\frac{5}{2}}}{2} & 0 \\
 -\frac{\sqrt{5}}{2} & \sqrt{3} & \frac{\sqrt{5}}{2} & 0 & \frac{\sqrt{5}}{2} & -\sqrt{3} & -\frac{\sqrt{5}}{2} \\
 0 & \frac{\sqrt{\frac{5}{2}}}{2} & 0 & \frac{\sqrt{5}}{2} & \sqrt{\frac{3}{2}} & \frac{\sqrt{\frac{5}{2}}}{2} & \sqrt{\frac{3}{2}} \\
 \frac{\sqrt{\frac{5}{2}}}{2} & 0 & -\frac{\sqrt{\frac{5}{2}}}{2} & \sqrt{3} & -\frac{\sqrt{\frac{5}{2}}}{2} & -\sqrt{6} & \frac{\sqrt{\frac{5}{2}}}{2} \\
 0 & \frac{\sqrt{\frac{5}{2}}}{2} & 0 & \frac{\sqrt{5}}{2} & -\sqrt{\frac{3}{2}} & \frac{\sqrt{\frac{5}{2}}}{2} & -\sqrt{\frac{3}{2}}
\end{array}
\right)
\end{equation}}
\begin{equation}\label{411}
 \mathcal{O}^4_{11} \, = \,   \left(
\begin{array}{lllllll}
 \frac{1}{2} & 0 & 0 & 0 & \frac{1}{2} & 0 & 0 \\
 0 & 0 & 0 & 0 & 0 & 0 & 0 \\
 0 & 0 & \frac{1}{2} & 0 & 0 & 0 & \frac{1}{2} \\
 0 & 0 & 0 & 0 & 0 & 0 & 0 \\
 -\frac{1}{2} & 0 & 0 & 0 & -\frac{1}{2} & 0 & 0 \\
 0 & 0 & 0 & 0 & 0 & 0 & 0 \\
 0 & 0 & -\frac{1}{2} & 0 & 0 & 0 & -\frac{1}{2}
\end{array}
\right)
\end{equation}
\begin{equation}\label{211}
 \mathcal{O}^2_{11} \, = \,    \left(
\begin{array}{lllllll}
 \frac{1}{2} & 0 & -\frac{1}{2} & 0 & 0 & 0 & 0 \\
 0 & 1 & 0 & \frac{1}{\sqrt{2}} & 0 & 0 & 0 \\
 \frac{1}{2} & 0 & -\frac{1}{2} & 0 & 0 & 0 & 0 \\
 0 & -\frac{1}{\sqrt{2}} & 0 & 0 & 0 & \frac{1}{\sqrt{2}} & 0 \\
 0 & 0 & 0 & 0 & \frac{1}{2} & 0 & -\frac{1}{2} \\
 0 & 0 & 0 & -\frac{1}{\sqrt{2}} & 0 & -1 & 0 \\
 0 & 0 & 0 & 0 & \frac{1}{2} & 0 & -\frac{1}{2}
\end{array}
\right)
\end{equation}
{
\begin{equation}\label{311}
  \mathcal{O}^3_{11}  \, = \,  \left(
\begin{array}{lllllll}
 1 & 0 & 0 & -\frac{1}{\sqrt{2}} & 0 & 0 & 0 \\
 0 & 1 & -\frac{1}{2} & 0 & \frac{1}{2} & 0 & 0 \\
 0 & \frac{1}{2} & 0 & 0 & 0 & -\frac{1}{2} & 0 \\
 \frac{1}{\sqrt{2}} & 0 & 0 & 0 & 0 & 0 & \frac{1}{\sqrt{2}} \\
 0 & -\frac{1}{2} & 0 & 0 & 0 & \frac{1}{2} & 0 \\
 0 & 0 & \frac{1}{2} & 0 & -\frac{1}{2} & -1 & 0 \\
 0 & 0 & 0 & -\frac{1}{\sqrt{2}} & 0 & 0 & -1
\end{array}
\right)
\end{equation}}
{
\begin{equation}\label{322}
  \mathcal{O}^3_{22} \, = \, \left(
\begin{array}{lllllll}
 1 & 0 & 0 & \frac{1}{\sqrt{2}} & 0 & 0 & 0 \\
 0 & 1 & -\frac{1}{2} & 0 & -\frac{1}{2} & 0 & 0 \\
 0 & \frac{1}{2} & 0 & 0 & 0 & \frac{1}{2} & 0 \\
 -\frac{1}{\sqrt{2}} & 0 & 0 & 0 & 0 & 0 & -\frac{1}{\sqrt{2}} \\
 0 & \frac{1}{2} & 0 & 0 & 0 & \frac{1}{2} & 0 \\
 0 & 0 & -\frac{1}{2} & 0 & -\frac{1}{2} & -1 & 0 \\
 0 & 0 & 0 & \frac{1}{\sqrt{2}} & 0 & 0 & -1
\end{array}
\right)
\end{equation}
}
{
\begin{equation}\label{321}
\mathcal{O}^3_{21}  \, = \,  \left(
\begin{array}{lllllll}
 -1 & 0 & 0 & \frac{1}{\sqrt{2}} & 0 & 0 & 0 \\
 0 & 0 & -\frac{1}{2} & 0 & -\frac{1}{2} & 0 & 0 \\
 0 & \frac{1}{2} & -1 & 0 & 0 & \frac{1}{2} & 0 \\
 -\frac{1}{\sqrt{2}} & 0 & 0 & 0 & 0 & 0 & -\frac{1}{\sqrt{2}} \\
 0 & \frac{1}{2} & 0 & 0 & 1 & \frac{1}{2} & 0 \\
 0 & 0 & -\frac{1}{2} & 0 & -\frac{1}{2} & 0 & 0 \\
 0 & 0 & 0 & \frac{1}{\sqrt{2}} & 0 & 0 & 1
\end{array}
\right)
\end{equation}
}
{
\begin{equation}\label{312}
 \mathcal{O}^3_{12} \, = \,  \left(
\begin{array}{lllllll}
 -1 & 0 & 0 & -\frac{1}{\sqrt{2}} & 0 & 0 & 0 \\
 0 & 0 & -\frac{1}{2} & 0 & \frac{1}{2} & 0 & 0 \\
 0 & \frac{1}{2} & -1 & 0 & 0 & -\frac{1}{2} & 0 \\
 \frac{1}{\sqrt{2}} & 0 & 0 & 0 & 0 & 0 & \frac{1}{\sqrt{2}} \\
 0 & -\frac{1}{2} & 0 & 0 & 1 & \frac{1}{2} & 0 \\
 0 & 0 & \frac{1}{2} & 0 & -\frac{1}{2} & 0 & 0 \\
 0 & 0 & 0 & -\frac{1}{\sqrt{2}} & 0 & 0 & 1
\end{array}
\right)
\end{equation}
}

\begin{table}
\begin{center}
{\scriptsize
$$
\begin{array}{||l||c|l|c|c|l||}
\hline
\hline
N & d_n &\alpha -\mbox{label} & \gamma\beta -\mbox{labels} & \mbox{Orbits} & \mathcal{W}_H -\mbox{classes}\\
\hline
\hline
1 & 7 &\mbox{[j=3]} &\gamma\beta_1 = \left\{8_1 4_1 0_1\right\} & \mathcal{O}_1^1& \left(\times,\gamma_1,\times\right)\\
 \hline
2 & 3 &\mbox{[j=1]$\times $2[j=1/2]} & \gamma\beta_1 = \left\{3_1 1_1 0_1 \right\}&\mathcal{O}_1^2 & \left(\gamma_1,\gamma_1,\times\right)\\
   \hline
7 & 3 &\mbox{2[j=1]$\times $[j=0]} & \begin{array}{lll}
                                    \gamma\beta_1& = & \left\{4_1 0_2 \right\} \\
                                    \gamma\beta_2 & =&\left\{ 2_2 0_1\right\} \\
                                  \end{array}&\begin{array}{l|cc}
                                                 & \beta_1 & \beta_2 \\
                                                \hline
                                                \gamma_1 &\mathcal{ O}^3_{1,1}& \mathcal{ O}^3_{1,2}  \\
                                                \gamma_2 & \mathcal{ O}^3_{2,1}& \mathcal{ O}^3_{2,2} \\
                                              \end{array}
                                  & \left(\gamma_1,\gamma_2,\gamma_2\right)\\
   \hline
4 &2 &\mbox{2[j=1/2]$\times $3 [j=0]} & \gamma\beta_1=\left\{1_2 0_1\right\}
  &\mathcal{O}_{1}^4 & \left(0,\gamma_1,\gamma_1\right)\\
  \hline
   \hline
\end{array}
$$
}
\caption{Classification of the nilpotent orbits of $\frac{\mathrm{G_{(2,2)}}}{\mathrm{SU(1,1)}\times \mathrm{SU(1,1)}}$. \label{tablettina}}
\end{center}
\end{table}
Note that, in some instances, these representatives  do not coincide with the representatives shown in our previous papers \cite{noig22}, \cite{titsblackholes}. The reason is that in the approach pursued in the present paper, it is no longer relevant to consider representatives possessing vanishing Taub-NUT charges. The vanishing of the Taub-NUT current will be anyhow implemented on the parametrization of the symmetric coset representative associated with the nilpotent orbit. So we rather prefer to choose the simplest representatives of each nilpotent orbit postponing the issue of the Taub-NUT charge at a later stage.
\par
Each orbit representative $\mathcal{ O}^\alpha_{\beta\gamma}$ identifies a standard triple $\{h,X,Y\}$ and hence an embedding of an $\slal(2,\mathbb{R})$ Lie algebra:
 \begin{equation}\label{gornayalavanda}
    \left [ h, X\right] \, = \,2 \,  X \quad ; \quad  \left [ h, Y\right] \, = \, -\, 2\, Y \quad ; \quad \left [ X, Y\right] \, = \, 2 \, h
 \end{equation}
  into $\mathfrak{g}_{(2,2)}$ in such a way that $h\in \mathbb{H}^\star$ and $X,Y \in \mathbb{K}^\star$. The triple is obtained by setting:
  \begin{equation}\label{cringollo}
    X_{\alpha|\beta\gamma} \, \equiv \, \mathcal{ O}^\alpha_{\beta\gamma} \quad ; \quad Y_{\alpha|\beta\gamma} \, \equiv \, X_{\alpha|\beta\gamma}^T \quad ; \quad  h_{\alpha|\beta\gamma}  \, \equiv \, \left [ X_{\alpha|\beta\gamma}, Y_{\alpha|\beta\gamma}\right]
  \end{equation}
The relevant item in the construction of solutions based on the integration of equations in the symmetric gauge is provided by the central element of the triple $ h_{\alpha|\beta\gamma}$ which defines the gradings. In the present example of the $S^3$ model, it turns out  the orbits having the same $\alpha$ and $ \gamma$ labels but different $\beta$-labels have the same central element, namely:
\begin{equation}\label{guinno}
    h_{\alpha|\beta\gamma}  \, = \, h_{\alpha|\beta^\prime\gamma}
\end{equation}
so that the solutions pertaining both to orbit $\mathcal{ O}^\alpha_{\beta\gamma}$ and to orbit $\mathcal{ O}^\alpha_{\beta^\prime\gamma}$ are obtained from the same construction and are distinguished only by different choices in the space of the available harmonic functions parameterizing the general solution.
\par
The explicit form of the central elements are the following ones:
\paragraph{Large Orbit $\mathcal{ O}^1_{11} $: central element}
\begin{eqnarray}\label{h111}
    h_{1|11}& =& \left(
\begin{array}{lllllll}
 0 & 0 & -1 & 0 & 5 & 0 & 0 \\
 0 & 0 & 0 & \sqrt{2} & 0 & 0 & 0 \\
 -1 & 0 & 0 & 0 & 0 & 0 & 5 \\
 0 & \sqrt{2} & 0 & 0 & 0 & \sqrt{2} & 0 \\
 5 & 0 & 0 & 0 & 0 & 0 & -1 \\
 0 & 0 & 0 & \sqrt{2} & 0 & 0 & 0 \\
 0 & 0 & 5 & 0 & -1 & 0 & 0
\end{array}
\right) \nonumber\\
\mbox{Eigenvalues} \, \left [\ft 12 \,  h_{1|11} \right ]& = & \{-3,3,-2,2,-1,1,0\}
\end{eqnarray}
\paragraph{Very Small Orbit $\mathcal{ O}^4_{11} $: central element}
\begin{eqnarray}\label{h411}
    h_{4|11}& =& \left(
\begin{array}{lllllll}
 0 & 0 & 0 & 0 & -1 & 0 & 0 \\
 0 & 0 & 0 & 0 & 0 & 0 & 0 \\
 0 & 0 & 0 & 0 & 0 & 0 & -1 \\
 0 & 0 & 0 & 0 & 0 & 0 & 0 \\
 -1 & 0 & 0 & 0 & 0 & 0 & 0 \\
 0 & 0 & 0 & 0 & 0 & 0 & 0 \\
 0 & 0 & -1 & 0 & 0 & 0 & 0
\end{array}
\right) \nonumber\\
\mbox{Eigenvalues} \, \left [\ft 12 \,  h_{4|11} \right ]& = &\left\{-\frac{1}{2},-\frac{1}{2},\frac{1}{2},\frac{1}{2},0,0,0\right\}
\end{eqnarray}
\paragraph{Small Orbit $\mathcal{ O}^2_{11} $: central element}
\begin{eqnarray}\label{h211}
    h_{2|11}& =&\left(
\begin{array}{lllllll}
 0 & 0 & 1 & 0 & 0 & 0 & 0 \\
 0 & 0 & 0 & -\sqrt{2} & 0 & 0 & 0 \\
 1 & 0 & 0 & 0 & 0 & 0 & 0 \\
 0 & -\sqrt{2} & 0 & 0 & 0 & -\sqrt{2} & 0 \\
 0 & 0 & 0 & 0 & 0 & 0 & 1 \\
 0 & 0 & 0 & -\sqrt{2} & 0 & 0 & 0 \\
 0 & 0 & 0 & 0 & 1 & 0 & 0
\end{array}
\right) \nonumber\\
\mbox{Eigenvalues} \, \left [\ft 12 \,  h_{2|11} \right ]& = &\left\{-1,1,-\frac{1}{2},-\frac{1}{2},\frac{1}{2},\frac{1}{2},0\right\}
\end{eqnarray}
\paragraph{Large BPS Orbit $\mathcal{ O}^3_{11} $: central element}
\begin{eqnarray}\label{h311}
    h_{3|11}\, = \, h_{3|21}& =&\left(
\begin{array}{lllllll}
 0 & 0 & 0 & \sqrt{2} & 0 & 0 & 0 \\
 0 & 0 & 1 & 0 & -1 & 0 & 0 \\
 0 & 1 & 0 & 0 & 0 & 1 & 0 \\
 \sqrt{2} & 0 & 0 & 0 & 0 & 0 & -\sqrt{2} \\
 0 & -1 & 0 & 0 & 0 & -1 & 0 \\
 0 & 0 & 1 & 0 & -1 & 0 & 0 \\
 0 & 0 & 0 & -\sqrt{2} & 0 & 0 & 0
\end{array}
\right)\nonumber\\
\mbox{Eigenvalues} \, \left [\ft 12 \,  h_{3|11} \right ]& = &\{-1,-1,1,1,0,0,0\}
\end{eqnarray}
\paragraph{Large non BPS Orbit $\mathcal{ O}^3_{22} $: central element}
\begin{eqnarray}\label{h311}
    h_{3|12}\, = \, h_{3|22}& =&\left(
\begin{array}{lllllll}
 0 & 0 & 0 & -\sqrt{2} & 0 & 0 & 0 \\
 0 & 0 & 1 & 0 & 1 & 0 & 0 \\
 0 & 1 & 0 & 0 & 0 & -1 & 0 \\
 -\sqrt{2} & 0 & 0 & 0 & 0 & 0 & \sqrt{2} \\
 0 & 1 & 0 & 0 & 0 & -1 & 0 \\
 0 & 0 & -1 & 0 & -1 & 0 & 0 \\
 0 & 0 & 0 & \sqrt{2} & 0 & 0 & 0
\end{array}
\right)\nonumber\\
\mbox{Eigenvalues} \, \left [\ft 12 \,  h_{3|22} \right ]& = &\{-1,-1,1,1,0,0,0\}
\end{eqnarray}
\section{Explicit construction of the multicenter Black Holes solutions of the $S^3$ model}
\label{examplu2}
Having enumerated the central elements for the independent orbits we proceed to the construction and discussion of the corresponding black hole solutions, whose properties are summarized in table (\ref{tablettina2}).
\begin{table}
\begin{center}
$$
\begin{array}{||c||c|c|c|c|c|c|}
\hline
\hline
\mbox{Name} & \mbox{pq} &\mbox{Quart. Inv.}&\mathbf{W}-\mbox{stab. group}& \mathrm{H}^\star-\mbox{stab. group}& \mbox{dim} \,  &\mbox{dim} \, \\
\mbox{of orbit} & \mbox{charges}&\mathfrak{I}_4  &\mathcal{S}_\mathbf{W} \subset \slal(2,\mathbb{R})&\mathfrak{S}_{\mathrm{H}^\star}\subset \widehat{\slal(2,\mathbb{R})}\oplus\slal(2,\mathbb{R})_{\mathrm{h}^\star}&\mathbb{N} & {\mathbb{N}}\bigcap \mathbb{K}^\star \\
\hline
\hline
\mathcal{O}^4_{11} &\left(\begin{array}{c}
                            0 \\
                            0 \\
                            0 \\
                            q
                          \end{array}
\right) & 0 & \left(\begin{array}{cc}
                      1 & 0 \\
                      c & 1
                    \end{array}
\right)& \underbrace{\mathrm{ISO(1,1)}}_{\mbox{3 gen.}} & 3& 3 \\
\hline
\mathcal{O}^2_{11} &\left(\begin{array}{c}
                            \sqrt{3} \, p \\
                            0 \\
                            0 \\
                            0
                          \end{array}
\right) & 0 & \mathbf{1}& \underbrace{\mathrm{SO(1,1)}\triangleright \mathbb{R}}_{\mbox {2 gen.}} & 4& 3 \\
\hline
\mathcal{O}^3_{11} &\left(\begin{array}{c}
                            0 \\
                            p\\
                            -\sqrt{3} q \\
                            0
                          \end{array}
\right) & 9\,p\,q^3 \, > \, 0 &\mathbb{Z}_3 & \underbrace{\mathbb{R}}_{\mbox {1 gen. $A^2=0$}} & 5& 4 \\
\hline
\mathcal{O}^3_{22} &\left(\begin{array}{c}
                            0 \\
                            p\\
                            \sqrt{3} q \\
                            0
                          \end{array}
\right) & - 9\,p\,q^3 \,< \, 0 &\mathbf{1} & \underbrace{\mathbb{R}}_{\mbox {1 gen. $A^3=0$}} & 3& 3 \\
\hline
\mathcal{O}^1_{11} &\left(\begin{array}{c}
                            \ft 12 \sqrt{\ft 32} p \\
                            0\\
                            \ft 76 p \\
                            \sqrt{2} q
                          \end{array}
\right) & \begin{array}{c}
            \ft {1}{128} p^3 \times \\
            (49 p +72 q)
          \end{array}
 & \mathbf{1} & \mathbf{1} &6& 4 \\
\hline
\end{array}
$$
\caption{Properties of the $\mathfrak{g}_{(2,2)}$ orbits in the $S^3$ model. The structure of the electromagnetic charge vector is that obtained for solutions with vanishing Taub-NUT current. The symbol $\triangleright$ is meant to denote semidirect product. $\mathcal{S}_\mathbf{W}$  denotes the subgroup of the $D=4$ duality group which leaves the charge vector invariant, while $\mathfrak{S}_{\mathrm{H}^\star}$ denotes the subgroup of the $\mathrm{H}^\star$ isotropy group of the $D=3$ sigma-model which leaves invariant the $X$ element of the standard triple. This latter is the Lax operator in the one-dimensional spherical symmetric approach.\label{tablettina2}}
\end{center}
\end{table}
\subsection{The Very Small Black Holes of $\mathcal{O}^4_{11}$}
We begin with the smallest orbits which, in a sense that will become clear further on, represent the elementary blocks in terms of which bigger black holes are constructed.
\par
Focusing on any orbit $\mathcal{O}^\alpha_{\beta\gamma}$ and considering the nilpotent element of the corresponding triple $X_{\alpha|\beta\gamma}\in \mathbb{K}^\star$ as a Lax operator $L_0$, we easily workout the electromagnetic charges by calculating  the  traces displayed below (see section \ref{KWseczia}, for more explanations)
\begin{equation}\label{chargiotte}
  \mathcal{Q}^\mathbf{w} \, = \, \mathrm{Tr}(X_{\alpha|\beta\gamma}\mathcal{T}^{\mathbf{w}} )
\end{equation}
\paragraph{$\mathbf{W}$-representation}
In the case of the orbit $\mathcal{O}^4_{11}$ we obtain:
\begin{equation}\label{chiavina}
   \mathcal{Q}^\mathbf{w}_{4|11} \, = \, \, = \, \left(0,0,0,1\right)
\end{equation}
Substituting such a result in the expression for the quartic symplectic invariant  (see \cite{noig22}):
\begin{equation}\label{invola}
   \mathfrak{I}_4 \, = \, \frac{1}{4} \left(4 \sqrt{3} Q_4 Q_1^3+3 Q_3^2 Q_1^2-18 Q_2 Q_3 Q_4 Q_1-Q_2 \left(4 \sqrt{3} Q_3^3+9 Q_2 Q_4^2\right)\right)
\end{equation}
of the $\mathbf{W}$ representation which happens to be the spin $\ft 32$ of $\slal(2,\mathbb{R})$ we find:
\begin{equation}\label{genorama}
    \mathfrak{I}_4 \, = \, 0
\end{equation}
The result is meaningful since, by calculating the trace $\mbox{Tr}(X_{4|11} L_+^E) \, = \,0$, we can also check that the Taub-NUT charge vanishes.
We can also address the question whether there are subgroups of the original duality group in four-dimensions $\mathrm{SL(2,\mathrm{R})}$ that leave the charge vector (\ref{chiavina}) invariant. Using the explicit form of the $j=\ft 32$ representation displayed in eq.(3.13) of \cite{noig22}, we realize that indeed such group exists and it is the parabolic subgroup described below:
\begin{equation}\label{parabolico}
   \forall \, c \, \in \, \mathbb{R} \quad : \quad  \left (\begin{array}{cc}
             1 & 0 \\
             c & 1
           \end{array}
     \right) \, \in \, \mathcal{S}_{4|11} \, \subset \, \mathrm{SL(2,\mathbb{R})}
\end{equation}
This stability subgroup together with the vanishing of the quartic invariant are the intrinsic definition of the $\mathbf{W}$-orbit pertaining to very small black holes.
\par
\paragraph{$\mathrm{H}^\star$-stability subgroup} In a parallel way we can pose the question what is the stability subgroup of the nilpotent element $X_{4|11}$ in $\mathrm{H}^\star \, = \, \widehat{\slal(2,\mathbb{R})}\oplus\slal(2,\mathbb{R})_{\mathrm{h}^\star}$ (For further explanations on $\mathrm{H}^\star$ and its structure see section \ref{hstarrostrucco}). The answer is the following:
\begin{equation}\label{gruppoS411}
    \mathfrak{S}_{4|11}  \, = \, \mathrm{ISO(1,1)}
\end{equation}
A generic element of the corresponding  Lie algebra is a linear combination of three generators $J,T_1,T_2$,
satisfying the commutation relations:
\begin{eqnarray}\label{gigio}
   \left[J \, , \, T_1\right] & = &  \frac{1}{\sqrt{2}} \, T_1 \, + \, \frac{3}{2\sqrt{6}} \, T_2 \nonumber\\
   \left[J \, , \, T_2\right] & = &  \frac{3}{2\sqrt{2}} \, T_1 \quad ; \quad  \left[T_1 \, , \, T_2\right]\, = \, 0 \nonumber\\
\end{eqnarray}
It is explicitly given by the following matrix:
\begin{equation}\label{formagruppo}
   \omega \, J \, + \, x \,T_1 \, + \, y \, T_2 \, = \,  \left(
\begin{array}{lllllll}
 0 & -\frac{x}{2 \sqrt{2}} & \frac{\omega }{2 \sqrt{2}} & -\frac{x}{2} & 0 & -\frac{1}{2} \sqrt{\frac{3}{2}} y & 0 \\
 \frac{x}{2 \sqrt{2}} & 0 & -\frac{1}{2} \sqrt{\frac{3}{2}} y & -\frac{\omega }{2} & \frac{x}{2 \sqrt{2}} & 0 & -\frac{1}{2} \sqrt{\frac{3}{2}} y \\
 \frac{\omega }{2 \sqrt{2}} & -\frac{1}{2} \sqrt{\frac{3}{2}} y & 0 & -\frac{x}{2} & 0 & -\frac{x}{2 \sqrt{2}} & 0 \\
 -\frac{x}{2} & -\frac{\omega }{2} & \frac{x}{2} & 0 & -\frac{x}{2} & -\frac{\omega }{2} & \frac{x}{2} \\
 0 & \frac{x}{2 \sqrt{2}} & 0 & \frac{x}{2} & 0 & \frac{1}{2} \sqrt{\frac{3}{2}} y & \frac{\omega }{2 \sqrt{2}} \\
 \frac{1}{2} \sqrt{\frac{3}{2}} y & 0 & -\frac{x}{2 \sqrt{2}} & -\frac{\omega }{2} & \frac{1}{2} \sqrt{\frac{3}{2}} y & 0 & -\frac{x}{2 \sqrt{2}} \\
 0 & \frac{1}{2} \sqrt{\frac{3}{2}} y & 0 & \frac{x}{2} & \frac{\omega }{2 \sqrt{2}} & \frac{x}{2 \sqrt{2}} & 0
\end{array}
\right)
\end{equation}
\paragraph{Nilpotent algebra $\mathbb{N}_{4|11}$ } Considering next the adjoint action of the central element $h_{4|11}$ on the subspace $\mathbb{K}^\star$ we find that its eigenvalues are the following ones:
\begin{equation}\label{eigetti}
    \mbox{Eigenvalues}_{4|11}^{\mathbb{K}^\star}\, = \, \left\{ -2,2,-1,-1,1,1,0,0\right\}
\end{equation}
Therefore the three eigenoperators $A_1,A_2,A_3$ corresponding to the positive eigenvalues $2,1,1$, respectively, form the restriction to $\mathbb{K}^\star$ of a nilpotent algebra $\mathbb{N}_{4|11}$. In this case $A_i$ commute among themselves so that $\mathbb{N}_{4|11}=\mathbb{N}_{4|11}\bigcap \mathbb{K}^\star$ and it is abelian. This structure of the nilpotent algebra implies that for the orbit $\mathcal{O}^4_{11}$ we have only three functions $\mathfrak{h}^{0}_i$ which will be harmonic and independent.
\par
Explicitly we set:
\begin{equation}\label{girollo}
    \mathfrak{H} (\mathfrak{h}_1,\mathfrak{h}_2,\mathfrak{h}_3) \, = \, \sum_{i=1}^3 \, \mathfrak{h}_i \, A_i \, = \, \left(
\begin{array}{lllllll}
 -\mathfrak{h} _1 & \mathfrak{h} _3 & 0 & -\sqrt{2} \mathfrak{h} _3 & -\mathfrak{h} _1 & -\mathfrak{h} _2 & 0 \\
 \mathfrak{h} _3 & 0 & -\mathfrak{h} _2 & 0 & \mathfrak{h} _3 & 0 & -\mathfrak{h} _2 \\
 0 & \mathfrak{h} _2 & -\mathfrak{h} _1 & \sqrt{2} \mathfrak{h} _3 & 0 & -\mathfrak{h} _3 & -\mathfrak{h} _1 \\
 \sqrt{2} \mathfrak{h} _3 & 0 & \sqrt{2} \mathfrak{h} _3 & 0 & \sqrt{2} \mathfrak{h} _3 & 0 & \sqrt{2} \mathfrak{h} _3 \\
 \mathfrak{h} _1 & -\mathfrak{h} _3 & 0 & \sqrt{2} \mathfrak{h} _3 & \mathfrak{h} _1 & \mathfrak{h} _2 & 0 \\
 -\mathfrak{h} _2 & 0 & \mathfrak{h} _3 & 0 & -\mathfrak{h} _2 & 0 & \mathfrak{h} _3 \\
 0 & -\mathfrak{h} _2 & \mathfrak{h} _1 & -\sqrt{2} \mathfrak{h} _3 & 0 & \mathfrak{h} _3 & \mathfrak{h} _1
\end{array}
\right)
\end{equation}
Considering $ \mathfrak{H} (\mathfrak{h}_1,\mathfrak{h}_2,\mathfrak{h}_3)$ as a Lax operator and calculating its Taub-NUT charge and electromagnetic charges we find:
\begin{equation}\label{fistl411}
    \mathbf{n}_{TN} \, = \, - 2\, \mathfrak{h}_2 \quad ; \quad \mathcal{Q} \, = \, \left( 0\, , \, 2\mathfrak{h}_2 \, , \, -2\sqrt{3} \mathfrak{h}_3 \, , \, -2\mathfrak{h}_1 \right )
\end{equation}
This implies that constructing the multi-centre solution with harmonic functions the condition $\mathfrak{h}_2 \, = \, 0$ should be sufficient to annihilate the Taub-NUT current.
\par
For later convenience let us change the normalization in the basis of harmonic functions as follows:
\begin{equation}\label{gignolo411}
    \mathfrak{h}_1^{(0)} \, = \, \ft 1{\sqrt{2}} \mathcal{H}_1 \quad ; \quad \mathfrak{h}_2^{(0)} \, = \, \ft 12 \left( 1 \, - \,\mathcal{H}_2\right) \quad ; \quad  \mathfrak{h}_3^{(0)} \, = \, \ft 1{\sqrt{2}} \mathcal{H}_3
\end{equation}
Implementing the symmetric coset construction with:
\begin{equation}\label{filodoro411}
    \mathcal{Y}\left(\mathcal{H}_1,\mathcal{H}_2,\mathcal{H}_3\right) \, \equiv \, \exp \left[  \mathfrak{H} \left(\ft 1{\sqrt{2}} \mathcal{H}_1,\ft 12 \left( 1 \, - \,\mathcal{H}_2\right),\ft 1{\sqrt{2}} \mathcal{H}_3\right)\right]
\end{equation}
and calculating the upper triangular coset representative $\mathbb{L}(\mathcal{Y})$ according to equations (\ref{solutioncosetrepr}) we find a relatively simple expression which, however, is still too large to be displayed. Yet the extraction of  the $\sigma$-model scalar fields produces a quite compact answer which we list below:
\begin{eqnarray}
  \exp\left[- \, U\right] &=&  \sqrt{\mathcal{H}_2^2-3 \mathcal{H}_3^2+\mathcal{H}_1}\label{u411}\\
  \mbox{Im} \, z &=&  \frac{\sqrt{\mathcal{H}_2^2-3 \mathcal{H}_3^2+\mathcal{H}_1}}{\mathcal{H}_2^2-\mathcal{H}_3^2+\mathcal{H}_1} \label{imz411}\\
\mbox{Re} \, z &=& -\frac{\sqrt{2} \mathcal{H}_3}{\mathcal{H}_2^2-\mathcal{H}_3^2+\mathcal{H}_1}  \label{rez411}
\end{eqnarray}
\begin{equation}\label{zz411}
  Z^M \, = \,   \left(
\begin{array}{l}
 \frac{\sqrt{6} \mathcal{H}_3^2}{\mathcal{H}_2^2-3 \mathcal{H}_3^2+\mathcal{H}_1} \\
 \frac{\left(\mathcal{H}_2-2 \mathcal{H}_3\right) \left(\mathcal{H}_2+\mathcal{H}_3\right)^2+\mathcal{H}_1 \mathcal{H}_2}{\sqrt{\left(\mathcal{H}_2^2-3
   \mathcal{H}_3^2+\mathcal{H}_1\right)^2}} \\
 -\frac{\sqrt{3} \mathcal{H}_3}{\mathcal{H}_2^2-3 \mathcal{H}_3^2+\mathcal{H}_1} \\
 \frac{\mathcal{H}_2^2-3 \mathcal{H}_3^2+\mathcal{H}_1-1}{\sqrt{2} \left(\mathcal{H}_2^2-3 \mathcal{H}_3^2+\mathcal{H}_1\right)}
\end{array}
\right)
\end{equation}
\begin{equation}\label{a411}
    a \, = \, \frac{\mathcal{H}_2^3+\left(-3 \mathcal{H}_3^2+\mathcal{H}_1+1\right) \mathcal{H}_2-2 \mathcal{H}_3^3}{\sqrt{2} \left(\mathcal{H}_2^2-3
   \mathcal{H}_3^2+\mathcal{H}_1\right)}
\end{equation}
\paragraph{The Taub-NUT current} Given this explicit result we can turn to the explicit oxidation formulae described in section \ref{genoxo} and calculate the Taub-NUT current which is the integrand of eq.(\ref{finkle}). We find:
\begin{equation}\label{cricco}
    j^{TN} \, = \,  \sqrt{2} \, \null^\star \nabla\, \mathcal{H}_2
\end{equation}
Hence the vanishing of the Taub-NUT current is guaranteed by the very simple condition:
\begin{equation}\label{finisco}
    \mathcal{H}_2 \, = \, \alpha \quad ; \quad \nabla\mathcal{H}_2 \,  =\,0
\end{equation}
where $\alpha$ is just a constant. This confirms the preliminary analysis obtained from the Lax operator which requires a vanishing component of the Lax along the second generator $A_2$ of the nilpotent algebra.
\paragraph{General form of the solution}
Imposing this condition we arrive at the following form of the solution depending on two harmonic functions $\mathcal{H}_1,\mathcal{H}_3$:
\begin{eqnarray}
  \exp[-U] &=& \sqrt{\alpha ^2-3 \mathcal{H}_3^2+\mathcal{H}_1} \\
  z &=& {\rm i} \, \frac{1}{\sqrt{\alpha ^2-3 \mathcal{H}_3^2+\mathcal{H}_1}} - \frac{\sqrt{2} \, \mathcal{H}_3}{{\alpha ^2-3 \mathcal{H}_3^2+\mathcal{H}_1}} \\
  j^{TN} &=& 0 \\
  j^{EM} &=&  \star \nabla \,\left(\begin{array}{c}
                     0 \\
                     0 \\
                     \sqrt{3} \, \mathcal{H}_3 \\
                     - \frac{1}{\sqrt{2}} \, \mathcal{H}_1
                   \end{array}
  \right)\label{filastrocca411}
\end{eqnarray}
Obviously the physical range of the solution is determined by the condition $(\alpha ^2-3 \mathcal{H}_3^2+\mathcal{H}_1) > 0 $ which can always be arranged, by tuning the parameters contained in the harmonic functions.
\par
To this effect let us discuss the nature of the black holes encompassed by this solution, that, by definition, are located at the poles of the harmonic functions $\mathcal{H}_1,\mathcal{H}_3$.
\par
According to the argument developed in section \ref{genepoles}, in the vicinity of each pole $|\vec{x} -\vec{x}_I| \, = \, r \, < \, \epsilon$ we can choose polar coordinates centered at $\vec{x}_\alpha$ and the behavior of the harmonic functions, for $\epsilon \rightarrow 0$ is the following one:
\begin{eqnarray}
  \mathcal{H}_1 & \sim & a_1  + \frac{b_1 }{r} \\
  \mathcal{H}_3 & \sim & a_3  + \frac{b_3 }{r}
\end{eqnarray}
which corresponds to the following behavior of the warp factor:
\begin{equation}\label{cruglo}
\exp[-U] \, \sim \,    \sqrt{\alpha ^2-3 a_3 ^2-\frac{3 b_3 ^2}{r^2}+a_1+\frac{b_1 }{r}-\frac{6
   a_3  b_3 }{r}}
\end{equation}
In order for the warp factor to be real for all values of $r\rightarrow 0$ we necessarily find
\begin{eqnarray}
  b_3 &=& 0 \nonumber\\
  b_1 & > & 0 \nonumber\\
  \alpha ^2-3 a_3^2 +a_1 &>&0 \label{ganimed}\\
\end{eqnarray}
Since conditions (\ref{ganimed}) hold true for each available pole, it means the harmonic function $\mathcal{H}_3 $ has actually no pole and is therefore equal to some constant. The boundary condition of asymptotic flatness fixes the value of such a constant:
\begin{equation}\label{craniotto}
    \lim_{r\rightarrow\infty} \exp[-U] \, = \, 1 \quad \Leftrightarrow \quad \mathcal{H}_3 \, = \, \frac{\sqrt{\alpha ^2+\mathcal{H}_1(\infty)-1}}{\sqrt{3}}
\end{equation}
Under such conditions in the vicinity of each pole $\vec{x}_\alpha$, the warp factor has the following behavior:
\begin{equation}\label{cagnolino}
    |\vec{x}-\vec{x}_\alpha|^2 \, \exp[-U] \, \stackrel{\vec{x} \rightarrow \vec{x}_\alpha}{\sim} \, \sqrt{b_1} \left|\vec{x}-\vec{x}_{\alpha }\right|^{3/2}+\mathcal{O}\left(\left|\vec{x}-\vec{x}_{\alpha
   }\right|^{5/2}\right)
\end{equation}
leading to a vanishing horizon area:
\begin{equation}\label{arietta411}
    \mbox{Area}_{\mathrm{H}_\alpha} \, = \, \lim_{\vec{x} \rightarrow \vec{x}_\alpha} \, |\vec{x}-\vec{x}_\alpha|^2 \, \exp[-U]\, =\, 0
\end{equation}
At the same time using the form of the electromagnetic current in eq.(\ref{filastrocca411}) and the behavior of the harmonic function in the vicinity of the poles we obtain the charge vector of each black hole encompassed by the solution:
\begin{equation}\label{caricucce411}
    \mathcal{Q}_\alpha \, = \, \int_{\mathbb{S}^2_\alpha} \, j^{EM} \, = \, \left(\begin{array}{c}
                                                                          0 \\
                                                                          0 \\
                                                                          0 \\
                                                                          - \frac{1}{\sqrt{2}} \, q_\alpha
                                                                        \end{array}
    \right) \quad ; \quad \mbox{where $q_\alpha \, = \, b_1$ for pole $\vec{x}_\alpha$}
\end{equation}
\paragraph{Summarizing} For the regular multicenter solutions associated with the orbit $4|11$ all blacks holes localized at each pole are of the same type, namely they are very small black holes with vanishing horizon area and a charge vector $\mathcal{Q}$ belonging to $\mathbf{W}$-orbit which is characterized by both a vanishing quartic invariant and the existence of a continuous parabolic stability subgroup of $\mathrm{SL(2,\mathbb{R})}$. Every black hole is a repetition in a different place of the spherical symmetric black hole which gives its name to the orbit.

\subsection{The Small Black Holes of $\mathcal{O}^2_{11}$}
Next let us consider the orbit $\mathcal{O}^2_{11}$.
\paragraph{$\mathbf{W}$-representation}
Applying the same strategy as in the previous case, from the general formula we obtain
\begin{equation}\label{chiavina211}
  \mathcal{Q}^\mathbf{w}_{2|11} \, = \, \mathrm{Tr}(X_{2|11}\mathcal{T}^{\mathbf{w}} )
 \, = \, \left(\sqrt{3},0,0,0\right)
\end{equation}
Substituting such a result in the expression for the quartic symplectic invariant (see eq.(\ref{invola}) we find:
\begin{equation}\label{genorama}
    \mathfrak{I}_4 \, = \, 0
\end{equation}
Just as before we stress that this result is meaningful since, by calculating the trace $\mbox{Tr}(X_{2|11} L_+^E) \, = \,0$, we can also check that the Taub-NUT charge vanishes.
Addressing the question whether there are subgroups of the original duality group in four-dimensions $\mathrm{SL(2,\mathrm{R})}$ that leave the charge vector (\ref{chiavina211}) invariant  we realize that such a group contains only the identity
\begin{equation}\label{parabolico}
   \mathrm{SL(2,\mathbb{R})} \, \supset \,  \mathcal{S}_{2|11}  \, = \, \mathbf{1}
\end{equation}
Hence we clearly establish the intrinsic difference between the two type of small black holes at the level of the $\mathbf{W}$-representation. Both have vanishing quartic invariant, yet only the orbit $4|11$ has a residual symmetry.
\par
\paragraph{$\mathrm{H}^\star$-stability subgroup} Considering next the stability subgroup of the nilpotent element $X_{2|11}$ in $\mathrm{H}^\star \, = \, \widehat{\slal(2,\mathbb{R})}\oplus\slal(2,\mathbb{R})_{\mathrm{h}^\star}$ we obtain:
\begin{equation}\label{gruppoS211}
    \mathfrak{S}_{2|11}  \, = \, \mathrm{SO(1,1)} \, \triangleright \, \mathbb{R}
\end{equation}
A generic element of the corresponding  Lie algebra is a linear combination of two generators $J,T $,
satisfying the commutation relations:
\begin{eqnarray}\label{gigio211}
   \left[J \, , \, T\right] & = &   \frac{3}{2\sqrt{6}} \, T  \nonumber\\
\end{eqnarray}
We do not give its explicit form which we do not use in the sequel.
\paragraph{Nilpotent algebra $\mathbb{N}_{4|11}$ } Considering next the adjoint action of the central element $h_{2|11}$ on the subspace $\mathbb{K}^\star$ we find that its eigenvalues are the following ones:
\begin{equation}\label{eigetti}
    \mbox{Eigenvalues}_{4|11}^{\mathbb{K}^\star}\, = \, \left\{ -3,3,-2,2,-1,1,0,0\right\}
\end{equation}
Therefore the three eigenoperators $A_3,A_2,A_1$ corresponding to the positive eigenvalues $3,2,1$, respectively, form the restriction to $\mathbb{K}^\star$ of a nilpotent algebra $\mathbb{N}_{2|11}$. In this case $A_i$ do not all commute among themselves so that, differently from the previous case we have $\mathbb{N}_{4|11}\neq\mathbb{N}_{4|11}\bigcap \mathbb{K}^\star$.
 In particular we find a new generator:
 \begin{equation}\label{b1genio}
 B \, \in \, \mathbb{H}^\star
 \end{equation}
 which completes a four-dimensional algebra with the following commutation relations:
 \begin{eqnarray}
   0 &=& \left[A_3 \, , \, A_2 \right] \, = \, \left[A_1 \, , \, A_3 \right]\\
   B &=& \left[A_2 \, , \, A_1 \right] \nonumber\\
   0 &=& \left[B \, , \, A_1 \right] \nonumber \\
   0 &=& \left[B \, , \, A_2 \right]  \nonumber \\
    0 &=& \left[B \, , \, A_3 \right]  \label{fortinopietra}
 \end{eqnarray}
As in the previous case, the structure of the nilpotent algebra implies that  for the orbit $\mathcal{O}^2_{11}$ we have only three functions $\mathfrak{h}^{0}_i$ which will be harmonic and independent. This is so because $\mathcal{D}^2 \mathbb{N}_{2|11} \,= \, 0$ and $\mathcal{D} \mathbb{N}_{2|11} \bigcap \mathbb{K}^\star \, = \, 0$.
\par
Explicitly we set:
\begin{eqnarray*}\label{girollo}
   & \mathfrak{H} (\mathfrak{h}_1,\mathfrak{h}_2,\mathfrak{h}_3) \, = \, \sum_{i=1}^3 \, \mathfrak{h}_i \, A_i \, = \, &\nonumber\\
   \end{eqnarray*}
{\scriptsize   \begin{eqnarray}
   &\left(
\begin{array}{lllllll}
 -\mathfrak{h}_2 & \mathfrak{h}_1-\mathfrak{h}_3 & \mathfrak{h}_2
   & -\sqrt{2} \mathfrak{h}_1-\sqrt{2} \mathfrak{h}_3 & 0 & -3
   \mathfrak{h}_1-\mathfrak{h}_3 & 0 \\
 \mathfrak{h}_1-\mathfrak{h}_3 & -2 \mathfrak{h}_2 &
   \mathfrak{h}_3-3 \mathfrak{h}_1 & -\sqrt{2} \mathfrak{h}_2 &
   \mathfrak{h}_1+\mathfrak{h}_3 & 0 & -3
   \mathfrak{h}_1-\mathfrak{h}_3 \\
 -\mathfrak{h}_2 & 3 \mathfrak{h}_1-\mathfrak{h}_3 &
   \mathfrak{h}_2 & \sqrt{2} \mathfrak{h}_1-\sqrt{2}
   \mathfrak{h}_3 & 0 & -\mathfrak{h}_1-\mathfrak{h}_3 & 0 \\
 \sqrt{2} \mathfrak{h}_1+\sqrt{2} \mathfrak{h}_3 & \sqrt{2}
   \mathfrak{h}_2 & \sqrt{2} \mathfrak{h}_1-\sqrt{2}
   \mathfrak{h}_3 & 0 & \sqrt{2} \mathfrak{h}_1-\sqrt{2}
   \mathfrak{h}_3 & -\sqrt{2} \mathfrak{h}_2 & \sqrt{2}
   \mathfrak{h}_1+\sqrt{2} \mathfrak{h}_3 \\
 0 & -\mathfrak{h}_1-\mathfrak{h}_3 & 0 & \sqrt{2}
   \mathfrak{h}_1-\sqrt{2} \mathfrak{h}_3 & -\mathfrak{h}_2 & 3
   \mathfrak{h}_1-\mathfrak{h}_3 & \mathfrak{h}_2 \\
 -3 \mathfrak{h}_1-\mathfrak{h}_3 & 0 &
   \mathfrak{h}_1+\mathfrak{h}_3 & \sqrt{2} \mathfrak{h}_2 &
   \mathfrak{h}_3-3 \mathfrak{h}_1 & 2 \mathfrak{h}_2 &
   \mathfrak{h}_1-\mathfrak{h}_3 \\
 0 & -3 \mathfrak{h}_1-\mathfrak{h}_3 & 0 & -\sqrt{2}
   \mathfrak{h}_1-\sqrt{2} \mathfrak{h}_3 & -\mathfrak{h}_2 &
   \mathfrak{h}_1-\mathfrak{h}_3 & \mathfrak{h}_2
\end{array}
\right)&\nonumber\\
\end{eqnarray}}
Considering $ \mathfrak{H} (\mathfrak{h}_1,\mathfrak{h}_2,\mathfrak{h}_3)$ as a Lax operator and calculating its Taub-NUT charge and electromagnetic charges we find:
\begin{equation}\label{fistl411}
    \mathbf{n}_{TN} \, = \, -2 \left(3 \mathfrak{h}_1+\mathfrak{h}_3\right) \quad ; \quad \mathcal{Q} \, = \, \left\{-2 \sqrt{3} \mathfrak{h}_2,6 \mathfrak{h}_1-2
   \mathfrak{h}_3,-2 \sqrt{3}
   \left(\mathfrak{h}_1+\mathfrak{h}_3\right),0\right\}
\end{equation}
This implies that constructing the multi-centre solution with harmonic functions the condition $\mathfrak{h}_3 \, = \, -\, 3 \, \mathfrak{h}_1$ might be sufficient to annihilate the Taub-NUT current.  We shall demonstrate that in this case the condition is slightly more complicated.
\par
For later convenience let us change the normalization in the basis of harmonic functions as follows:
\begin{equation}\label{gignolo}
    \mathfrak{h}_1^{(0)} \, = \, \ft 14 \mathcal{H}_3 \quad ; \quad \mathfrak{h}_2^{(0)} \, = \, \ft 12 \left( 1 \, - \,\mathcal{H}_2\right) \quad ; \quad  \mathfrak{h}_3^{(0)} \, = \, \ft 14 \mathcal{H}_1
\end{equation}
Implementing the symmetric coset construction with:
\begin{equation}\label{filodoro}
    \mathcal{Y}\left(\mathcal{H}_3,\mathcal{H}_2,\mathcal{H}_1\right) \, \equiv \, \exp \left[  \mathfrak{H} \left(\ft 14 \mathcal{H}_3,\ft 12 \left( 1 \, - \,\mathcal{H}_2\right),\ft 14 \mathcal{H}_1\right)\right]
\end{equation}
calculating the upper triangular coset representative $\mathbb{L}(\mathcal{Y})$ according to equations (\ref{solutioncosetrepr}) and extracting   the $\sigma$-model scalar fields we obtain the answer which we list below:
\begin{eqnarray}
  \exp\left[- \, U\right] &=& \frac{1}{2} \sqrt{-\mathcal{H}_3^2+\left(4 \mathcal{H}_1^3+6 \mathcal{H}_2 \mathcal{H}_1\right) \mathcal{H}_3+\mathcal{H}_2^2 \left(3 \mathcal{H}_1^2+4
   \mathcal{H}_2\right)} \label{u211}\\
  \mbox{Im} \, z &=& \frac{\sqrt{-\mathcal{H}_3^2+\left(4 \mathcal{H}_1^3+6 \mathcal{H}_2 \mathcal{H}_1\right) \mathcal{H}_3+\mathcal{H}_2^2 \left(3 \mathcal{H}_1^2+4 \mathcal{H}_2\right)}}{2
   \left(\mathcal{H}_1^2+\mathcal{H}_2\right)} \label{imz211}\\
\mbox{Re} \, z &=& \frac{\mathcal{H}_3-\mathcal{H}_2 \mathcal{H}_1}{2 \left(\mathcal{H}_1^2+\mathcal{H}_2\right)} \label{rez211}
\end{eqnarray}
\begin{equation}\label{zz211}
  Z^M \, = \,  \left(
\begin{array}{l}
 \frac{\sqrt{\frac{3}{2}} \left(\mathcal{H}_3^2-2 \mathcal{H}_1 \left(2 \mathcal{H}_1^2+3 \mathcal{H}_2-1\right) \mathcal{H}_3+\mathcal{H}_2 \left(-4
   \mathcal{H}_2^2+\left(4-3 \mathcal{H}_1^2\right) \mathcal{H}_2+2 \mathcal{H}_1^2\right)\right)}{\mathcal{H}_3^2-2 \left(2 \mathcal{H}_1^3+3 \mathcal{H}_2
   \mathcal{H}_1\right) \mathcal{H}_3-\mathcal{H}_2^2 \left(3 \mathcal{H}_1^2+4 \mathcal{H}_2\right)} \\
 \frac{\sqrt{2} \left(2 \mathcal{H}_1^3+3 \mathcal{H}_2 \mathcal{H}_1-\mathcal{H}_3\right)}{-\mathcal{H}_3^2+\left(4 \mathcal{H}_1^3+6 \mathcal{H}_2 \mathcal{H}_1\right)
   \mathcal{H}_3+\mathcal{H}_2^2 \left(3 \mathcal{H}_1^2+4 \mathcal{H}_2\right)} \\
 \frac{\sqrt{6} \left(\mathcal{H}_1 \mathcal{H}_2^2+\mathcal{H}_3 \left(2 \mathcal{H}_1^2+\mathcal{H}_2\right)\right)}{\mathcal{H}_3^2-2 \left(2 \mathcal{H}_1^3+3
   \mathcal{H}_2 \mathcal{H}_1\right) \mathcal{H}_3-\mathcal{H}_2^2 \left(3 \mathcal{H}_1^2+4 \mathcal{H}_2\right)} \\
 \frac{4 \mathcal{H}_3 \mathcal{H}_1^3+3 \mathcal{H}_2^2 \mathcal{H}_1^2+\mathcal{H}_3^2}{\sqrt{2} \left(-\mathcal{H}_3^2+\left(4 \mathcal{H}_1^3+6 \mathcal{H}_2
   \mathcal{H}_1\right) \mathcal{H}_3+\mathcal{H}_2^2 \left(3 \mathcal{H}_1^2+4 \mathcal{H}_2\right)\right)}
\end{array}
\right)
\end{equation}
\begin{equation}\label{a211}
    a \, = \,\frac{\mathcal{H}_3 \left(-6 \mathcal{H}_1^2-3 \mathcal{H}_2+1\right)-\mathcal{H}_1 \left(3 \mathcal{H}_2^2+3 \mathcal{H}_2+2 \mathcal{H}_1^2\right)}{\mathcal{H}_3^2-2
   \left(2 \mathcal{H}_1^3+3 \mathcal{H}_2 \mathcal{H}_1\right) \mathcal{H}_3-\mathcal{H}_2^2 \left(3 \mathcal{H}_1^2+4 \mathcal{H}_2\right)}
\end{equation}
\paragraph{The Taub-NUT current} Given this explicit result we can turn to the explicit oxidation formulae described in section \ref{genoxo} and calculate the Taub-NUT current which is the integrand of eq.(\ref{finkle}). We find:
\begin{equation}\label{cricco211}
    j^{TN} \, = \, \ft 12 \left( \null^\star\nabla \mathcal{H}_3 + 3\, \left (\mathcal{H}_2\, \null^\star\nabla \mathcal{H}_1 \, - \,
    \mathcal{H}_1\, \null^\star\nabla \mathcal{H}_2\right)\right)
\end{equation}
Analyzing eq.(\ref{cricco211}) we see that there are just two possible solutions to the condition $j^{TN} \, = \,0$:
\begin{description}
  \item[case a)] $\mathcal{H}_3 \, = \, \beta \, = \, \mbox{const} \quad ; \quad \mathcal{H}_1 \, = \,0$. With this condition we obtain:
      \begin{eqnarray}
        \exp[-U] &=& \frac{1}{2} \sqrt{4 \mathcal{H}_2^3-\beta ^2}\\
        z &=& \frac{\beta \,+ \,{\rm i} \, \sqrt{4 \mathcal{H}_2^3-\beta ^2}}{2
   \mathcal{H}_2} \\
        j^{EM} &=& \star \nabla \, \left(
\begin{array}{l}
 -\sqrt{\frac{3}{2}} \mathcal{H}_2 \\
 0 \\
 0 \\
 0
\end{array}
\right)
      \end{eqnarray}
  \item[case b)] $\mathcal{H}_3 \, = \, \beta \, = \, \mbox{const} \quad ; \quad \mathcal{H}_2 \, = \,0$
  \begin{eqnarray}
    \exp[-U] &=& \frac{1}{2} \sqrt{\beta  \left( 4
   \mathcal{H}_1^3 -\beta\right)} \\
   z &=& \frac{\beta \, + \, {\rm i} \, \sqrt{\beta  \left( 4
   \mathcal{H}_1^3 -\beta\right)}}{2 \mathcal{H}_3^2}\\
   j^{EM} &=& \left(
\begin{array}{l}
 0 \\
 0 \\
 -\sqrt{\frac{3}{2}} \mathcal{H}_1 \\
 0
\end{array}
\right)
  \end{eqnarray}
\end{description}
It might seem that these two solutions correspond to different types of black holes but this is not the case, as we now show. From the asymptotic flatness boundary condition we find that the value of $\beta$ is fixed in terms of the value at infinity of the corresponding harmonic function $\mathcal{H}_{1,2}$, which of course must satisfy the necessary condition for reality of the solution $\mathcal{H}_{1,2}(\infty)\, \geq \, 1$ :
\begin{equation}\label{gurlone}
    \left \{ \begin{array}{cccc}
               \beta & = & 2 \sqrt{\left[\mathcal{H}_2(\infty)\right]^3-1} & \mbox{case a} \\
               \beta & = & 2 \left(\left[\mathcal{H}_1(\infty)\right]^3+\sqrt{\left[\mathcal{H}_1(\infty)\right]^6-1}\right) & \mbox{case b} \\
             \end{array}\right.
\end{equation}
In the vicinity of a pole by means of the usual argument we obtain the following behavior of the warp factor:
\begin{equation}\label{cagnolino211}
    |\vec{x}-\vec{x}_\alpha|^2 \, \exp[-U] \, \stackrel{\vec{x} \rightarrow \vec{x}_\alpha}{\sim} \,
    \left \{ \begin{array}{ccc}
    \sqrt{b_2^3} \sqrt{\left|\vec{x}-\vec{x}_{\alpha }\right|}+\mathcal{O}\left(\left|\vec{x}-\vec{x}_{\alpha }\right|^{3/2}\right) & : & \mbox{case a} \\
   \sqrt{\beta \, b_1^3} \sqrt{\left|\vec{x}-\vec{x}_{\alpha }\right|}+\mathcal{O}\left(\left|\vec{x}-\vec{x}_{\alpha }\right|^{3/2}\right) & : & \mbox{case b}
    \end{array}\right.
\end{equation}
Hence in both cases the horizon area vanishes at all poles $\vec{x}_\alpha$ and the reality conditions are satisfied choosing the appropriate sign of $b_{1,2}$. The charge vector has the same structure for all black holes encompassed in the first or in the second solution, namely:
\begin{equation}\label{craccolini211}
    \mathcal{Q}_\alpha \, = \, \left \{ \begin{array}{ccccc}
                                \left \{- \sqrt{\ft 32 } \, p_\alpha \, , \, 0 \, , \, 0 \, , \, 0 \right\} & : & p_\alpha & = & b_2  \quad \mbox{ for pole $\alpha$}\\
                                \left \{0 \, , \, 0 \, , \, - \,\sqrt{\ft 32 } \, q_\alpha \, , \,  0 \right\} & : & q_\alpha & = & b_1 \quad \mbox{ for pole $\alpha$}
                              \end{array}
    \right.
\end{equation}
In both cases the quartic invariant $\mathfrak{I}_4$ is zero for all black holes in the solutions, yet one might still doubt whether the $\mathbf{W}$-orbit for the two cases might be different. It is not so, since a direct calculation shows that the image in the $j=\ft 32$ representation $\Lambda[\mathfrak{A}]$\footnote{See \cite{noig22} for details, in particular eq.(3.13) of that reference for the explicit form of the spin $\ft 32$ matrices.} of the following $\mathrm{SL(2,\mathbb{R})}$ element:
\begin{equation}\label{corneato}
 \mathfrak{A} \, = \,   \left(
\begin{array}{ll}
 0 & \frac{p}{q} \\
 -\frac{q}{p} & 0
\end{array}
\right)
\end{equation}
maps the charge vector $\mathcal{Q}_{[q]}\, = \,\left \{0 \, , \, 0 \, , \, - \,q \, , \,  0 \right\}$, into the charge vector $\mathcal{Q}_{[p]}\, = \,\left \{p \, , \, 0 \, , \, 0 \, , \,  0 \right\}$, namely we have $\Lambda[\mathfrak{A}]\,\mathcal{Q}_{[q]} \, = \, \mathcal{Q}_{[p]}$. Hence the two solutions we have here discussed  simply give different representatives of the same $\mathbf{W}$-orbit.
\paragraph{Summary} Just as in the previous case for a multicenter solution associated with the $\mathcal{O}^2_{11}$ orbit all the black holes included in one solution are of the same type, namely small black holes with  the same identical properties.
\subsection{The Large BPS Black Holes of $\mathcal{O}^3_{11}$}
Next let us consider the orbit $\mathcal{O}^3_{11}$, which in the spherical symmetric case leads to BPS Black holes with a finite horizon area.
\paragraph{$\mathbf{W}$-representation}
In order to better appreciate the structure of these solutions,   let us slightly generalize our orbit representative, writing the following nilpotent matrix that depends on two parameters $(p,q)$ to be interpreted later as the magnetic and the electric charge of the hole:
\begin{equation}\label{xpq}
  X_{3|11}(p,q) \, = \, \left(
\begin{array}{lllllll}
 q & 0 & 0 & -\frac{q}{\sqrt{2}} & 0 & 0 & 0 \\
 0 & \frac{p+q}{2} & -\frac{p}{2} & 0 & \frac{q}{2} & 0 & 0 \\
 0 & \frac{p}{2} & \frac{q-p}{2} & 0 & 0 & -\frac{q}{2} & 0 \\
 \frac{q}{\sqrt{2}} & 0 & 0 & 0 & 0 & 0 & \frac{q}{\sqrt{2}}
   \\
 0 & -\frac{q}{2} & 0 & 0 & \frac{p-q}{2} & \frac{p}{2} & 0 \\
 0 & 0 & \frac{q}{2} & 0 & -\frac{p}{2} & \frac{1}{2} (-p-q) &
   0 \\
 0 & 0 & 0 & -\frac{q}{\sqrt{2}} & 0 & 0 & -q
\end{array}
\right)
\end{equation}
The standard triple representative mentioned in eq.(\ref{311}) is just the particular case $X_{3|11}(1,1)$.
Applying the same strategy as in the previous case, from the general formula we obtain
\begin{equation}\label{chiavina311}
  \mathcal{Q}^\mathbf{w}_{3|11} \, = \, \mathrm{Tr}(X_{3|11}(p,q)\mathcal{T}^{\mathbf{w}} )
 \, = \, \left(0,p,-\sqrt{3} q,0\right)
\end{equation}
Substituting such a result in the expression for the quartic symplectic invariant (see eq.(\ref{invola})) we find:
\begin{equation}\label{genorama11}
    \mathfrak{I}_4 \, = \, 9 \, p \, q^3 \, > \, 0 \quad \mbox{if $p$ and $q$ have the same sign}
\end{equation}
Just as before we stress that this result is meaningful since, by calculating the trace $\mbox{Tr}(X_{3|11} L_+^E) \, = \,0$, we can also check that the Taub-NUT charge vanishes. Furthermore we note that the condition that $p$ and $q$ have the same sign was singled out in \cite{noig22} as the defining condition of the orbit $O^{3}_{11}$ which, in the spherical symmetry approach leads to regular BPS solutions. The choice of opposite signs was proved in \cite{noig22} to correspond to a different $\mathrm{H}^\star$ orbit, the non diagonal $O^{3}_{21}$ which instead contains only singular solutions. Here we will show another important and intrinsically four dimensional  reason to separate the two cases.
\par
Addressing the question whether there are subgroups of the original duality group in four-dimensions $\mathrm{SL(2,\mathrm{R})}$ that leave the charge vector (\ref{chiavina311}) invariant  we realize that such a subgroup exists and is the finite  cyclic group of order three \footnote{The authors express their gratitude to Alessio Marrani who attracted their attention, after the first appearance of the present paper in the arXive to paper \cite{Borsten:2011ai} where the existence of a discrete stability subgroup for the charges of the regular BPS orbit in the $S^3$ model had already been found. }:
\begin{equation}\label{parabolico}
   \mathrm{SL(2,\mathbb{R})} \, \supset \,  \mathcal{S}_{3|11}  \, = \, \mathbb{Z}_3
\end{equation}
$\mathcal{S}_{3|11}$ is made by the following three elements:
\begin{eqnarray}
  \mathbf{1} &=& \left(\begin{array}{cc}
           1 & 0 \\
           0 & 1
         \end{array}\right)
   \\
  \mathfrak{B} &=& \left(
\begin{array}{ll}
 -\frac{1}{2} & -\frac{\sqrt{3} }{2 } \,\sqrt{\frac p q}\\
 \frac{\sqrt{3} }{2 }\,\sqrt{\frac q p} & -\frac{1}{2}
\end{array}
\right) \\
\mathfrak{B}^2 &=& \left(
\begin{array}{ll}
 -\frac{1}{2} & \frac{\sqrt{3} }{2 } \,\sqrt{\frac p q}\\
- \frac{\sqrt{3} }{2 }\,\sqrt{\frac q p} & -\frac{1}{2}
\end{array}
\right) \quad ; \quad \mathfrak{B}^3 \, = \, \mathbf{1}
\end{eqnarray}
It is evident that such a $\mathbb{Z}_3$ subgroup exists if and only if the two charges $p,q$ have the same sign. Otherwise the corresponding matrices develop imaginary elements and migrate to $\mathrm{SL(2,\mathbb{C})}$.
The existence of this isotropy group $\mathbb{Z}_3$ can be considered the very definition of the $\mathbf{W}$-orbit corresponding to BPS black holes. Indeed let us name $\lambda=\sqrt{\frac p q}$ and consider the algebraic condition imposed on a generic charge vector:
$\mathcal{Q}=\left\{Q_1,Q_2,Q_3,Q_4\right\}$ by the request that it should admit the above described $\mathbb{Z}_3$ stability group:
\begin{equation}\label{vecchiomerletto}
    \Lambda[\mathfrak{B}] \mathcal{Q} \, = \, \mathcal{Q} \quad \Leftrightarrow \quad \mathcal{Q} \, = \,\left(\sqrt{3} \lambda ^2 Q_4,-\frac{\lambda ^2
   Q_3}{\sqrt{3}},Q_3,Q_4\right)
\end{equation}
It is evident from the above explicit result that the charge vectors having this symmetry depend only on three parameters $(\lambda^2, Q_3, Q_4)$. The very relevant fact is that substituting this restricted charge vector in the general formula (\ref{invola}) for the quartic invariant we obtain:
\begin{equation}\label{invaspec311}
   \mathfrak{J}_4 \, = \, \lambda ^2 \left(Q_3^2+3 \lambda ^2 Q_4^2\right)^2 \,> \, 0
\end{equation}
Hence the $\mathbb{Z}_3$ guarantees that the quartic invariant is a perfect square and hence positive. It is an intrinsic restriction characterizing the $\mathbb{W}$-orbit.
\par
\paragraph{$\mathrm{H}^\star$-stability subgroup} Considering next the stability subgroup of the nilpotent element $X_{3|11}(1,1)$ in $\mathrm{H}^\star \, = \, \widehat{\slal(2,\mathbb{R})}\oplus\slal(2,\mathbb{R})_{\mathrm{h}^\star}$ we obtain:
\begin{equation}\label{gruppoS211}
    \mathfrak{S}_{3|11}  \, = \, \mathbb{R}
\end{equation}
the group being generated by a matrix $\mathbb{A}_{3|11}$ of nilpotency degree $2$:
\begin{eqnarray}\label{gigio311}
   \mathbb{A}_{3|11} ^2\, = \, \mathbf{0}
\end{eqnarray}
We do not give its explicit form which we do not use in the sequel.
\paragraph{Nilpotent algebra $\mathbb{N}_{3|11}$ } Considering next the adjoint action of the central element $h_{3|11}$ on the subspace $\mathbb{K}^\star$ we find that its eigenvalues are the following ones:
\begin{equation}\label{eigetti}
    \mbox{Eigenvalues}_{3|11}^{\mathbb{K}^\star}\, = \, \{-2,-2,-2,-2,2,2,2,2\}
\end{equation}
Therefore the four eigenoperators $A_1,A_2,A_3,A_4$ corresponding to the four positive eigenvalues $2$, respectively, form the restriction to $\mathbb{K}^\star$ of a nilpotent algebra $\mathbb{N}_{3|11}$. Also in this case the $A_i$ do not all commute among themselves so that,  we have $\mathbb{N}_{3|11}\neq\mathbb{N}_{3|11}\bigcap \mathbb{K}^\star$.
 In particular we find a new generator:
 \begin{equation}\label{b1genio311}
 B \, \in \, \mathbb{H}^\star
 \end{equation}
 which completes a five-dimensional algebra with the following commutation relations:
 \begin{eqnarray}
 \left[A_i \, , \, A_j \right] & = & \Omega_{ij} \, B \nonumber\\
  \left[B \, , \, A_i \right]&=&  0\nonumber\\
   B& = & \left(
\begin{array}{llll}
 0 & 0 & -1 & 1 \\
 0 & 0 & -1 & -1 \\
 1 & 1 & 0 & 0 \\
 -1 & 1 & 0 & 0
\end{array}
\right) \label{fortino311}
 \end{eqnarray}
The structure of the nilpotent algebra implies that  for the orbit $\mathcal{O}^3_{11}$ we have only four functions $\mathfrak{h}^{0}_i$ which will be harmonic and independent. This is so because $\mathcal{D}^2 \mathbb{N}_{3|11} \,= \, 0$ and $\mathcal{D} \mathbb{N}_{3|11} \bigcap \mathbb{K}^\star \, = \, 0$.
\par
Explicitly we set:
\begin{eqnarray*}\label{girollo}
   & \mathfrak{H} (\mathfrak{h}_1,\mathfrak{h}_2,\mathfrak{h}_3,\mathfrak{h}_4) \, = \, \sum_{i=1}^4 \, \mathfrak{h}_i \, A_i \, = \, &\nonumber\\
   \end{eqnarray*}
  {\small \begin{eqnarray}
&  \left(
\begin{array}{lllllll}
 2 \mathfrak{h}_3 & \mathfrak{h}_1-2 \mathfrak{h}_2 & 2 \mathfrak{h}_1-\mathfrak{h}_2 & -\sqrt{2} \mathfrak{h}_3 & -3 \mathfrak{h}_2 & -3
   \mathfrak{h}_1 & 0 \\
 \mathfrak{h}_1-2 \mathfrak{h}_2 & \mathfrak{h}_3-\mathfrak{h}_4 & \mathfrak{h}_4 & \sqrt{2} \mathfrak{h}_2-2 \sqrt{2} \mathfrak{h}_1 & \mathfrak{h}_3
   & 0 & -3 \mathfrak{h}_1 \\
 \mathfrak{h}_2-2 \mathfrak{h}_1 & -\mathfrak{h}_4 & \mathfrak{h}_3+\mathfrak{h}_4 & \sqrt{2} \mathfrak{h}_1-2 \sqrt{2} \mathfrak{h}_2 & 0
   & -\mathfrak{h}_3 & -3 \mathfrak{h}_2 \\
 \sqrt{2} \mathfrak{h}_3 & 2 \sqrt{2} \mathfrak{h}_1-\sqrt{2} \mathfrak{h}_2 & \sqrt{2} \mathfrak{h}_1-2
   \sqrt{2} \mathfrak{h}_2 & 0 & \sqrt{2} \mathfrak{h}_1-2 \sqrt{2} \mathfrak{h}_2 & \sqrt{2}
   \mathfrak{h}_2-2 \sqrt{2} \mathfrak{h}_1 & \sqrt{2} \mathfrak{h}_3 \\
 3 \mathfrak{h}_2 & -\mathfrak{h}_3 & 0 & \sqrt{2} \mathfrak{h}_1-2 \sqrt{2} \mathfrak{h}_2 & -\mathfrak{h}_3-\mathfrak{h}_4 &
   -\mathfrak{h}_4 & 2 \mathfrak{h}_1-\mathfrak{h}_2 \\
 -3 \mathfrak{h}_1 & 0 & \mathfrak{h}_3 & 2 \sqrt{2} \mathfrak{h}_1-\sqrt{2} \mathfrak{h}_2 & \mathfrak{h}_4 &
   \mathfrak{h}_4-\mathfrak{h}_3 & \mathfrak{h}_1-2 \mathfrak{h}_2 \\
 0 & -3 \mathfrak{h}_1 & 3 \mathfrak{h}_2 & -\sqrt{2} \mathfrak{h}_3 & \mathfrak{h}_2-2 \mathfrak{h}_1 & \mathfrak{h}_1-2 \mathfrak{h}_2 &
   -2 \mathfrak{h}_3
\end{array}
\right)&\nonumber\\
\end{eqnarray}}
Considering $ \mathfrak{H} (\mathfrak{h}_1,\mathfrak{h}_2,\mathfrak{h}_3,\mathfrak{h}_4)$ as a Lax operator and calculating its Taub-NUT charge and electromagnetic charges we find:
\begin{equation}\label{fistl411}
    \mathbf{n}_{TN} \, = \, -6  \mathfrak{h}_1 \quad ; \quad \mathcal{Q} \, = \, \left\{2 \sqrt{3} \left(\mathfrak{h}_2-2 \mathfrak{h}_1\right),-2 \mathfrak{h}_4,-2 \sqrt{3}
   \mathfrak{h}_3,-6 \mathfrak{h}_2\right\}
\end{equation}
This implies that constructing the multi-centre solution with harmonic functions the condition $\mathfrak{h}_1 \, = 0$ might be sufficient to annihilate the Taub-NUT current.  We shall demonstrate that also in this case the condition is slightly more complicated. This emphasizes the difference between the Lax operator one-dimensional approach and the multicenter construction based on harmonic functions.
\par
For later convenience let us change the normalization in the basis of harmonic functions as follows:
\begin{equation}\label{gignolo311}
    \mathfrak{h}_1^{(0)} \, = \, \ft {1}{\sqrt{12}} \mathcal{H}_1 \quad ; \quad \mathfrak{h}_2^{(0)} \, = \, \ft {1}{\sqrt{12}}  \,\mathcal{H}_2 \quad ; \quad  \mathfrak{h}_3^{(0)} \, = \, \ft 12 \left(\mathcal{H}_3-1\right)\quad ; \quad \mathfrak{h}_4^{(0)} \, = \, \ft 12 \left(\mathcal{H}_4+1\right)
\end{equation}
Implementing the symmetric coset construction with:
\begin{equation}\label{filodoro311}
    \mathcal{Y}\left(\mathcal{H}_1,\mathcal{H}_2,\mathcal{H}_3,\mathcal{H}_4\right) \, \equiv \,
    \exp \left[  \mathfrak{H} \left(\ft {1}{\sqrt{12}} \mathcal{H}_1,\ft {1}{\sqrt{12}}  \,\mathcal{H}_2,\ft 12 \left(\mathcal{H}_3-1\right),\ft 12 \left(\mathcal{H}_4+1\right)\right)\right]
\end{equation}
calculating the upper triangular coset representative $\mathbb{L}(\mathcal{Y})$ according to equations (\ref{solutioncosetrepr}) and extracting   the $\sigma$-model scalar fields we obtain an explicit but rather messy  answer which we present in the appendix in eq.s (\ref{fullyfilli1}, \ref{fullyfilli2}, \ref{fullyfilli3}). In particular we obtain the Taub-NUT current in the following form:
\begin{equation}\label{grignolo311}
    j^{TN} \,=\,\sum_{i=1}^{4}\, \mathfrak{R}_i\left(\mathcal{H}\right) \,\nabla \mathcal{H}_i
\end{equation}
where $\mathfrak{R}_i\left(\mathcal{H}\right)$ are rational functions of the four harmonic functions, the maximal degree of involved polynomials being  16. A priori, imposing the vanishing of the Taub-NUT current   is a problem without guaranteed solutions. In the $4$-dimensional linear space of the harmonic functions we can introduce $r$-linear relations of the form:
\begin{eqnarray}
\label{ugonotto311}
  0 &=& V_\alpha^i \, \mathcal{H}_i  \quad ; \quad \alpha \, = \, 1,\dots , r
\end{eqnarray}
Let $U_a^i$ be a set of $4-r$ linear independent $4$-vectors orthogonal to the vectors $V_\alpha^i$. Then it must happen that on the locus defined by eq.s (\ref{ugonotto311}), the following rational functions should also vanish
\begin{equation}\label{raniato311}
0\, = \, \mathfrak{P}_a(\mathcal{H}) \, \equiv \, U_a^i \, \mathfrak{R}_i\left(\mathcal{H}\right) \quad ; \quad (a=1,\dots ,r-4)
\end{equation}
For generic rational functions this will never happen, yet we know that for our system such solutions should exist and  in want of a clear cut algorithm it is a matter of ingenuity to find them. We do not find any solution with $r=1$ but we find two nice solutions with $r=2$. They are the following ones:
\begin{description}
  \item[a)] $\mathcal{H}_1 \,= \, \mathcal{H}_2 \, = \, 0$. The complete form of the supergravity solution corresponding to this choice is:
      \begin{eqnarray}
        \exp[\,-\,U] &=& \sqrt{-\mathcal{H}_3^3 \mathcal{H}_4} \\
        z &=& {\rm i} \,\frac{ \sqrt{-\mathcal{H}_3^3
   \mathcal{H}_4}}{\mathcal{H}_3^2} \\
        j^{TN} &=& 0 \\
        j^{EM} &=& \star \nabla \, \left(
\begin{array}{l}
 0 \\
 \frac{ \mathcal{H}_4}{\sqrt{2}} \\
 \sqrt{\frac{3}{2}}  \mathcal{H}_3 \\
 0
\end{array}
\right) \label{soluz311atomica}
      \end{eqnarray}
  \item[b)] $\mathcal{H}_1 \,= \,  0$, $\mathcal{H}_3 \, = \, - \, \mathcal{H}_4$.The complete form of the supergravity solution corresponding to this choice is:
      \begin{eqnarray}
        \exp[\,-\,U] &=& \sqrt{-\frac{\mathcal{H}_2^4}{3}-2 \mathcal{H}_4^2
   \mathcal{H}_2^2+\mathcal{H}_4^4} \\
        z &=& \frac{2 \mathcal{H}_2 \mathcal{H}_4-{\rm i} \,
   \sqrt{-\mathcal{H}_2^4-6 \mathcal{H}_4^2 \mathcal{H}_2^2+3
   \mathcal{H}_4^4}}{\sqrt{3}
   \left(\mathcal{H}_2^2-\mathcal{H}_4^2\right)} \\
        j^{TN} &=& 0 \\
        j^{EM} &=& \star \nabla \, \left(
\begin{array}{l}
 -\frac{  \mathcal{H}_2}{\sqrt{2}} \\
 \frac{  \mathcal{H}_4}{\sqrt{2}} \\
 -\sqrt{\frac{3}{2}}   \mathcal{H}_4 \\
 \sqrt{\frac{3}{2}}   \mathcal{H}_2
\end{array}
\right) \label{soluz311molecola}
      \end{eqnarray}
\end{description}
We can now make some comments about the two solutions. First of all both in case a) and in case b) we have to fix the asymptotic value of the harmonic functions at spatial infinity $r=\infty$, in such a way as to obtain asymptotic flatness. This is quite easy and we do not dwell on it. Secondly we have to fix the parameters of the harmonic functions in such a way that the warp factor is always real on the whole physical range. These conditions are also easily spelled out:
\begin{equation}\label{barabano}
    \begin{array}{cccc}
       a) & -\mathcal{H}_3 \mathcal{H}_4 & > & 0 \\
       b) & -\frac{\mathcal{H}_2^4}{3}-2 \mathcal{H}_4^2
   \mathcal{H}_2^2+\mathcal{H}_4^4& > & 0
     \end{array}
\end{equation}
and in a multicenter solution can be easily arranged adjusting the coefficients of each pole.
 Thirdly we can comment about the structure of the charge vector that we obtain at each pole:
\begin{equation}\label{grimaldello}
    \mathcal{H}_i \,  \sim \, a_i \, + \, \frac{Q_i}{|x - x_\alpha|}
\end{equation}
In case a) and b) we respectively obtain:
\begin{eqnarray}
  \mathcal{Q}_\alpha &=& \left(
\begin{array}{l}
 0 \\
 \frac{ Q_4}{\sqrt{2}} \\
 \sqrt{\frac{3}{2}}  Q_3 \\
 0
\end{array}
\right) \\
  \mathcal{Q}_\alpha &=& \left(\begin{array}{l}
 -\frac{  Q_2}{\sqrt{2}} \\
 \frac{  Q_4}{\sqrt{2}} \\
 -\sqrt{\frac{3}{2}}   Q_4 \\
 \sqrt{\frac{3}{2}}   Q_2
\end{array}
\right)
\end{eqnarray}
Comparing with eq.s (\ref{vecchiomerletto},\ref{invaspec311}) we see that in both cases the structure of these charges is that imposed by the $\mathbb{Z}_3$ invariance which characterizes BPS black holes. The necessary choice of signs in the case a)
\begin{equation}\label{fruttolonato311}
    \frac{Q_4}{Q_3}\, < \, 0
\end{equation}
is the same which is required by the reality of the warp factor. Hence in  case b) all the black holes encompassed by the solution at each pole are finite area BPS black holes. In case a) the same is true for all the poles common to the harmonic function $\mathcal{H}_3$ and $\mathcal{H}_4$: they are finite area BPS black holes. Yet we can envisage the situation where some poles of
$\mathcal{H}_3$ are not shared by $\mathcal{H}_4$ and viceversa. In this case the pole of $\mathcal{H}_4$ defines a very small black hole, while the pole of $\mathcal{H}_3$ defines a small black hole. This is confirmed by the fact that a charge vector of type
$\{0,p,0,0\}$ is mapped into $\{0,0,0,p\}$ by $\Lambda\left[\left(\begin{array}{cc}
                                                                    0 & -1 \\
                                                                    1 & 0
                                                                  \end{array}
\right)\right]$ and as such admits a parabolic subgroup of stability $\Lambda\left[\left(\begin{array}{cc}
                                                                    1 & b\\
                                                                    0 & 1
                                                                  \end{array}
\right)\right]$.
\paragraph{Summary} For a multicenter solution associated with the $\mathcal{O}^3_{11}$ orbit there are two possibilities either all  the black holes included in one solution are regular, finite  area, BPS black holes, either we have a mixture of very small and small black holes. A finite area BPS black hole emerges when the center of a very small black hole coincides with the center of a small one. This provides the challenging suggestion that a BPS black hole can be considered quantum mechanically as a composite object where the "quarks" are small and very small black holes.
\subsection{BPS Kerr-Newman solution}
\label{kerronumano}
Next we want to show how this orbit encompasses also the BPS Kerr-Newman solution that was found by Luest et al in \cite{lustokerro}.
\par
To this effect we go back to the general formulae (\ref{fullyfilli1}-\ref{fullyfilli3}) for the scalar fields in this orbit and we make the following reduction from four to two independent harmonic functions:
\begin{equation}\label{reduzio311}
    \mathcal{H}_2 \, = \, 0 \quad ; \quad \mathcal{H}_4 \, = \, -\ft 13 \, \mathcal{H}_3
\end{equation}
With such a choice the expressions for all the scalar fields dramatically simplify and we obtain:
\begin{eqnarray}
  \mathfrak{W} &=& \frac{\sqrt{3}}{\mathcal{H}_1^2+\mathcal{H}_3^2} \\
  z  &=& {\rm i} \, \frac{1}{\sqrt{3}} \\
 Z &=& \left(
\begin{array}{l}
 -\frac{3 \mathcal{H}_1}{\sqrt{2} \left(\mathcal{H}_1^2+\mathcal{H}_3^2\right)} \\
 \frac{\mathcal{H}_1^2+\left(\mathcal{H}_3-3\right) \mathcal{H}_3}{\sqrt{2} \left(\mathcal{H}_1^2+\mathcal{H}_3^2\right)} \\
 -\frac{\sqrt{\frac{3}{2}} \left(\mathcal{H}_1^2+\left(\mathcal{H}_3-1\right) \mathcal{H}_3\right)}{\mathcal{H}_1^2+\mathcal{H}_3^2} \\
 -\frac{\mathcal{H}_1}{\sqrt{6} \left(\mathcal{H}_1^2+\mathcal{H}_3^2\right)}
\end{array}
\right)\\
 a &=& \frac{5 \mathcal{H}_1}{\sqrt{3} \left(\mathcal{H}_1^2+\mathcal{H}_3^2\right)}
\end{eqnarray}
Utilizing the above expressions in the final oxidation formulae we obtain the following result for the Taub-Nut current and for the electromagnetic currents:
\begin{equation}\label{gualtiero}
    j^{TN} \, = \,\frac{2 \left(\star\nabla \mathcal{H}_1 \mathcal{H}_3\, -\, \star\nabla \mathcal{H}_3 \mathcal{H}_1\right)}{\sqrt{3}}
\end{equation}
\begin{equation}\label{gromada}
    j^{EM} \, = \, \left(
\begin{array}{l}
 \frac{2 \, \star\nabla \mathcal{H}_3 \mathcal{H}_1 \left(\mathcal{H}_1^2+\left(\mathcal{H}_3-2\right) \mathcal{H}_3\right)-\, \star\nabla \mathcal{H}_1 \left(\left(2
   \mathcal{H}_3+1\right) \mathcal{H}_1^2+\mathcal{H}_3^2 \left(2 \mathcal{H}_3-3\right)\right)}{\sqrt{2} \left(\mathcal{H}_1^2+\mathcal{H}_3^2\right)} \\
 \frac{\, \star\nabla \mathcal{H}_3 \left(3 \mathcal{H}_1^2-\mathcal{H}_3^2\right)-4 \, \star\nabla \mathcal{H}_1 \mathcal{H}_1 \mathcal{H}_3}{3 \sqrt{2}
   \left(\mathcal{H}_1^2+\mathcal{H}_3^2\right)} \\
 \frac{\sqrt{\frac{3}{2}} \left(4 \, \star\nabla \mathcal{H}_1 \mathcal{H}_1 \mathcal{H}_3+\, \star\nabla \mathcal{H}_3 \left(\mathcal{H}_3^2-3
   \mathcal{H}_1^2\right)\right)}{\mathcal{H}_1^2+\mathcal{H}_3^2} \\
 \frac{2 \, \star\nabla \mathcal{H}_3 \mathcal{H}_1 \left(\mathcal{H}_1^2+\left(\mathcal{H}_3-6\right) \mathcal{H}_3\right)-\, \star\nabla \mathcal{H}_1 \left(\left(2
   \mathcal{H}_3+3\right) \mathcal{H}_1^2+\mathcal{H}_3^2 \left(2 \mathcal{H}_3-9\right)\right)}{\sqrt{6} \left(\mathcal{H}_1^2+\mathcal{H}_3^2\right)}
\end{array}
\right)
\end{equation}
Next identifying the two harmonic functions with those introduced in eq.s(\ref{frantaccio1}-\ref{frantaccio2}), according to:
\begin{equation}\label{carizzoli}
    \mathcal{H}_1 \, = \, 3^{\ft 14} \, \left(1\, + \, m\, \mathcal{P}\right ) \quad ; \quad \mathcal{H}_3 \, = \,3^{\ft 14} \,  \, m\, \mathcal{R}
\end{equation}
we obtain the following result for the warp-factor:
\begin{equation}\label{zorrone}
    \exp[U] \, = \, \frac{(m+r)^2+\alpha ^2 \cos ^2(\theta )}{r^2+\alpha ^2 \cos ^2(\theta )}
\end{equation}
and for the Kaluza-Klein vector:
\begin{equation}\label{grommo}
   \mathbf{A}^{[KK]} \, = \,  \omega \,  \equiv \,\frac{m (m+2 r) \alpha  \sin ^2(\theta )}{r^2+\alpha ^2 \cos ^2(\theta )} \, {d\phi }
\end{equation}
Indeed one can easily check that, in the spheroidal coordinates (\ref{spheroidal}) with flat metric \ref{flateuclimet} we have:
\begin{equation}\label{cagliostro}
    2 m \, \left( \star \nabla \mathcal{P} \, \mathcal{R} \, - \,  \mathcal{P} \, \star \nabla \mathcal{R}\right ) \, = \, \mathrm{d} \omega
\end{equation}
where $\star\nabla$ denotes the Hodge dual of the exterior derivative $d$. Writing  the corresponding final form of the metric::
\begin{equation}\label{BPSKN}
    ds_{BPSKN}^2 \, = \, - \, \exp[U] \left ( dt \, + \, \omega \right)^2 \, + \, \exp[-U] \, d\Omega^2_{spheroidal}
\end{equation}
we can easily check that it  is just the Kerr-Newman metric (\ref{KNmetra}) with $q=m$. The only necessary step, in order to verify such an identity is a redefinition of the coordinate $r$. If in the metric (\ref{KNmetra}) one replaces $r \rightarrow r+m$, then (\ref{KNmetra}) becomes identical to (\ref{BPSKN}).
\par
It is interesting to consider the expressions for the vector field strengths that solve the Maxwell-Einstein system together with the BPS Kerr-Newmann metric. For the first two field strengths (magnetic), from eq.(\ref{gromada}) we find:
\begin{eqnarray}
  F^1 &=& -\frac{1}{\sqrt{2}
   \left(r^2+\alpha ^2 \cos ^2\theta\right)^2 \left((m+r)^2+\alpha ^2 \cos ^2\theta\right)} \, \left( \sqrt[4]{3} m \alpha  \sin \theta \left(\left(\left(-3+2 \sqrt[4]{3}\right) \alpha ^4 \cos ^4\theta \right.\right.\right.\nonumber\\
   &&\left.\left.\left. +m \left(2 \sqrt[4]{3} m+m+2 \left(1+\sqrt[4]{3}\right)
   r\right) \alpha ^2 \cos ^2\theta-r (m+r)^2 \left(2 \sqrt[4]{3} m+\left(-3+2 \sqrt[4]{3}\right) r\right)\right) \sin \theta  {dr}\wedge  {d \phi  }\right.\right.\nonumber\\
   &&\left.\left.+2
   \left(r^2+\alpha ^2\right) \cos \theta \left(\left(\left(-2+\sqrt[4]{3}\right) m+\left(-3+2 \sqrt[4]{3}\right) r\right) \alpha ^2 \cos ^2\theta+(m+r)
   \left(\sqrt[4]{3} m^2\right.\right.\right.\right.\nonumber\\
   &&\left.\left.\left.\left.+\left(-1+3 \sqrt[4]{3}\right) r m+\left(-3+2 \sqrt[4]{3}\right) r^2\right)\right)  {d \theta  }\wedge  {d \phi  }\right)\right) \\
  F^2 &=& \frac{1}{\sqrt{2} 3^{3/4} \left(r^2+\alpha ^2 \cos ^2\theta\right)^2
   \left((m+r)^2+\alpha ^2 \cos ^2\theta\right)}\, \left (m \sin \theta \left(\alpha ^2 \left(-2 \cos \theta \sin \theta r^3 \right.\right.\right.\nonumber\\
   &&\left.\left. \left.  + m^2 \sin 2\theta r -2 (2 m+r) \alpha ^2 \cos ^3\theta \sin \theta\right)
    {dr}\wedge  {d \phi  }\right.\right.\nonumber\\
    &&\left.\left.-\frac{1}{8} \left(r^2+\alpha ^2\right) \left(8 r^4+16 m r^3+8 m^2 r^2+\alpha ^4 \right.\right.\right.\nonumber\\
    &&\left.\left.\left.-8 \alpha ^2 \left(-3 m^2-6 r m+\alpha ^2\right) \cos
   ^2\theta-\alpha ^4 \cos (4 \theta )\right)  {d \theta  }\wedge  {d \phi  }\right)\right) \label{magneticata}
\end{eqnarray}
while for the second two we get:
\begin{eqnarray}
 G^3 &=& \frac{1}{\sqrt{2} \left(r^2+\alpha ^2 \cos ^2\theta\right)^2
   \left((m+r)^2+\alpha ^2 \cos ^2\theta\right)} \, \left(3^{3/4} m \sin \theta \left(\left(\sin 2\theta r^3-2 m^2 \cos \theta \sin \theta r \right.\right.\right.\nonumber\\
   &&\left.\left.\left. +2 (2 m+r) \alpha ^2 \cos ^3\theta \sin \theta\right)
    {dr}\wedge  {d \phi  } \alpha ^2\right.\right.\nonumber\\
    &&\left.\left.+\frac{1}{8} \left(r^2+\alpha ^2\right) \left(8 r^4+16 m r^3+8 m^2 r^2+\alpha ^4-8 \alpha ^2 \left(-3 m^2-6 r m+\alpha ^2\right)
   \cos ^2\theta-\alpha ^4 \cos (4 \theta )\right)  {d \theta  }\wedge  {d \phi  }\right) \right) \nonumber\\
   \null & \null & \null \\
 G^4 &=& -\frac{1}{\sqrt{2} \left(r^2+\alpha ^2 \cos ^2\theta\right)^2
   \left((m+r)^2+\alpha ^2 \cos ^2\theta\right)}\, \left(m \alpha  \sin \theta \left(\left(-\left(-2+3 3^{3/4}\right) \alpha ^4 \cos ^4\theta \right.\right.\right. \nonumber\\
   &&\left.\left.\left.+m \left(\left(2+3^{3/4}\right) m+2 \left(1+3^{3/4}\right) r\right) \alpha
   ^2 \cos ^2\theta+r (m+r)^2 \left(\left(-2+3 3^{3/4}\right) r-2 m\right)\right) \sin \theta  {dr}\wedge  {d \phi  }\right.\right.\nonumber\\
   &&\left.\left.-2 \left(r^2+\alpha ^2\right) \cos
   \theta \left(-m^3+\left(-4+3^{3/4}\right) r m^2+\left(-5+4 3^{3/4}\right) r^2 m+\left(-2+3 3^{3/4}\right) r^3\right.\right.\right.\nonumber\\
   &&\left.\left.\left.+\left(\left(-1+2 3^{3/4}\right) m+\left(-2+3
   3^{3/4}\right) r\right) \alpha ^2 \cos ^2\theta\right)  {d \theta  }\wedge  {d \phi  }\right)\right)
\end{eqnarray}
The above expressions are rather formidable, yet considering them in some limit their meaning can be decoded. First of all we recall that in the limit
$\alpha \rightarrow 0$ the metric (\ref{BPSKN}) becomes the Reissner-Nordstrom metric. Correspondingly in the same limit the above four-vector of field strengths degenerates into:
\begin{equation}\label{alfadegen}
    \left(\begin{array}{c}
            F^1 \\
            F^2 \\
            G^3 \\
            G^4
          \end{array}
    \right) \, \stackrel{\alpha \to 0}{\Longrightarrow} \, \left(
\begin{array}{l}
 0 \\
 -\frac{m \sin (\theta )  {d \theta  }\wedge  {d \phi  }}{\sqrt{2} 3^{3/4}} \\
 \frac{3^{3/4} m \sin (\theta )  {d \theta  }\wedge  {d \phi  }}{\sqrt{2}} \\
 0
\end{array}
\right)
\end{equation}
showing that the black hole charges $\left(0, - \, \frac{m}{\sqrt{2} \, 3^{1/4}},\frac{m \, 3^{1/4}}{\sqrt{2}} ,0\right)$ have the correct form for a BPS black hole and are endowed with the characteristic $\mathbb{Z}_3$ symmetry.
\par
Also in the $\alpha \ne 0$ we can easily determine the black hole charges by integrating the field strengths on a two-sphere of very large radius $r \to \infty$. For this purpose it is important to evaluate the asymptotic expansion of the field strengths for large radius. We find:
\begin{equation}\label{rgrande}
    \left(\begin{array}{c}
            F^1 \\
            F^2 \\
            G^3 \\
            G^4
          \end{array}
    \right) \, \stackrel{r \to \infty}{\simeq} \, \left(
\begin{array}{l}
 -\frac{\sqrt{2} \sqrt[4]{3} \left(-3+2 \sqrt[4]{3}\right) m \alpha  \cos \theta \sin \theta  {d \theta  }\wedge  {d \phi
    }}{r}+\mathcal{O}\left(\frac{1}{r^2}\right) \\
 -\frac{m \sin \theta  {d \theta  }\wedge  {d \phi  }}{\sqrt{2} 3^{3/4}}+\mathcal{O}\left(\frac{1}{r^2}\right) \\
 \frac{3^{3/4} m \sin \theta  {d \theta  }\wedge  {d \phi  }}{\sqrt{2}}+\mathcal{O}\left(\frac{1}{r^2}\right) \\
 \frac{\sqrt{2} \left(-2+3 3^{3/4}\right) m \alpha  \cos \theta \sin \theta  {d \theta  }\wedge  {d \phi }}{r}+\mathcal{O}\left(\frac{1}{r^2}\right)
\end{array}
\right)
\end{equation}
and the integration on the angular variables produces the same result as for the corresponding  Reissner-Nordstrom black hole:
\begin{equation}\label{gomotarro}
    \mathcal{Q}_{BPSKN} \, = \, \left(0, - \, \frac{m}{\sqrt{2} \, 3^{1/4}},\frac{m \, 3^{1/4}}{\sqrt{2}} ,0\right)
\end{equation}
In conclusion the BPS Kerr-Newman solution is a deformation of the Reissner-Nordstrom BPS black hole. It is extremal in the $\sigma$-model sense and for this reason could be retrieved from the nilpotent orbit construction. However it is not extremal in the sense of General Relativity since the mass is less than $\sqrt{q^2 +\alpha^2}$ being equal to $m$. For this reason we are below the limit of the cosmic censorship, there is no horizon and we have instead a naked singularity.
\par
The important message is that, notwithstanding the deformation and the presence of a Kaluza-Klein vector, the structure of the charges is that pertaining to the orbit where the solution has been constructed, namely the BPS orbit $\mathcal{O}^3_{11}$.
\subsection{The Large non BPS Black Holes of $\mathcal{O}^3_{22}$}
Next let us consider the orbit $\mathcal{O}^3_{22}$, which in the spherical symmetric case leads to non BPS Black holes with a finite horizon area.
\paragraph{$\mathbf{W}$-representation}
As in the previous case, in order to better appreciate the structure of these solutions, let us slightly generalize our orbit representative, writing the following nilpotent matrix that depends on two parameters $(p,q)$
\begin{equation}\label{xpq322}
  X_{3|22}(p,q) \, = \left(
\begin{array}{lllllll}
 q & 0 & 0 & \frac{q}{\sqrt{2}} & 0 & 0 & 0 \\
 0 & \frac{p+q}{2} & -\frac{p}{2} & 0 & -\frac{q}{2} & 0 & 0
   \\
 0 & \frac{p}{2} & \frac{q-p}{2} & 0 & 0 & \frac{q}{2} & 0 \\
 -\frac{q}{\sqrt{2}} & 0 & 0 & 0 & 0 & 0 & -\frac{q}{\sqrt{2}}
   \\
 0 & \frac{q}{2} & 0 & 0 & \frac{p-q}{2} & \frac{p}{2} & 0 \\
 0 & 0 & -\frac{q}{2} & 0 & -\frac{p}{2} & \frac{1}{2} (-p-q)
   & 0 \\
 0 & 0 & 0 & \frac{q}{\sqrt{2}} & 0 & 0 & -q
\end{array}
\right)
\end{equation}
The standard triple representative mentioned in eq.(\ref{322}) is just the particular case $X_{3|22}(1,1)$.
Applying the usual strategy from the general formula we obtain
\begin{equation}\label{chiavina322}
  \mathcal{Q}^\mathbf{w}_{3|22} \, = \, \mathrm{Tr}(X_{3|22}(p,q)\mathcal{T}^{\mathbf{w}} )
 \, = \, \left(0,p,\sqrt{3} q,0\right)
\end{equation}
Substituting such a result in the expression for the quartic symplectic invariant (see eq.(\ref{invola}) we find:
\begin{equation}\label{genorama322}
    \mathfrak{I}_4 \, = \, - \, 9 \, p \, q^3 \, < \, 0 \quad \mbox{if $p$ and $q$ have the same sign}
\end{equation}
This result is meaningful since, by calculating the trace $\mbox{Tr}(X_{3|22} L_+^E) \, = \,0$, we find that the Taub-NUT charge vanishes. Furthermore we note that the condition that $p$ and $q$ have the same sign was singled out in \cite{noig22} as the defining condition of the orbit $O^{3}_{22}$ which, in the spherical symmetry approach leads to regular non BPS solutions. The choice of opposite signs was proved in \cite{noig22} to correspond to a different $\mathrm{H}^\star$ orbit, the non diagonal $O^{3}_{12}$ which instead contains only singular solutions.
\par
Addressing the question  of stability subgroups of the original duality group in four-dimensions $\mathrm{SL(2,\mathrm{R})}$, we realize that for the charge vector (\ref{chiavina322}) this subgroup is just trivial:
\begin{equation}\label{paraboliconiente}
   \mathrm{SL(2,\mathbb{R})} \, \supset \,  \mathcal{S}_{3|22}  \, = \,\mathbf{ 1}
\end{equation}
\par
\paragraph{$\mathrm{H}^\star$-stability subgroup} Considering next the stability subgroup of the nilpotent element $X_{3|22}(1,1)$ in $\mathrm{H}^\star \, = \, \widehat{\slal(2,\mathbb{R})}\oplus\slal(2,\mathbb{R})_{\mathrm{h}^\star}$ we obtain:
\begin{equation}\label{gruppoS211}
    \mathfrak{S}_{3|22}  \, = \, \mathbb{R}
\end{equation}
the group being generated by a matrix $\mathbb{A}_{3|22}$ of nilpotency degree $2$:
\begin{eqnarray}\label{gigio311}
   \mathbb{A}_{3|22} ^3\, = \, \mathbf{0}
\end{eqnarray}
We do not give its explicit form which we do not use in the sequel.
\paragraph{Nilpotent algebra $\mathbb{N}_{3|22}$ } Considering next the adjoint action of the central element $h_{3|22}$ on the subspace $\mathbb{K}^\star$ we find that its eigenvalues are the following ones:
\begin{equation}\label{eigetti}
    \mbox{Eigenvalues}_{3|22}^{\mathbb{K}^\star}\, = \, \{-4,4,-2,-2,2,2,0,0\}
\end{equation}
Therefore the three eigenoperators $A_1,A_2,A_3$ corresponding to the three positive eigenvalues $4,2,2$, respectively, form the restriction to $\mathbb{K}^\star$ of a nilpotent algebra $\mathbb{N}_{3|22}$. In this case the $A_i$ do  all commute among themselves so that  we have $\mathbb{N}_{3|22}\, = \, \mathbb{N}_{3|22}\bigcap \mathbb{K}^\star$ and it is abelian.
The abelian structure of the nilpotent algebra implies that  for the orbit $\mathcal{O}^3_{22}$ we have only three functions $\mathfrak{h}^{0}_i$ which will be harmonic and independent. This is so because $\mathcal{D} \mathbb{N}_{3|22} \,= \, 0$
\par
Explicitly we set:
\begin{eqnarray*}\label{girollo322}
   & \mathfrak{H} (\mathfrak{h}_1,\mathfrak{h}_2,\mathfrak{h}_3) \, = \, \sum_{i=1}^3 \, \mathfrak{h}_i \, A_i \, = \, &\nonumber\\
   \end{eqnarray*}
  {\small \begin{eqnarray}
&\left(
\begin{array}{lllllll}
 2 \mathfrak{h}_3 & \mathfrak{h}_1-2 \mathfrak{h}_2 & 2 \mathfrak{h}_1-\mathfrak{h}_2 & -\sqrt{2} \mathfrak{h}_3 & -3 \mathfrak{h}_2 & -3
   \mathfrak{h}_1 & 0 \\
 \mathfrak{h}_1-2 \mathfrak{h}_2 & \mathfrak{h}_3 & 0 & \sqrt{2} \mathfrak{h}_2-2 \sqrt{2} \mathfrak{h}_1 & \mathfrak{h}_3 & 0 &
   -3 \mathfrak{h}_1 \\
 \mathfrak{h}_2-2 \mathfrak{h}_1 & 0 & \mathfrak{h}_3 & \sqrt{2} \mathfrak{h}_1-2 \sqrt{2} \mathfrak{h}_2 & 0 & -\mathfrak{h}_3
   & -3 \mathfrak{h}_2 \\
 \sqrt{2} \mathfrak{h}_3 & 2 \sqrt{2} \mathfrak{h}_1-\sqrt{2} \mathfrak{h}_2 & \sqrt{2} \mathfrak{h}_1-2
   \sqrt{2} \mathfrak{h}_2 & 0 & \sqrt{2} \mathfrak{h}_1-2 \sqrt{2} \mathfrak{h}_2 & \sqrt{2}
   \mathfrak{h}_2-2 \sqrt{2} \mathfrak{h}_1 & \sqrt{2} \mathfrak{h}_3 \\
 3 \mathfrak{h}_2 & -\mathfrak{h}_3 & 0 & \sqrt{2} \mathfrak{h}_1-2 \sqrt{2} \mathfrak{h}_2 & -\mathfrak{h}_3 & 0 & 2
   \mathfrak{h}_1-\mathfrak{h}_2 \\
 -3 \mathfrak{h}_1 & 0 & \mathfrak{h}_3 & 2 \sqrt{2} \mathfrak{h}_1-\sqrt{2} \mathfrak{h}_2 & 0 & -\mathfrak{h}_3 &
   \mathfrak{h}_1-2 \mathfrak{h}_2 \\
 0 & -3 \mathfrak{h}_1 & 3 \mathfrak{h}_2 & -\sqrt{2} \mathfrak{h}_3 & \mathfrak{h}_2-2 \mathfrak{h}_1 & \mathfrak{h}_1-2 \mathfrak{h}_2 &
   -2 \mathfrak{h}_3
\end{array}
\right)  &\nonumber\\
\label{girollo322}
\end{eqnarray}}
Considering $ \mathfrak{H} (\mathfrak{h}_1,\mathfrak{h}_2,\mathfrak{h}_3)$ as a Lax operator and calculating its Taub-NUT charge and electromagnetic charges we find:
\begin{equation}\label{fistl322}
    \mathbf{n}_{TN} \, = \, -6  \mathfrak{h}_1 \quad ; \quad \mathcal{Q} \, = \, \left\{2 \sqrt{3} \left(\mathfrak{h}_2-2 \mathfrak{h}_1\right),0,-2 \sqrt{3} \mathfrak{h}_3,-6
   \mathfrak{h}_2\right\}
\end{equation}
 This implies that constructing the multi-centre solution with harmonic functions the condition $\mathfrak{h}_1 \, = 0$ might be sufficient to annihilate the Taub-NUT current.  In this case we will be lucky and such a condition suffices.
\par
For later convenience let us change the normalization in the basis of harmonic functions as follows:
\begin{equation}\label{gignolo322}
    \mathfrak{h}_1^{(0)} \, = \,  \mathcal{H}_1 \quad ; \quad  \mathfrak{h}_2^{(0)} \, = \, \ft 12 \left(1-\mathcal{H}_2\right)\quad ; \quad \mathfrak{h}_3^{(0)} \, = \, \ft 12 \left(1-\mathcal{H}_3\right)
\end{equation}
Implementing the symmetric coset construction with:
\begin{equation}\label{filodoro}
    \mathcal{Y}\left(\mathcal{H}_1,\mathcal{H}_2,\mathcal{H}_3\right) \, \equiv \,
    \exp \left[  \mathfrak{H} \left(\mathcal{H}_1 ,\ft 12 \left(1-\mathcal{H}_2\right),\ft 12 \left(1-\mathcal{H}_3\right)\right)\right]
\end{equation}
calculating the upper triangular coset representative $\mathbb{L}(\mathcal{Y})$ according to equations (\ref{solutioncosetrepr}) and extracting   the $\sigma$-model scalar fields we obtain an explicit expression which is sufficiently simple to be displayed:
\begin{eqnarray}
  \exp\left[- \, U\right] &=& \sqrt{\mathcal{H}_2 \mathcal{H}_3^3-4 \mathcal{H}_1^2} \label{u322}\\
  \mbox{Im} \, z &=& \frac{\sqrt{\mathcal{H}_2 \mathcal{H}_3^3-4
   \mathcal{H}_1^2}}{\mathcal{H}_3^2} \label{imz322}\\
\mbox{Re} \, z &=& -\frac{2 \mathcal{H}_1}{\mathcal{H}_3^2} \label{rez322}
\end{eqnarray}
\begin{equation}\label{zz322}
  Z^M \, = \,  \left(
\begin{array}{l}
 -\frac{\sqrt{6} \mathcal{H}_1 \mathcal{H}_3}{4
   \mathcal{H}_1^2-\mathcal{H}_2 \mathcal{H}_3^3} \\
 \frac{4 \mathcal{H}_1^2-\left(\mathcal{H}_2-1\right)
   \mathcal{H}_3^3}{\sqrt{2} \left(4
   \mathcal{H}_1^2-\mathcal{H}_2 \mathcal{H}_3^3\right)} \\
 \frac{\sqrt{\frac{3}{2}} \left(4
   \mathcal{H}_1^2-\mathcal{H}_2 \left(\mathcal{H}_3-1\right)
   \mathcal{H}_3^2\right)}{4 \mathcal{H}_1^2-\mathcal{H}_2
   \mathcal{H}_3^3} \\
 \frac{\sqrt{2} \mathcal{H}_1 \mathcal{H}_2}{4
   \mathcal{H}_1^2-\mathcal{H}_2 \mathcal{H}_3^3}
\end{array}
\right)
\end{equation}
\begin{equation}\label{a322}
    a \, = \,-\frac{\mathcal{H}_1 \left(\mathcal{H}_2+3
   \mathcal{H}_3-2\right)}{4 \mathcal{H}_1^2-\mathcal{H}_2
   \mathcal{H}_3^3}
\end{equation}
Using these results  we easily obtain the Taub-NUT current in the following form:
\begin{equation}\label{grignolo322}
    j^{TN} \,=\,2 \, \star\nabla \mathcal{H}_1
\end{equation}
In this case the predicted condition $\mathcal{H}_1\,=\,0$ is sufficient to annihilate the Taub-NUT current and we obtain an extremely simple result\footnote{Actually even the condition $\mathcal{H}_1 \, = \, \mbox{const}$ suffices to annihilate the Taub-NUT charge allowing for a non trivial real part of the $z$-field. However in this section we analyze the case $\mathcal{H}_1 \, = \, 0$ for its remarkable simplicity}. The complete form of the supergravity solution corresponding to this choice is:
      \begin{eqnarray}
        \exp[\,-\,U] &=& \sqrt{\mathcal{H}_3^3 \mathcal{H}_2} \\
        z &=& {\rm i} \,\frac{ \sqrt{\mathcal{H}_3^3
   \mathcal{H}_2}}{\mathcal{H}_3^2} \\
        j^{TN} &=& 0 \\
        j^{EM} &=& \star \nabla \, \left(
\begin{array}{l}
 0 \\
 - \, \frac{ \mathcal{H}_2}{\sqrt{2}} \\
 - \, \sqrt{\frac{3}{2}}  \mathcal{H}_3 \\
 0
\end{array}
\right) \label{soluz322atomica}
      \end{eqnarray}
Comparing with the case of the large BPS orbit we see that the only difference is the relative sign of the harmonic functions in the electromagnetic current. What we said for the BPS black holes extends to the non BPS ones in the same way.
\paragraph{Summary} For a multicenter solution associated with the $\mathcal{O}^3_{22}$ orbit  we have a mixture of very small and small black holes as in the case of the orbit $\mathcal{O}^3_{22}$.  Also here a finite area non BPS black hole emerges when the center of a very small black comes to coincides with the center of a small one. The only difference is the relative sign of the two charges. With equal signs we construct a non BPS state, while with opposite charges we construct a BPS one. This reinforces  the conjecture that at the quantum level finite black holes can be interpreted as composite states.
\par
This conjecture is also supported by an angular momentum analysis. Looking at the representations in table \ref{tablettina}, we see that the representation $2(j=1)+(j=0)$ that corresponds to BPS and non BPS large black holes can be obtained by summing the representation $(j=1)+2(j=\ft 12 ) $ that corresponds to small black holes with the representation $3(j=0)+2(j=\ft 12 ) $ that corresponds to very small black holes. Consider the following table:
$$\begin{array}{|l|l|l|l|l|l|l|}
\hline
 1 & \frac{1}{2} & \frac{1}{2} & 0 & -\frac{1}{2} &
   -\frac{1}{2} & -1 \\
   \hline
 0 & \frac{1}{2} & -\frac{1}{2} & 0 & \frac{1}{2} &
   -\frac{1}{2} & 0 \\
   \hline
   \hline
 1 & 1 & 0 & 0 & 0 & -1 & -1\\
 \hline
\end{array}$$
the numbers in the first line are the eigenvalues of the central element $h$ in the triplet $(h,X,Y)$ characterizing the orbit $\mathcal{O}^4_{11}$. The second line contains the eigenvalues for the central element of the triplet of the orbit $\mathcal{O}^4_{11}$. In the last line we have the eigenvalues for the $h$ in the triplet characterizing  the orbit $\mathcal{O}^3_{i,j}$. We realize that the coincidence of centres correspond to the identification of a new $\mathrm{SL(2,\mathbb{}R)}$ subgroup which is the direct sum of the original two associated with the two small black holes.
\subsection{The Largest orbit $\mathcal{O}^1_{11}$}
Next let us consider the orbit $\mathcal{O}^1_{11}$, which in the spherical symmetric case leads only to singular solutions.
\paragraph{$\mathbf{W}$-representation}
Applying the usual strategy from the general formula we obtain a charge vector
\begin{equation}\label{chiavina111}
  \mathcal{Q}^\mathbf{w}_{1|11} \, = \, \mathrm{Tr}(X_{1|11}(p,q)\mathcal{T}^{\mathbf{w}} )
\end{equation}
which has no invariance:
\begin{equation}\label{paraboliconiente111}
   \mathrm{SL(2,\mathbb{R})} \, \supset \,  \mathcal{S}_{1|11}  \, = \,\mathbf{ 1}
\end{equation}
and yields a quartic invariant generically different from zero:
\begin{equation}\label{genorama322}
    \mathfrak{I}_4 \, \neq \, 0
\end{equation}
Because of our simplified choice of the representative the Taub-NUT charge is not zero and only later we will enforce the vanishing of the Taub-NUT current on the harmonic function parameterized solution.
\par
\paragraph{$\mathrm{H}^\star$-stability subgroup} Considering next the stability subgroup of the nilpotent element $X_{1|11}$ in $\mathrm{H}^\star \, = \, \widehat{\slal(2,\mathbb{R})}\oplus\slal(2,\mathbb{R})_{\mathrm{h}^\star}$ we obtain that it is trivial:
\begin{equation}\label{gruppoS211}
    \mathfrak{S}_{1|11}  \, = \, \mathbf{1}
\end{equation}
\paragraph{Nilpotent algebra $\mathbb{N}_{1|11}$ } Considering next the adjoint action of the central element $h_{1|11}$ on the subspace $\mathbb{K}^\star$ we find that its eigenvalues are the following ones:
\begin{equation}\label{eigetti}
    \mbox{Eigenvalues}_{3|22}^{\mathbb{K}^\star}\, = \, \{-5,5,-3,3,-1,-1,1,1\}
\end{equation}
Therefore the four eigenoperators $A_1,A_2,A_3,A_4$ corresponding to the four positive eigenvalues $5,3,1,1$, respectively, form the restriction to $\mathbb{K}^\star$ of a nilpotent algebra $\mathbb{N}_{1|11}$. In this case the $A_i$ do not all commute among themselves so that  we have $\mathbb{N}_{1|11}\, \neq \, \mathbb{N}_{1|11}\bigcap \mathbb{K}^\star$. The full algebra involves also two operators $B_1,B_2 \, \in \, \mathbb{H}^\star$ and the full set of commutation relations is the following one:
\begin{eqnarray}
  0 &=& \left [ A_1\, , \, A_2\right] \, = \, \left [ A_1\, , \, A_3\right]\, = \, \left [ A_1\, , \, A_4\right]\nonumber\\
  0 &=& \left [ A_2\, , \, A_3\right] \nonumber\\
  0 &=& \left [ B_1\, , \, B_2\right]\, = \,\left [ B_1\, , \, A_1\right]\, = \,\left [ B_1\, , \, A_2\right]\nonumber\\
  0 &=& \left [ B_1\, , \, A_4\right]\, = \,\left [ B_2\, , \, A_1\right]\, = \,\left [ B_2\, , \, A_3\right]\nonumber\\
  B_1 &=& \left [ A_2\, , \, A_4\right]\nonumber\\
  B_2 &=& \left [ A_3\, , \, A_4\right] \nonumber\\
  -16 \, A_1 &=& \left [ B_1\, , \, A_3\right]\nonumber\\
  -16 \, A_1& = & \left [ B_2\, , \, A_1\right]\nonumber\\
  24 \, A_2 & = & \left [ B_2\, , \, A_4\right] \label{gomorra}
\end{eqnarray}
By inspection of eq.s(\ref{gomorra}) we easily see that:
\begin{eqnarray}
  \mathcal{D}\mathbb{N}_{1|11} &=& \mbox{span} \left\{B_1,B_2,A_1,A_2 \right\}
  \quad ; \quad \mathcal{D}\mathbb{N}_{1|11}\bigcap \mathbb{K}^\star \, = \, \mbox{span} \left\{A_1,A_2 \right\} \\
  \mathcal{D}^2\mathbb{N}_{1|11}&=& \mbox{span} \left\{A_1 \right\} \, = \, \mathcal{D}^2\mathbb{N}_{1|11}\bigcap \mathbb{K}^\star \end{eqnarray}
This structure of the nilpotent algebra implies that  for the orbit $\mathcal{O}^1_{11}$ we have only two functions $\mathfrak{h}^{0}_3,\mathfrak{h}^{0}_4 $ which are harmonic and independent. The other two functions
$\mathfrak{h}^{2}_1,\mathfrak{h}^{1}_2 $,  obey instead equations in which the previous two play the role of sources.
Not surprisingly $\mathfrak{h}^{2}_1,\mathfrak{h}^{1}_2 $ correspond to the higher gradings $5$ and $3$, while $\mathfrak{h}^{0}_3,\mathfrak{h}^{0}_4 $ correspond to the gradings $1,1$. More precisely $\mathfrak{h}^{1}_2 $ receives source contributions only from $\mathfrak{h}^{0}_3,\mathfrak{h}^{0}_4 $, while $\mathfrak{h}^{2}_1$ receives source contributions from
$\mathfrak{h}^{1}_2,\mathfrak{h}^{0}_3,\mathfrak{h}^{0}_4 $
\par
Explicitly we set:
\begin{eqnarray*}
   & \mathfrak{H} (\mathfrak{h}_1,\dots,\mathfrak{h}_4) \, = \, \sum_{i=1}^4 \, \mathfrak{h}_i \, A_i \, = \, &\nonumber\\
   \end{eqnarray*}
  {\small \begin{eqnarray}
&\left(
\begin{array}{lllllll}
 \mathfrak{h}_1+\mathfrak{h}_4 & \frac{\mathfrak{h}_2}{3}-\mathfrak{h}_3 & \mathfrak{h}_4 & \frac{\sqrt{2}
   \mathfrak{h}_2}{3}-\sqrt{2} \mathfrak{h}_3 & -\mathfrak{h}_1 & -\mathfrak{h}_2-\mathfrak{h}_3 & 0 \\
 \frac{\mathfrak{h}_2}{3}-\mathfrak{h}_3 & 2 \mathfrak{h}_4 & \mathfrak{h}_2+\mathfrak{h}_3 & -\sqrt{2} \mathfrak{h}_4 &
   \mathfrak{h}_3-\frac{\mathfrak{h}_2}{3} & 0 & -\mathfrak{h}_2-\mathfrak{h}_3 \\
 -\mathfrak{h}_4 & -\mathfrak{h}_2-\mathfrak{h}_3 & \mathfrak{h}_1-\mathfrak{h}_4 & \frac{\sqrt{2} \mathfrak{h}_2}{3}-\sqrt{2}
   \mathfrak{h}_3 & 0 & \frac{\mathfrak{h}_2}{3}-\mathfrak{h}_3 & -\mathfrak{h}_1 \\
 \sqrt{2} \mathfrak{h}_3-\frac{\sqrt{2} \mathfrak{h}_2}{3} & \sqrt{2} \mathfrak{h}_4 &
   \frac{\sqrt{2} \mathfrak{h}_2}{3}-\sqrt{2} \mathfrak{h}_3 & 0 & \frac{\sqrt{2}
   \mathfrak{h}_2}{3}-\sqrt{2} \mathfrak{h}_3 & -\sqrt{2} \mathfrak{h}_4 & \sqrt{2}
   \mathfrak{h}_3-\frac{\sqrt{2} \mathfrak{h}_2}{3} \\
 \mathfrak{h}_1 & \frac{\mathfrak{h}_2}{3}-\mathfrak{h}_3 & 0 & \frac{\sqrt{2} \mathfrak{h}_2}{3}-\sqrt{2}
   \mathfrak{h}_3 & \mathfrak{h}_4-\mathfrak{h}_1 & -\mathfrak{h}_2-\mathfrak{h}_3 & \mathfrak{h}_4 \\
 -\mathfrak{h}_2-\mathfrak{h}_3 & 0 & \mathfrak{h}_3-\frac{\mathfrak{h}_2}{3} & \sqrt{2} \mathfrak{h}_4 & \mathfrak{h}_2+\mathfrak{h}_3 &
   -2 \mathfrak{h}_4 & \frac{\mathfrak{h}_2}{3}-\mathfrak{h}_3 \\
 0 & -\mathfrak{h}_2-\mathfrak{h}_3 & \mathfrak{h}_1 & \frac{\sqrt{2} \mathfrak{h}_2}{3}-\sqrt{2} \mathfrak{h}_3 &
   -\mathfrak{h}_4 & \frac{\mathfrak{h}_2}{3}-\mathfrak{h}_3 & -\mathfrak{h}_1-\mathfrak{h}_4
\end{array}
\right)  &\nonumber\\
\label{girollono111}
\end{eqnarray}}
Considering $ \mathfrak{H} (\mathfrak{h}_1,\dots,\mathfrak{h}_4)$ as a Lax operator and calculating its Taub-NUT charge and electromagnetic charges we find:
\begin{equation}\label{fistl322}
    \mathbf{n}_{TN} \, = \, -2  (\mathfrak{h}_2 + \mathfrak{h}_3) \quad ; \quad \mathcal{Q} \, = \, \left\{-2 \sqrt{3} \mathfrak{h}_4,-2 \left(\mathfrak{h}_2+\mathfrak{h}_3\right),\frac{2
   \left(\mathfrak{h}_2-3 \mathfrak{h}_3\right)}{\sqrt{3}},-2 \mathfrak{h}_1\right\}
\end{equation}
 This implies that constructing the multi-centre solution with harmonic functions the condition $\mathfrak{h}_2 \, = \,  - \,\mathfrak{h}_3$ might be sufficient to annihilate the Taub-NUT current.
\par
Implementing the symmetric coset construction with:
\begin{equation}\label{filodoro}
    \mathcal{Y}\left(\mathfrak{h}_1,\dots,\mathfrak{h}_4\right) \, \equiv \,
    \exp \left[  \mathfrak{H} \left(\mathfrak{h}_1,\dots,\mathfrak{h}_4\right)\right]
\end{equation}
and imposing the field equations (\ref{equatata}) we obtain the following conditions:
\begin{eqnarray}
  0 &=& \frac{224}{5} \nabla \mathfrak{h}_3\circ \nabla \mathfrak{h}_3
   \,\mathfrak{h}_4^3-\frac{16}{5} \,\mathfrak{h}_3 \Delta
   \,\mathfrak{h}_3 \,\mathfrak{h}_4^3-\frac{416}{5} \nabla
   \,\mathfrak{h}_3\circ \nabla \mathfrak{h}_4 \,\mathfrak{h}_3
   \,\mathfrak{h}_4^2+\frac{16}{5} \,\mathfrak{h}_3^2 \Delta
   \,\mathfrak{h}_4 \,\mathfrak{h}_4^2\nonumber\\
   &&+\frac{192}{5} \nabla
   \,\mathfrak{h}_4\circ \nabla \mathfrak{h}_4 \,\mathfrak{h}_3^2
   \,\mathfrak{h}_4+\frac{32}{3} \nabla \mathfrak{h}_2\circ
   \nabla \mathfrak{h}_3 \,\mathfrak{h}_4-\frac{8}{3}
   \,\mathfrak{h}_3 \Delta\mathfrak{h}_2
   \,\mathfrak{h}_4-\frac{8}{3} \,\mathfrak{h}_2 \Delta
   \,\mathfrak{h}_3 \,\mathfrak{h}_4\nonumber\\
   &&-\frac{16}{3} \nabla
   \,\mathfrak{h}_3\circ \nabla \mathfrak{h}_4
   \,\mathfrak{h}_2-\frac{16}{3} \nabla \mathfrak{h}_2\circ
   \nabla \mathfrak{h}_4 \,\mathfrak{h}_3+\Delta
   \,\mathfrak{h}_1+\frac{16}{3} \,\mathfrak{h}_2 \,\mathfrak{h}_3
   \Delta\mathfrak{h}_4\nonumber\\
 0 &=& 4 \Delta\mathfrak{h}_3 \,\mathfrak{h}_4^2-8 \nabla
   \,\mathfrak{h}_3\circ \nabla \mathfrak{h}_4 \,\mathfrak{h}_4-4
   \,\mathfrak{h}_3 \Delta\mathfrak{h}_4 \,\mathfrak{h}_4+8
   \nabla \mathfrak{h}_4\circ \nabla \mathfrak{h}_4
   \,\mathfrak{h}_3+\Delta\mathfrak{h}_2\nonumber\\
 0 &=& \Delta\mathfrak{h}_3 \nonumber\\
 0 &=& \Delta\mathfrak{h}_4 \label{equaziatofili}
\end{eqnarray}
Solutions of the above system can be quite complicated and can encompass many different types of behaviors, yet what is generically true is that the contributions from the source term introduces in $\mathfrak{h}_1$ and $\mathfrak{h}_2$ poles $1/r^p$ stronger than $p=1$ , while $\mathfrak{h}_3$ and $\mathfrak{h}_4$ have only simple poles.  Hence if the structure of the polynomials in the functions $\mathfrak{h}_{1,2,3,4}$ is such that at simple poles the divergence of the inverse warp factor is already too strong or the coefficient already becomes imaginary, introducing stronger poles can only make the situation worse. For this reason we confine ourselves to analyze  solutions encompassed in this orbit in which the source terms vanish identically upon the implementation of some identifications.
\par
There are few different reductions with such a property and we choose just one that has also the additional feature of annihilating the Taub-NUT current. It is the following one:
\begin{equation}\label{reduzionna}
    \mathfrak{h}_3\, = \, \mathfrak{h}_4 \, = \, -\, \mathfrak{h}_2 \, \equiv \, \mathfrak{h}
\end{equation}
The reader can easily check that with the choice (\ref{reduzionna}) the system of equations (\ref{equaziatofili})  reduces to:
\begin{equation}\label{curlandia}
    \Delta\mathfrak{h} \, = \,  \Delta\mathfrak{h}_1 \, = \, 0
\end{equation}
For later convenience let us change the normalization in the basis of harmonic functions as follows:
\begin{equation}\label{gignolo111}
    \mathfrak{h}_4 \, = \, \ft 14 \,\mathcal{H} \quad ; \quad \mathfrak{h}_3 \, = \, \ft 14 \,\mathcal{H}\quad ; \quad \mathfrak{h}_2 \, = \, -\, \ft 14 \,\mathcal{H}\quad ; \quad \mathfrak{h}_1 \, = \, -\, \ft 14 \,+ \, \mathcal{W}
\end{equation}
calculating the upper triangular coset representative $\mathbb{L}(\mathcal{Y})$ according to equations (\ref{solutioncosetrepr}) and extracting   the $\sigma$-model scalar fields we obtain explicit expressions which are sufficiently simple to be displayed:
\begin{eqnarray}
  \exp\left[\, U\right] &=& \frac{8 \sqrt{15}}{\sqrt{-(\mathcal{H}+2)^3 \left(\mathcal{H}^5+10 \mathcal{H}^4+40 \mathcal{H}^3+80 \mathcal{H}^2-60 (4 \mathcal{W }+1)\right)}} \label{u111}\\
  \mbox{Im} \, z &=& \frac{3 \sqrt{15} (\mathcal{H}+2)}{\sqrt{-\frac{\mathcal{H}+2}{\mathcal{H}^2 (\mathcal{H} (\mathcal{H} (\mathcal{H}+10)+40)+80)-60 (4 \mathcal{W} +1)}} \left((\mathcal{H}
   (\mathcal{H} (\mathcal{H}+10)+20)-40) \mathcal{H}^2+90 (4 \mathcal{W} +1)\right)}\nonumber\\
   \null & \null & \null \label{imz111}\\
\mbox{Re} \, z &=& \frac{15 \mathcal{H} (\mathcal{H}+2) (\mathcal{H}+4)}{\mathcal{H}^5+10 \mathcal{H}^4+20 \mathcal{H}^3-40 \mathcal{H}^2+360 \mathcal{W} +90} \label{rez111}
\end{eqnarray}
We skip the form of the $Z$ fields and of $a$ but we mention their consequences, namely the Taub-NUT current
\begin{equation}\label{grignolo111}
    j^{TN} \,=\, 0
\end{equation}
and the electromagnetic currents
\begin{eqnarray}
        j^{EM} &=& \star \nabla \, \left\{\frac{1}{2} \sqrt{\frac{3}{2}}  \mathcal{H},0,\frac{7  \mathcal{H}}{6},\sqrt{2}  \mathcal{W} \right\} \label{soluz111atomica}
      \end{eqnarray}
This shows that a black hole belonging to this orbit has a charge vector $\mathcal{Q} \, = \, \left\{\frac{1}{2} \sqrt{\frac{3}{2}}  p,0,\frac{7  p}{6},\sqrt{2}  q \right\}$, whose quartic invariant is:
\begin{equation}\label{ragliando}
    \mathfrak{I}_4 \, = \, \frac{1}{128}  p^3 (49  p+72 q )
\end{equation}
This latter can be positive or negative depending on the choices for $p$ and $q$. The problem, however, is that this solution is always singular around all poles of $\mathcal{H}$. Indeed setting:
\begin{equation}\label{gartollo}
    \mathcal{H} \, \sim \, \frac{p}{r} \quad ; \quad \mathcal{W} \, \sim \, \frac{q}{r}
\end{equation}
we find that for $r\to 0$ the inverse warp factor behaves as follows:
\begin{equation}\label{carneade}
  \exp[-U] \, \sim \,   \frac{\sqrt{-p^8}}{8 \sqrt{15} r^4}+\frac{\sqrt{-p^8}}{\sqrt{15} p r^3}+\frac{\sqrt{\frac{3}{5}} \sqrt{-p^8}}{p^2 r^2}+\frac{4 \sqrt{-p^8}}{\sqrt{15} p^3
   r}+\frac{\sqrt{\frac{3}{5}} p^3 (p+5 q)}{\sqrt{-p^8}}+\mathcal{O}\left(r\right)
\end{equation}
The coefficient $\sqrt{-p^8}$ indicates that approaching the pole the warp factor becomes  imaginary at a finite distance from it and the would be horizon $r \, = \, 0$ is never reached. If it were reached, the divergence $\frac{1}{r^4}$ would imply an infinite area of the horizon.
As we know from our general discussion the Riemann tensor diverges if the warp factor goes to zero faster than $r^2$ so that the would be horizon would actually be a singularity. Yet since the warp factor becomes imaginary at a finite distance from the pole it remains open the question if solutions of this type  can be prolonged by suitably changing the coordinate system.  In that case they might acquire a physical meaning. So far such a question has not been tackled but it deserves to be.
\section{Classification of the sugra-relevant symmetric spaces and discussion of their general properties}
\label{sugrarelevant}
As we highlighted in the introduction there is a general group-theoretical framework underlying the construction of supergravity black holes which allows both for
\begin{description}
  \item[1)] a classification of the relevant symmetric spaces,
  \item[2)] a general description of their structures which are relevant to the black hole solutions.
\end{description}
The presentation of both items in the above list is the goal of
the present section. To achieve such a goal we need to emphasize a
few general aspects of the decomposition (\ref{gendecompo}) that
relate to the underlying root systems and Dynkin diagrams. In the
following we heavily rely on results presented several years ago
in \cite{Fre':2005si}. Indeed from the algebraic view-point a
crucial property of the general decomposition in
eq.(\ref{gendecompo}) is encoded into the following statements
which are true for all the cases \footnote{An apparent exception
is given by the case of $\mathcal{N}=3$ supergravity. The extra
complicacy, there, is that the duality algebra in $D=3$, namely
$\mathbb{U}_{D=3}$ has rank $r+2$, rather than $r+1$ with respect
to the rank of the algebra $\mathbb{U}_{D=4}$. Actually in this
case there is an extra $\mathrm{U(1)_Z}$ factor that is active on
the vectors, but not on the scalars and which is responsible for
the additional complications. It happens in this case that there
are two vector roots, one for the complex representation to which
the vectors are assigned and one for its conjugate. They have
opposite charges under $\mathrm{U(1)_Z}$. This case together with
that of $\mathcal{N}=5$ supergravity and with one of the series of
$\mathcal{N}=2$ theories completes the list of three
\textit{exotic models} which are anomalous also from the point of
view of the Tits Satake projection (see below).}:
\begin{enumerate}
  \item The  $A_1$ root-system associated with the
  ${\slal(2,\mathbb{R})_E}$ algebra in the decomposition
  (\ref{gendecompo}) is made of $\pm \, \psi$ where $\psi$ is the
  highest root of $\mathbb{U}_{D=3}$.
  \item Out of the $r$ simple roots $\alpha_i$ of $\mathbb{U}_{D=3}$
  there are $r-1$ that have grading zero with respect to $\psi$ and
  just one $\alpha_{W}$ that has grading $1$:
\begin{eqnarray}
  \left ( \psi \, ,\, \alpha_i\right ) & = & 0 \quad \quad i \ne W
  \nonumber\\
\left ( \psi \, ,\, \alpha_W\right ) & = & 1
\label{bingopsi}
\end{eqnarray}
  \item The only simple root $\alpha_W$ that has non vanishing grading with
  respect $\psi$ is just the highest weight of the symplectic
  representation $\mathbf{W}$ of $\mathbb{U}_{D=4}$ to which the vector fields
  are assigned.
  \item The Dynkin diagram of $\mathbb{U}_{D=4}$ is obtained from that of
  $\mathbb{U}_{D=3}$ by removing the dot corresponding to the special
  root $\alpha_W$.
  \item Hence we can arrange a basis   for the simple roots
  of the rank $r$ algebra $\mathbb{U}_{D=3}$ such that:
\begin{equation}
  \begin{array}{rcl}
     \alpha_i & = & \left\{ \overline{\alpha}_i , 0 \right\}  \quad ; \quad i \ne W\\
      \alpha_W & = & \left\{ \overline{\mathbf{w}}_h , \frac{1}{\sqrt{2}} \right\}   \\
     \psi  & = & \left\{  \mathbf{0}  , {\sqrt{2}} \right\} \
  \end{array}
\label{lucullus}
\end{equation}
where $\overline{\alpha}_i$ are $(r-1)$--component vectors representing
a basis of simple roots for the Lie algebra $\mathbb{U}_{D=4}$,
$\overline{\mathbf{w}}_h$ is also an $(r-1)$--vector representing the \textit{highest
weight} of the representation $\mathbf{W}$.
\end{enumerate}
This means that the entire root system and the Cartan subalgebra of the $\mathbb{U}_{D=3}$ Lie algebra can be organized as follows:
\begin{equation}
\begin{array}{ccccccc}
  \pm\psi & = & \pm \left( \mathbf{0}\, , \, \sqrt{2}\right) & ;  & \null &\null &2 \\
  \pm \hat{\alpha} & = & \pm \, \left( \alpha \, , \, \sqrt{2}\right) & ; & 2 \, \times\, \# \, \mbox{ of roots} & = & 2 \, n_r\\
  \pm \,\hat w & = & \pm \, \left( w \, , \, \frac{\sqrt{2}}{2}\right) & ; &  2 \, \times\, \# \, \mbox{ of weights} & = & 2 \, \times \, \mbox{dim} \, \mathbf{W} \\
  \mathcal{H}^i & \in \mathrm{CSA}\subset\mathbb{U}_{D=4} &  & ; & \mbox{rank}\mathbb{U}_{D=4} & = & r \\
   \mathcal{H}^\psi & \null & \null & \null & \null &\null & 1 \\
   \hline
   \null & \null & \null & \null  & \mbox{dim}\mathbb{U}_{D=4} & = & 3 + \mbox{dim}\mathbb{U}_{D=3} \, + \, 2 \, \times \, \mbox{dim} \mathbf{W}\\
   \end{array}
\end{equation}
This organization of the Lie algebra is very important, as it was
thoroughly discussed in \cite{Fre':2005si},  for the systematics
of the Ka\v c Moody extension which occurs when stepping down from
D=3 to D=2 dimensions,  but it is equally important in the present
context to analyze the structure of the $H^\star$-subalgebra and
the Tits Satake projection.
\subsection{Tits Satake projection}
In most cases of lower
supersymmetry, neither the algebra $\mathbb{U}_{D=4}$ nor the
algebra $\mathbb{U}_{D=3}$ are \textbf{maximally split}. In short
this means that the non-compact rank $r_{nc} < r $ is less than
the rank of $\mathbb{U}$, namely not all the Cartan generators are
non-compact. Rigorously $r_{nc}$ is defined as follows:
\begin{equation}
  r_{nc}\, = \, \mbox{rank} \left( \mathrm{U/H}\right)  \, \equiv \, \mbox{dim} \,
  \mathcal{H}^{n.c.} \quad ; \quad \mathcal{H}^{n.c.} \, \equiv \,
  \mbox{CSA}_{\mathbb{U}(\mathbb{C})} \, \bigcap \, \mathbb{K}
\label{rncdefi}
\end{equation}
When this happens it means
that, just as the billiard dynamics, also the structure of black hole solutions is effectively determined by a
\textit{maximally split subalgebra} $\mathbb{U}^{TS} \subset
\mathbb{U}$ named the \textit{Tits Satake} subalgebra of
$\mathbb{U}$, whose rank is equal to $r_{nc}$. Effectively determined
does not mean that  solutions of the big system
coincide with those
of the smaller system  rather it means that the former can be obtained from the
latter by means of rotations of a compact subgroup of the big algebra
$\mathrm{G_{paint} \subset U}$ which we name the \textit{paint
group}, for whose precise definition we refer the reader to  \cite{titsusataku}.
Here we just emphasize few important
facts, relevant for our goals. To this effect we recall that the Tits Satake algebra is obtained from the
original algebra via a projection of the root system of $\mathbb{U}$ onto the subspace orthogonal to the compact part of the Cartan subalgebra
of $\mathbb{U}^{TS}$:
\begin{equation}
  \Pi^{TS} \quad ; \quad \Delta_\mathbb{U} \,\mapsto \,
  \overline{\Delta}_{\mathbb{U}^{TS}}
\label{Tsproj}
\end{equation}
In euclidian geometry $\overline{\Delta}_{\mathbb{U}^{TS}}$ is
just a collection of vectors in $r_{nc}$ dimensions; a priori
there is no reason why it should be the root system of another Lie
algebra. Yet in almost all cases,
$\overline{\Delta}_{\mathbb{U}^{TS}}$ turns out to be a Lie
algebra root system and the maximal split Lie algebra
corresponding to it, $\mathbb{U}^{TS}$, is, by definition, the
Tits Satake subalgebra of the original non maximally split Lie
algebra: $\mathbb{U}^{TS} \subset \mathbb{U}$. Such algebras
$\mathbb{U}$ are called \textit{non-exotic}. The \textit{exotic}
non compact algebras are those for which the system
$\overline{\Delta}_{\mathbb{U}^{TS}}$ is not an admissible root
system. In such cases there is no Tits Satake subalgebra
$\mathbb{U}^{TS}$. Exotic algebras are very few and in
supergravity they appear only in three instances that display
additional pathologies relevant also for the black hole solutions.
For the non exotic models  we have that the decomposition
(\ref{gendecompo}) commutes with the projection, namely:
\begin{equation}
\begin{array}{rcl}
\mbox{adj}(\mathbb{U}_{D=3}) &=&
\mbox{adj}(\mathbb{U}_{D=4})\oplus\mbox{adj}({\slal(2,\mathbb{R})_E})\oplus
W_{(2,W)} \\
\null &\Downarrow & \null\\
\mbox{adj}(\mathbb{U}^{TS}_{D=3}) &=&
\mbox{adj}(\mathbb{U}^{TS}_{D=4})\oplus\mbox{adj}({\slal(2,\mathbb{R})_E})\oplus
W_{(2,W^{TS})} \\
\end{array}
\label{gendecompo2}
\end{equation}
In other words the projection leaves the $A_1$ Ehlers subalgebra untouched and has a non trivial effect only
on the duality algebra $\mathbb{U}_{D=4}$. Furthermore the image under the projection of the highest root
of $\mathbb{U}$ is the highest root of $\mathbb{U}^{TS}$:
\begin{equation}
  \Pi^{TS} \quad : \quad \psi \, \rightarrow \, \psi^{TS}
\label{Pionpsi}
\end{equation}
The reason why the Tits Satake projection is relevant to us was pointed out in \cite{titsblackholes} where we advocated that the classification of nilpotent orbits and hence of extremal black hole solutions depends only on the Tits Satake subalgebra and therefore is universal for all members of the same Tits Satake universality class. By this name we mean all algebras who share the same Tits Satake projection.
\par
Having clarified these points we can proceed with the classification of homogeneous symmetric spaces relevant to supergravity models and to black hole solutions.
\subsection{Classification of the sugra-relevant symmetric spaces}
\begin{table}
\begin{center}
{\tiny
\begin{tabular}{|l|c|c||c|c||c|c||c||}
  \hline
  \null & TS & TS & coset &coset &  Paint & subP &  susy\\
  $\#$ & D=4 & D=3 & D=4 & D=3 &  Group & Group &  \\
  \hline
  \hline
1&\null & \null & \null & \null &\null & \null & \null \\
\null \null & $ \frac{E_{7(7)}}{\mathrm{SU(8)}}$ & $ \frac{\mathrm{E_{8(8)}}}{\mathrm{SO^\star(16)}}$  & $ \frac{\mathrm{E_{7(7)}}}{\mathrm{SU(8)}}$ & $ \frac{\mathrm{E_{8(8)}}}{\mathrm{SO^\star(16)}}$  & $1$ & $1$ & $\mathcal{N}=8$ \\
\null &\null & \null & \null & \null &\null& \null & \null \\
\hline
2 &\null & \null & \null & \null &\null & \null & \null \\
\null & $ \frac{\mathrm{SU(1,1)}}{\mathrm{U(1)}}$ & $ \frac{\mathrm{G_{2(2)}}}{\mathrm{SL(2,R)\times SL(2,R)}}$  & $ \frac{\mathrm{SU(1,1)}}{\mathrm{U(1)}}$ & $ \frac{\mathrm{G_{2(2)}}}{\mathrm{SL(2,R)\times SL(2,R)}}$  & $1$ & $1$ & $\mathcal{N}=2$ \\
\null &\null & \null & \null & \null &\null & \null & n=1 \\
\hline
\hline
3 &\null & \null & \null & \null &\null & \null & \null \\
\null & \null & \null & $ \frac{\mathrm{Sp(6,R)}}{\mathrm{SU(3)\times  U(1)}}$ & $ \frac{\mathrm{F_{4(4)}}}{\mathrm{Sp(6,R)\times SL(2,R)}}$  & $1$ & $1$  & $\mathcal{N}=2$ \\
\null &\null & \null & \null & \null &\null & \null & $n=6$ \\
\cline{1-1} \cline{4-8}
4 &\null & \null & \null & \null &\null & \null & \null \\
\null & \null & \null & $ \frac{\mathrm{SU(3,3)}}{\mathrm{SU(3)\times SU(3) \times U(1)}}$ & $ \frac{\mathrm{E_{6(2)}}}{\mathrm{SU(3,3)\times SL(2,R)}}$  & $\mathrm{SO(2)\times SO(2)}$ & $1$  & $\mathcal{N}=2$ \\
\null &\null & \null & \null & \null &\null & \null & $n=9$ \\
\cline{1-1} \cline{4-8}
5 &\null & \null & \null & \null &\null & \null & $\mathcal{N}=6$ \\
\null & $ \frac{\mathrm{Sp(6,R)}}{\mathrm{SU(3)\times  U(1)}}$ & $ \frac{\mathrm{F_{4(4)}}}{\mathrm{Sp(6,R)\times SL(2,R)}}$ & $ \frac{\mathrm{SO^\star(12)}}{\mathrm{SU(6)\times U(1)}}$ & $ \frac{\mathrm{E_{7(-5)}}}{\mathrm{SO^\star(12)\times SL(2,R)}}$  & $\mathrm{SO(3)\times SO(3)}$ & $\mathrm{SO(3)_{d}}$  & $\mathcal{N}=2$ \\
\null &\null & \null & \null & \null &$\mathrm{\times SO(3)}$ & \null & n=16 \\
\cline{1-1} \cline{4-8}
6 &\null & \null & \null & \null &\null & \null & \null \\
\null & \null & \null & $ \frac{\mathrm{E_{7(-25)}}}{\mathrm{E_{6(-78)} \times U(1)}}$ & $ \frac{\mathrm{E_{8(-24)}}}{\mathrm{E_{7(-25)}\times SL(2,R)}}$  & $\mathrm{SO(8)}$ & $G_{2(-14)}$  & $\mathcal{N}=2$ \\
\null &\null & \null & \null & \null &\null & \null & $n=27$ \\
\hline
\hline
\hline
7 & \null \null & \null & \null & \null &\null & \null & \null \\
\null & $ \frac{\mathrm{SL(2,\mathbb{R})}}{\mathrm{O(2)}}\times\frac{\mathrm{SO(2,1)}}{\mathrm{SO(2)}}$ & $ \frac{\mathrm{SO(4,3)}}{\mathrm{SO(2,2)\times SO(2,1)}}$  & $ \frac{\mathrm{SL(2,\mathbb{R})}}{\mathrm{O(2)}}\times\frac{\mathrm{SO(6,1)}}{\mathrm{SO(6)}}$ & $ \frac{\mathrm{SO(8,3)}}{\mathrm{SO(6,2)\times SO(2,1)}}$ & $\mathrm{SO(5)}$ & $\mathrm{SO(4)}$ & $\mathcal{N}=4$ \\
\null &\null & \null & \null & \null &\null & \null & n=1 \\
\hline
8 & \null \null & \null & \null & \null &\null & \null & \null \\
\null & $ \frac{\mathrm{SL(2,\mathbb{R})}}{\mathrm{O(2)}}\times\frac{\mathrm{SO(3,2)}}{\mathrm{SO(3)\times SO(2)}}$ & $ \frac{\mathrm{SO(5,4)}}{\mathrm{SO(3,2)\times SO(2,2)}}$  & $ \frac{\mathrm{SL(2,\mathbb{R})}}{\mathrm{O(2)}}\times\frac{\mathrm{SO(6,2)}}{\mathrm{SO(6)\times SO(2)}}$ & $ \frac{\mathrm{SO(8,4)}}{\mathrm{SO(6,2)\times SO(2,2)}}$ & $\mathrm{SO(4)}$ & $\mathrm{SO(3)}$ & $\mathcal{N}=4$ \\
\null &\null & \null & \null & \null &\null & \null & n=2 \\
\hline
9 & \null \null & \null & \null & \null &\null & \null & \null \\
\null & $ \frac{\mathrm{SL(2,\mathbb{R})}}{\mathrm{O(2)}}\times\frac{\mathrm{SO(4,3)}}{\mathrm{SO(4)\times SO(3)}}$ & $ \frac{\mathrm{SO(6,5)}}{\mathrm{SO(4,2)\times SO(2,3)}}$  & $ \frac{\mathrm{SL(2,\mathbb{R})}}{\mathrm{O(2)}}\times\frac{\mathrm{SO(6,3)}}{\mathrm{SO(6)\times SO(3)}}$ & $ \frac{\mathrm{SO(8,5)}}{\mathrm{SO(6,2)\times SO(2,3)}}$ & $\mathrm{SO(3)}$ & $\mathrm{SO(2)}$ & $\mathcal{N}=4$ \\
\null &\null & \null & \null & \null &\null & \null & n=3 \\
\hline
10 & \null \null & \null & \null & \null &\null & \null & \null \\
\null & $ \frac{\mathrm{SL(2,\mathbb{R})}}{\mathrm{O(2)}}\times\frac{\mathrm{SO(5,4)}}{\mathrm{SO(5)\times SO(4)}}$ & $ \frac{\mathrm{SO(7,6)}}{\mathrm{SO(5,2)\times SO(2,4)}}$  & $ \frac{\mathrm{SL(2,\mathbb{R})}}{\mathrm{O(2)}}\times\frac{\mathrm{SO(6,4)}}{\mathrm{SO(6)\times SO(4)}}$ & $ \frac{\mathrm{SO(8,6)}}{\mathrm{SO(6,2)\times SO(2,4)}}$ & $\mathrm{SO(2)}$ & $\mathrm{1}$ & $\mathcal{N}=4$ \\
\null &\null & \null & \null & \null &\null & \null & n=4 \\
\hline
11 & \null \null & \null & \null & \null &\null & \null & \null \\
\null & $ \frac{\mathrm{SL(2,\mathbb{R})}}{\mathrm{O(2)}}\times\frac{\mathrm{SO(6,5)}}{\mathrm{SO(6)\times SO(5)}}$ & $ \frac{\mathrm{SO(8,7)}}{\mathrm{SO(6,2)\times SO(2,5)}}$  & $ \frac{\mathrm{SL(2,\mathbb{R})}}{\mathrm{O(2)}}\times\frac{\mathrm{SO(6,5)}}{\mathrm{SO(6)\times SO(5)}}$ & $ \frac{\mathrm{SO(8,7)}}{\mathrm{SO(6,2)\times SO(2,5)}}$ & $\mathrm{1}$ & $\mathrm{1}$ & $\mathcal{N}=4$ \\
\null &\null & \null & \null & \null &\null & \null & n=5 \\
\hline
12 & \null \null & \null & \null & \null &\null & \null & \null \\
\null & $ \frac{\mathrm{SL(2,\mathbb{R})}}{\mathrm{O(2)}}\times\frac{\mathrm{SO(6,6)}}{\mathrm{SO(6)\times SO(6))}}$ & $ \frac{\mathrm{SO(8,8)}}{\mathrm{SO(6,2)\times SO(2,6)}}$  & $ \frac{\mathrm{SL(2,\mathbb{R})}}{\mathrm{O(2)}}\times\frac{\mathrm{SO(6,6)}}{\mathrm{SO(6)\times SO(6))}}$ & $ \frac{\mathrm{SO(8,8)}}{\mathrm{SO(6,2)\times SO(2,6)}}$ & $1$ & $1$ & $\mathcal{N}=4$ \\
\null &\null & \null & \null & \null &\null & \null & n=6 \\
\hline
13 & \null \null & \null & \null & \null &\null & \null & \null \\
\null & $ \frac{\mathrm{SL(2,\mathbb{R})}}{\mathrm{O(2)}}\times\frac{\mathrm{SO(6,7)}}{\mathrm{SO(6)\times SO(7))}}$ & $ \frac{\mathrm{SO(8,9)}}{\mathrm{SO(6,2)\times SO(2,7)}}$  & $ \frac{\mathrm{SL(2,\mathbb{R})}}{\mathrm{O(2)}}\times\frac{\mathrm{SO(6,6+p)}}{\mathrm{SO(6)\times SO(6+p))}}$ & $ \frac{\mathrm{SO(8,8+p)}}{\mathrm{SO(6,2)\times SO(2,6+p)}}$ & $\mathrm{SO(p)}$ & $\mathrm{SO(p-1)}$ & $\mathcal{N}=4$ \\
\null &\null & \null & \null & \null &\null & \null & n=6+p \\
\hline
\hline
\hline
14 & \null \null & \null & \null & \null &\null & \null & \null \\
\null & $ \frac{\mathrm{SL(2,\mathbb{R})}}{\mathrm{O(2)}}\times\frac{\mathrm{SO(2,1)}}{\mathrm{SO(2)}}$ & $ \frac{\mathrm{SO(4,3)}}{\mathrm{SO(2,2)\times SO(2,1)}}$  & $ \frac{\mathrm{SL(2,\mathbb{R})}}{\mathrm{O(2)}}\times\frac{\mathrm{SO(2,1)}}{\mathrm{SO(2)}}$ & $ \frac{\mathrm{SO(4,3)}}{\mathrm{SO(2,2)\times SO(2,1)}}$ & $\mathrm{1}$ & $\mathrm{1}$ & $\mathcal{N}=2$ \\
\null &\null & \null & \null & \null &\null & \null & n=2 \\
\hline
15 & \null \null & \null & \null & \null &\null & \null & \null \\
\null & $ \frac{\mathrm{SL(2,\mathbb{R})}}{\mathrm{O(2)}}\times\frac{\mathrm{SO(2,2)}}{\mathrm{SO(2)\times SO(2)}}$ & $ \frac{\mathrm{SO(4,4)}}{\mathrm{SO(2,2)\times SO(2,2)}}$  & $ \frac{\mathrm{SL(2,\mathbb{R})}}{\mathrm{O(2)}}\times\frac{\mathrm{SO(2,2)}}{\mathrm{SO(2)\times SO(2)}}$ & $ \frac{\mathrm{SO(4,4)}}{\mathrm{SO(2,2)\times SO(2,2)}}$ & $\mathrm{1}$ & $\mathrm{1}$ & $\mathcal{N}=2$ \\
\null &\null & \null & \null & \null &\null & \null & n=3 \\
\hline
16 & \null \null & \null & \null & \null &\null & \null & \null \\
\null & $ \frac{\mathrm{SL(2,\mathbb{R})}}{\mathrm{O(2)}}\times\frac{\mathrm{SO(2,3)}}{\mathrm{SO(2)\times SO(3)}}$ & $ \frac{\mathrm{SO(4,5)}}{\mathrm{SO(2,2)\times SO(2,3)}}$  & $ \frac{\mathrm{SL(2,\mathbb{R})}}{\mathrm{O(2)}}\times\frac{\mathrm{SO(2,2+p)}}{\mathrm{SO(2)\times SO(2+p)}}$ & $ \frac{\mathrm{SO(4,4+p)}}{\mathrm{SO(2,2)\times SO(2,2+p)}}$ & $\mathrm{SO(p)}$ & $\mathrm{SO(p-1)}$ & $\mathcal{N}=2$ \\
\null &\null & \null & \null & \null &\null & \null & n=3+p \\
\hline
\end{tabular}
} \caption{The 16 instances  of \textit{non-exotic} homogeneous
symmetric scalar manifolds appearing in $D=4$ supergravity. Non
exotic means that the Tits Satake projection of the root system is
a standard Lie Algebra root system. The 16 models are grouped
according to their Tits Satake Universality classes. The time-like
dimensional reduction is listed side by side. Within each class
the models are distinguished by the different structure of the
Paint Group and of its subPaint subgroup. The Paint group is the
same in D=4 and in D=3 \label{homomodels}}
\end{center}
\end{table}
The classification of the symmetric coset based supergravity models is exhaustive and it is presented in tables \ref{homomodels} and
\ref{exohomomodels}. There are 16 universality classes of non-exotic models and 3 exceptional instances of exotic models which appear in the second table.
\par
In the tables we have also listed the Paint groups and the subpaint groups. These latter are always compact and their different structures is what distinguishes the different elements belonging to the same class. As it was shown in \cite{titsusataku}, these groups are dimensional reduction invariant, namely they are the same in $D=4$ and in $D=3$. Hence the representation $\mathbf{W}$, which in particular contains the electromagnetic charges of the hole, can be decomposed with respect to the Tits Satake subalgebra and the Paint group revealing a regularity structure inside each Tits Satake universality class which is at the heart of the classification of \textit{charge orbits}. The same decomposition can be given also for the $\mathbb{K}^\star$ representation and this is at the heart of the classification of black holes according to nilpotent orbits.
\par
Focusing on the non-exotic models, we note that the 16 classes have a quite different type of population. There are six one element classes whose single member is maximally split. They are the following ones and all have  a distinguished standpoint within the panorama of supergravity theories:
\begin{enumerate}
  \item The $\mathcal{N}=8$ supergravity theory, which is the maximal one in $D=4$, (model $1$).
  \item The $\mathcal{N}=2$ supergravity theory with a single vector multiplet and non-vanishing Yukawa (model $2$).
  \item The $\mathcal{N}=4$ supergravity theory with $5$ vector multiplets (model $11$).
  \item The $\mathcal{N}=4$ supergravity theory with $6$ vector multiplets which is obtained compactifying a type II theory on
  a $\mathrm{T}^6/\mathbb{Z}_2$ orbifold (model $12$).
  \item The $\mathcal{N}=2$ theory with two vector multiplets and non vanishing Yukava couplings, usually called the $st$-model (model $14$).
  \item The $\mathcal{N}=2$ theory with three vector multiplets and non vanishing Yukava couplings, usually called the $stu$-model (model $15$).
\end{enumerate}
Next we have two universality classes, each containing an infinite number of elements. They are
\begin{enumerate}
  \item The $\mathcal{N}=4$ supergravity theory with $n=6+p$ vector multiplets ($p\ge1$), (model $13$).
  \item The $\mathcal{N}=2$ supergravity theory with $n=3+p$ vector multiplets ($p\ge1$) and non vanishing Yukawa couplings (model $16$).
\end{enumerate}
We still have the very interesting 4-element universality class whose maximally split representative corresponds to the maximally split
special K\"ahler manifold $\frac{\mathrm{Sp(4,\mathbb{R})}}{\mathrm{SU(3)\times U(1)}}$. This class contains the models $3,4,5,6$ distinguished by quite peculiar Paint groups. We will thoroughly analyze the structure of this class.
\par
Finally we have the three exotic models whose common feature is that their  group and subgroup all belong to the pseudo-unitary series $\mathrm{SU(p,q)}$. The general decomposition (\ref{gendecompo}) still holds true, but the Tits Satake projection looses its significance.
\begin{table}
\begin{center}
{\small
\begin{tabular}{|l|c|c||c|c||c|c||c||}
  \hline
  \null & TS & TS & coset &coset &  Paint & subP &  susy\\
  $\#$ & D=4 & D=3 & D=4 & D=3 &  Group & Group &  \\
  \hline
  \hline
$1_e$&\null & \null & \null & \null &\null & \null & \null \\
\null \null & $ bc_1$ & $ bc_2$  & $ \frac{\mathrm{SU(p+1,1)}}{\mathrm{SU(p+1)\times U(1)}}$ & $ \frac{\mathrm{SU(p+2,2)}}{\mathrm{SU(p+1,1)\times SL(2,R)_{h^\star}}}$  & $\mathrm{U(1)\times U(1) \times U(p)}$ & $\mathrm{U(p-1)}$ & $\mathcal{N}=2$ \\
\null &\null & \null & \null & \null &\null& \null & n=p+1 \\
\hline
$2_e$&\null & \null & \null & \null &\null & \null & \null \\
\null \null & $ bc_3$ & $ bc_4$  & $ \frac{\mathrm{SU(p+1,3)}}{\mathrm{SU(p+1)\times SU(3) \times U(1)}}$ & $ \frac{\mathrm{SU(p+2,4)}}{\mathrm{SU(p+1,2)\times SU(1,2) \times U(1)}}$  & $\mathrm{U(1)\times U(1) \times U(p)}$ & $\mathrm{U(p-1)}$ & $\mathcal{N}=3$ \\
\null &\null & \null & \null & \null &\null& \null & n=p+1 \\
\hline
$3_e$&\null & \null & \null & \null &\null & \null & \null \\
\null \null & $ bc_1$ & $ bc_2$  & $ \frac{\mathrm{SU(5,1)}}{\mathrm{SU(5)\times U(1)}}$ & $ \frac{\mathrm{E_{6(-14)}}}{\mathrm{SO^\star(10)\times SO(2)}}$  & $\mathrm{U(1)\times U(1) \times U(4)}$ & $\mathrm{U(3)}$ & $\mathcal{N}=5$ \\
\null &\null & \null & \null & \null &\null& \null & \null \\
\hline
\end{tabular}
}
\caption{The 3 instances  of \textit{exotic} homogenous symmetric scalar manifolds appearing in $D=4$ supergravity. Exotic means that the Tits Satake projection of the root system is not a standard Lie Algebra root system. Notwithstanding this anomaly the concept of Paint Group, according to its definition as group of external automorphisms of the solvable Lie algebra generating the non compact coset manifold still exists.  The Paint group is the same in D=4 and in D=3 \label{exohomomodels}}
\end{center}
\end{table}
\subsection{Dynkin diagram analysis of the principal models}
\label{dynkino}
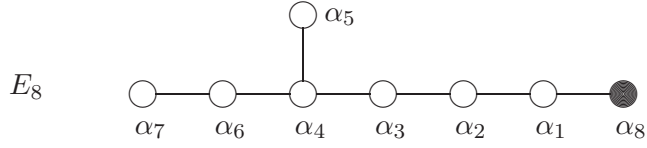
\begin{figure}
\centering
\begin{picture}(100,70)
      \put (-70,35){$E_8$} \put (-20,35){\circle {10}} \put
(-23,20){$\alpha_7$} \put (-15,35){\line (1,0){20}} \put
(10,35){\circle {10}} \put (7,20){$\alpha_6$} \put (15,35){\line
(1,0){20}} \put (40,35){\circle {10}} \put (37,20){$\alpha_4$}
\put (40,65){\circle {10}} \put (48,62.8){$\alpha_5$} \put
(40,40){\line (0,1){20}} \put (45,35){\line (1,0){20}} \put
(70,35){\circle {10}} \put (67,20){$\alpha_{3}$} \put
(75,35){\line (1,0){20}} \put (100,35){\circle {10}} \put
(97,20){$\alpha_{2}$} \put (105,35){\line (1,0){20}} \put
(130,35){\circle {10}} \put (127,20){$\alpha_1$} \put
(135,35){\line (1,0){20}} \put (160,35){\circle{10}}\put (160,35){\circle{9}}
\put (160,35){\circle{8}}\put (160,35){\circle{7}}\put (160,35){\circle{6}}
\put (160,35){\circle{5}}\put (160,35){\circle{4}} \put (160,35){\circle{3}}
\put (160,35){\circle{2}}\put (160,35){\circle{1}}\put
(157,20){$\alpha_8$}
\end{picture}
$$\begin{array}{l}\psi = 3\alpha_1 +4\alpha_2 +5\alpha_3+6\alpha_4 +3\alpha_5+4\alpha_6
+2\alpha_7 +2\alpha_8 \\(\psi \, , \,\alpha_8) = 1 \quad; \quad  (\psi \, , \, \alpha_i ) = 0 \quad i \ne 8 \
\end{array}$$
\vskip 1cm \caption{The Dynkin diagram of $E_{8(8)}$. The only
simple root which has grading one with respect to the highest root
$\psi$ is $\alpha_8$ (painted black). With respect to the algebra
$\mathbb{U}_{D=4}=E_{7(7)}$ whose Dynkin diagram is obtained by
removal of the black circle, $\alpha_8$ is the highest weight of
the symplectic representation of the vector fields, namely
$\mathbf{W}=\mathbf{56}$. } \label{except2}
\end{figure}

Next we analyze the form of the root systems of the $\mathbb{U}_{\mathrm{D=3}}$ algebras in relation with the decomposition (\ref{gendecompo}).
\vskip 0.2cm
\paragraph{$\mathcal{N}$=8}
\vskip 0.2cm
This is the case of maximal supersymmetry and it is illustrated by
fig. \ref{except2}.
\par
In this case all the involved Lie algebras are maximally split and we
have
\begin{equation}
  \mbox{adj} \, \mathrm{E_{8(8)}} = \mbox{adj} \, \mathrm{E_{7(7)}} \oplus
  \mbox{adj} \, \mathrm{SL(2,\mathbb{R})_E} \oplus \left( \mathbf{2},\mathbf{56}\right)
\label{urgeE8}
\end{equation}
The highest root of $\mathrm{E_{8(8)}}$ is
\begin{equation}
  \psi = 3\alpha_1 +4\alpha_2 +5\alpha_3+6\alpha_4 +3\alpha_5+4\alpha_6
+2\alpha_7 +2\alpha_8
\label{highE8}
\end{equation}
and the unique simple root not orthogonal to $\psi$ is $\alpha_8 =
\alpha_W$, according to the labeling of roots as in fig.
\ref{except2}. This root is the highest weight of the fundamental
$\mathbf{56}$-representation  of $E_{7(7)}$.
\par
The well adapted basis of simple  $E_{8}$  roots is
constructed as follows:
\begin{equation}
\begin{array}{rclcl}
\alpha_1 & = & \{ 1,-1,0,0,0,0,0,0\} & = &
\{\overline{\alpha}_1,0\} \\
\alpha_2  & = & \{ 0,1,-1,0,0,0,0,0\}& = &
\{\overline{\alpha}_2,0\} \\
\alpha_3  & = & \{ 0,0,1,-1,0,0,0,0\} & = &
\{\overline{\alpha}_3,0\}\\
\alpha_4  & = & \{ 0,0,0,1,-1,0,0,0\} & = &
\{\overline{\alpha}_4,0\}\\
\alpha_5  & = & \{ 0,0,0,0,1,-1,0,0\}& = &
\{\overline{\alpha}_5,0\} \\
\alpha_6  & = & \{ 0,0,0,0,1,1,0,0\}& = &
\{\overline{\alpha}_6,0\} \\
\alpha_7  & = & \{ - \frac{1}{2}  ,
  - \frac{1}{2}  ,
  - \frac{1}{2}  ,
  - \frac{1}{2}  ,
  - \frac{1}{2}  ,
  - \frac{1}{2}  ,
  \frac{1}{{\sqrt{2}}},0\}& = &
\{\overline{\alpha}_7, 0\}\\
\alpha_8  & = & \{ - 1,0,0,
  0,0,0,- \frac{1}{{\sqrt{2}}} , \frac{1}{{\sqrt{2}}}   \}
  & = & \{ \mathbf{w}_h,\frac{1}{{\sqrt{2}}}\}\
  \end{array}
\label{alfeE8}
\end{equation}
In this basis we recognize that the seven $7$-vectors
$\bar{\alpha}_i$ constitute a simple root basis for the $E_7$ root
system, while:
\begin{equation}
 \mathbf{w}_h \, = \, \{ -1,0,0,
  0,0,0,- \frac{1}{{\sqrt{2}}} \}
\label{ierunda}
\end{equation}
is the highest weight of the fundamental $\mathbf{56}$ dimensional
representation. Finally in this basis the highest root $\psi$
defined by eq.(\ref{highE8}) takes the expected form:
\begin{equation}
  \psi = \{ 0,0,0,0,0,0,0,\sqrt{2}\}
\label{expepsi8}
\end{equation}
\paragraph{N=6}
\vskip 0.2cm
In this case the $D=4$ duality algebra is
$\mathbb{U}_{D=4}=\mathrm{SO^\star(12)}$, whose maximal compact subgroup is $\mathrm{H=SU(6) \times U(1)}$. The
scalar manifold:
\begin{equation}
  \mathcal{SK}_{N=6} \equiv \frac{\mathrm{SO^\star(12)}}{\mathrm{SU(6) \times U(1)}}
\label{SK6}
\end{equation}
is an instance of special K\"ahler manifold which can also be
utilized in an $\mathcal{N}=2$ supergravity context. The $D=3$
algebra is $\mathbb{U}_{D=3}=E_{7(-5)}$. The $16$ vector fields of
$D=4~\mathcal{N}=6$ supergravity with their electric and magnetic
field strengths fill the spinor representation $\mathbf{32}_s$ of
$\mathrm{SO^\star(12)}$, so that the decomposition
(\ref{gendecompo}), in this case becomes:
\begin{equation}
  \mbox{adj} \, \mathrm{E_{7(-5)}} = \mbox{adj} \, \mathrm{SO^\star(12)} \oplus
  \mbox{adj} \, \mathrm{SL(2,\mathbb{R})_E} \oplus \left( \mathbf{2},\mathbf{32}_s\right)
\label{decompoN6}
\end{equation}
\begin{figure}
\centering
\begin{picture}(100,70)
      \put (-180,35){$E_{7(-5)}$} \put (-20,35){\circle {10}} \put (-20,35){\circle {9}}
      \put (-20,35){\circle {8}}\put (-20,35){\circle {7}}
\put (-20,35){\circle {6}}\put (-20,35){\circle {5}}\put (-20,35){\circle {4}}\put (-20,35){\circle {3}}
\put (-20,35){\circle {2}}
\put (-20,35){\circle {1}}\put
(-23,20){$\alpha_7$} \put (-15,35){\line (1,0){20}} \put
(10,35){\circle {10}} \put (7,20){$\alpha_6$} \put (15,35){\line
(1,0){20}} \put (40,35){\circle {10}} \put (37,20){$\alpha_4$}
\put (40,65){\circle {10}} \put (48,62.8){$\alpha_5$} \put
(40,40){\line (0,1){20}} \put (45,35){\line (1,0){20}} \put
(70,35){\circle {10}} \put (67,20){$\alpha_{3}$} \put
(75,35){\line (1,0){20}} \put (100,35){\circle {10}} \put
(97,20){$\alpha_{2}$} \put (105,35){\line (1,0){20}} \put
(130,35){\circle {10}} \put (127,20){$\alpha_1$}
\end{picture}
$$\begin{array}{l}\psi = \alpha_1 +2\alpha_2 +3\alpha_3+4\alpha_4 +2\alpha_5+3\alpha_6
+2\alpha_7  \\(\psi \, , \,\alpha_7) = 1 \quad; \quad  (\psi \, , \, \alpha_i ) = 0 \quad i \ne 7 \
\end{array}$$
\vskip 1cm \caption{The Dynkin diagram of $E_{7(-5)}$.
The only simple root which has grading one
with respect to the highest root $\psi$ is $\alpha_4$ (painted black).
With respect to the algebra $\mathbb{U}_{D=4}=\mathrm{SO^\star(12)}$ whose Dynkin diagram is
obtained by removal of the black circle, $\alpha_7$ is the highest weight of the symplectic representation of
the vector fields, namely the $\mathbf{W}=\mathbf{32}_s$.}
\label{except3}
\end{figure}
The simple root $\alpha_W$  is $\alpha_7$ and the highest root is:
\begin{equation}
\begin{array}{l}\psi = \alpha_1 +2\alpha_2 +3\alpha_3+4\alpha_4 +2\alpha_5+3\alpha_6
+2\alpha_7  \
\end{array}
\label{highestE7}
\end{equation}
A well adapted basis of simple  $E_{7}$  roots can be
written as follows:
\begin{equation}
\begin{array}{rclcl}
\alpha_1 & = & \{ 1,-1,0,0,0,0,0\} & = &
\{\overline{\alpha}_1,0\} \\
\alpha_2  & = & \{ 0,1,-1,0,0,0,0\}& = &
\{\overline{\alpha}_2,0\} \\
\alpha_3  & = & \{ 0,0,1,-1,0,0,0\} & = &
\{\overline{\alpha}_3,0\}\\
\alpha_4  & = & \{ 0,0,0,1,-1,0,0\} & = &
\{\overline{\alpha}_4,0\}\\
\alpha_5  & = & \{ 0,0,0,0,1,-1,0\}& = &
\{\overline{\alpha}_5,0\} \\
\alpha_6  & = & \{ 0,0,0,0,1,1,0\}& = &
\{\overline{\alpha}_6,0\} \\
\alpha_7  & = & \{ - \frac{1}{2}  ,
  - \frac{1}{2}  ,
  - \frac{1}{2}  ,
  - \frac{1}{2}  ,
  - \frac{1}{2}  ,
  - \frac{1}{2}  ,
  \frac{1}{{\sqrt{2}}}\}& = &
\{\overline{\mathbf{w}}_h,\frac{1}{{\sqrt{2}}} \}\
  \end{array}
\label{alfeE7}
\end{equation}
In this basis we recognize that the six $6$-vectors
$\bar{\alpha}_i$ ($i=1,\dots,6$) constitute a simple root
basis for the $D_6\simeq \mathrm{SO^\star(12)}$ root
system, while:
\begin{equation}
 \mathbf{w}_h \, = \, \{ - \frac{1}{{{2}}}  ,-
 \frac{1}{{{2}}},- \frac{1}{{{2}}},- \frac{1}{{{2}}},
  - \frac{1}{{{2}}},- \frac{1}{{{2}}},- \frac{1}{{{2}}},
  - \frac{1}{{{2}}} \}
\label{ierunda2}
\end{equation}
is the highest weight of the spinor $\mathbf{32}$-dimensional
representation of $\mathrm{SO^\star(12)}$. Finally in this basis
the highest root $\psi$
defined by eq.(\ref{highestE7}) takes the expected form:
\begin{equation}
  \psi = \{ 0,0,0,0,0,0,\sqrt{2}\}
\label{expepsi7}
\end{equation}
\par
In this case, as in most cases of lower
supersymmetry, neither the algebra $\mathbb{U}_{D=4}$ nor the
algebra $\mathbb{U}_{D=3}$ are \textbf{maximally split}. The Tits Satake projection of $\mathrm{E_{7(-5)}}$ is
$\mathrm{F_{4(4)}}$ and the explicit form of eq.(\ref{gendecompo2}) is the following one:
\begin{equation}
\begin{array}{rcl}
\mbox{adj}(\mathrm{E_{7(-5)}}) &=&
\mbox{adj}(\mathrm{SO^\star(12)})\oplus\mbox{adj}(\mathrm{SL(2,\mathbb{R})_E})\oplus
(\mathbf{2},\mathbf{32}_s)  \\
\null &\Downarrow & \null\\
\mbox{adj}(\mathrm{F_{4(4)}}) &=&
\mbox{adj}(\mathrm{Sp(6,\mathbb{R})}\oplus\mbox{adj}(\mathrm{SL(2,\mathbb{R})_E})\oplus
(\mathbf{2},\mathbf{14}^\prime)  \
\end{array}
\label{Tsdecompo6}
\end{equation}
The representation $\mathbf{14}^\prime$ of
$\mathrm{Sp(6,\mathbb{R})}$ is that of an antisymmetric symplectic
traceless tensor:
\begin{equation}
\begin{array}{ccc}
 \mbox{dim}_{\mathrm{Sp(6,\mathbb{R})}} & \begin{array}{c}
  \vspace{-0.6cm} \\ \widetilde{\yng(1,1,1)}\
 \end{array} & \, = \, \mathbf{14}^\prime
\end{array}
\label{14repre}
\end{equation}
The Dynkin diagram of the Tits Satake subalgebra $\mathfrak{f}_{4(4)}$ is discussed in fig.\ref{F4dynk}.
\begin{figure}
\centering
\begin{picture}(90,50)
      \put (-70,35){$F_4$}  \put
(10,35){\circle {10}} \put (7,20){$\varpi_4$} \put (15,35){\line
(1,0){20}} \put (40,35){\circle {10}} \put (37,20){$\varpi_3$}\put (45,38){\line (1,0){20}}
\put (55,35){\line (1,1){10}} \put (55,35){\line (1,-1){10}}\put (45,33){\line (1,0){20}} \put
(70,35){\circle {10}} \put (67,20){$\varpi_{2}$} \put
(75,35){\line (1,0){20}} \put (100,35){\circle {10}} \put (100,35){\circle {9}}\put (100,35){\circle {8}}\put (100,35){\circle {10}}\put (100,35){\circle {7}}
\put (100,35){\circle {6}}\put (100,35){\circle {5}}\put (100,35){\circle {4}}\put (100,35){\circle {3}}\put (100,35){\circle {2}}\put (100,35){\circle {1}}\put
(97,20){$\varpi_{1}$}
\end{picture}
$$ \begin{array}{l}\psi = 2\varpi_1 +3\varpi_2 +4\varpi_3+2\varpi_4   \\(\psi \, , \,\varpi_1) = 2
\quad; \quad  (\psi \, , \, \varpi_i ) = 0 \quad i \ne 1 \
\end{array}     $$
\vskip 1cm \caption{The Dynkin diagram of $F_{4(4)}$. The only root which is not orthogonal
to the highest root is $\varpi_V = \varpi_1$. In the Tits Satake projection
$\Pi^{TS}$ the highest root $\psi$ of $F_{4(4)}$ is the image of the highest root of $E_{7(-5)}$ and the root $\varpi_V = \varpi_1 = \Pi^{TS} \left( \alpha_7\right)$
is the image of the root associated with the vector fields. \label{F4dynk}}
\end{figure}
\vskip 0.2cm
\paragraph{N=5}
\vskip 0.2cm The case of $\mathcal{N}=5$ supergravity is described by
fig.\ref{except4} and it is one of the three exotic models whose  Tits - Satake projection does not produce a Lie algebra root system.
\par
In the $\mathcal{N}=5$ theory the scalar manifold is a complex coset of rank
$r=1$,
\begin{equation}
  \mathcal{{M}}_{N=5,D=4}=\frac{\mathrm{SU(1,5)}}{\mathrm{SU(5) \times U(1)}}
\label{MN5D4}
\end{equation}
and there are $\mathbf{10}$ vector fields whose electric and magnetic field strengths are assigned to the
$\mathbf{20}$-dimensional representation of $\mathrm{SU(1,5)}$, which
is that of an antisymmetric three-index tensor
\begin{equation}\begin{array}{ccc}
  \mbox{dim}_{\mathrm{SU(1,5)}} \, & \begin{array}{c}
     \vspace{-0.4cm} \\
     \yng(1,1,1) \
  \end{array} & \, = \, 20
\end{array}
\label{orpo}
\end{equation}
The decomposition (\ref{gendecompo}) takes the explicit form:
\begin{equation}
 \mbox{adj}(\mathrm{E_{6(-14)}}) =\mbox{adj}(\mathrm{SU(1,5})\oplus\mbox{adj}(\mathrm{SL(2,\mathbb{R})_E})\oplus
(\mathbf{2},\mathbf{20})
\label{decompe6}
\end{equation}
and we have that the highest root of $\mathrm{E_6}$, namely
\begin{equation}
  \psi =\alpha_1 +2\alpha_2 +3\alpha_3+2\alpha_4 +2\alpha_5+\alpha_6
\label{psie6}
\end{equation}
has non vanishing scalar product only with the root $\alpha_4$
 in the form depicted in fig.\ref{except4}.
\begin{figure}
\centering
\begin{picture}(100,50)
      \put (-180,35){$E_{6(-14)}$} \put (-20,35){\circle {10}}
\put(-23,20){$\alpha_6$} \put (-15,35){\line (1,0){20}} \put
(10,35){\circle {10}} \put (7,20){$\alpha_5$} \put (15,35){\line
(1,0){20}} \put (40,35){\circle {10}} \put (37,20){$\alpha_3$}
\put (40,65){\circle {10}}\put (40,65){\circle {9}}\put (40,65){\circle {8}}\put (40,65){\circle {7}}\put (40,65)
{\circle {6}}\put (40,65){\circle {5}}\put (40,65){\circle {4}}\put (40,65){\circle {3}}\put (40,65){\circle {2}}
\put (40,65){\circle {1}}
\put (48,62.8){$\alpha_4$} \put
(40,40){\line (0,1){20}} \put (45,35){\line (1,0){20}} \put
(70,35){\circle {10}} \put (67,20){$\alpha_{2}$} \put
(75,35){\line (1,0){20}} \put (100,35){\circle {10}} \put
(97,20){$\alpha_{1}$}
\end{picture}
$$\begin{array}{l}\psi = \alpha_1 +2\alpha_2 +3\alpha_3+2\alpha_4 +2\alpha_5+\alpha_6
  \\(\psi \, , \,\alpha_4) = 1 \quad; \quad  (\psi \, , \, \alpha_i ) = 0 \quad i \ne 4 \
\end{array}$$
\vskip 1cm \caption{The Dynkin diagram of $E_{6(-14)}$.
The only simple root which has grading one
with respect to the highest root $\psi$ is $\alpha_7$ (painted black).
With respect to the algebra $\mathbb{U}_{D=4}=\mathrm{SU(5,1))}$ whose Dynkin diagram is
obtained by removal of the black circle, $\alpha_4$ is the highest weight of the symplectic representation of
the vector fields, namely the $\mathbf{W}=\mathbf{20}$. }
\label{except4}
\end{figure}
\par
Writing a well adapted basis of $E_6$ roots is a little bit more laborious
but it can be done. We find:
\begin{equation}
\begin{array}{rclcl}
\alpha_1 & = & \left \{ 0,0,- \frac{{\sqrt{3}}}{2},
  \frac{1}{2\,{\sqrt{5}}},
  {\sqrt{\frac{6}{5}}},0\right \} & = &
\left \{\overline{\alpha}_1,0\right \} \\
\alpha_2  & = & \left \{ - \frac{1}
      {{\sqrt{2}}}  ,
  \frac{1}{{\sqrt{6}}},
  \frac{2}{{\sqrt{3}}},0,0,
  0\right \}& = &
\left \{\overline{\alpha}_2,0\right \} \\
\alpha_3  & = & \left \{ {\sqrt{2}},0,0,0,0,0\right \} & = &
\left \{\overline{\alpha}_3,0\right \}\\
\alpha_4  & = & \left \{ - \frac{1}
      {{\sqrt{2}}}  ,
  \frac{1}{{\sqrt{6}}},
  - \frac{1}
      {{\sqrt{3}}}  ,
  \frac{1}{{\sqrt{5}}},
  -{\sqrt{\frac{3}{10}}},
  \frac{1}{{\sqrt{2}}}\right \} & = &
\left \{\overline{\mathbf{w}}_h, \frac{1}{{\sqrt{2}}} \right \}\\
\alpha_5  & = & \left \{ - \frac{1}
      {{\sqrt{2}}}  ,
  -{\sqrt{\frac{3}{2}}},0,0,0,
  0\right \}& = &
\left \{\overline{\alpha}_4,0\right \} \\
\alpha_6  & = & \left \{ 0,{\sqrt{\frac{2}{3}}},
 - \frac{1}{2\,{\sqrt{3}}},
  -\frac{\sqrt{5}}{2},0,0\right \}& = &
\left \{\overline{\alpha}_5,0\right \} \\
  \end{array}
\label{alfeE6}
\end{equation}
In this basis we can check that the five $5$-vectors
$\bar{\alpha}_i$ ($i=1,\dots,5$) constitute a simple root
basis for the $A_5\simeq \mathrm{SU(1,5)}$ root
system, namely:
\begin{equation}
 \langle \bar{\alpha}_i \, , \, \bar{\alpha}_j \rangle = \left(\matrix{ 2 & -1 & 0 & 0 & 0
 \cr -1 & 2 & -1 & 0 & 0 \cr 0 & -1 & 2 &
     -1 & 0 \cr 0 & 0 & -1 & 2 & -1 \cr 0 & 0 & 0 & -1 & 2 \cr  }
     \right)\, = \, \mbox{Cartan matrix of $A_5$}
\label{a5carta}
\end{equation}
while:
\begin{equation}
 \mathbf{w}_h \, = \, \left \{ - \frac{1}
      {{\sqrt{2}}}  ,
  \frac{1}{{\sqrt{6}}},
  - \frac{1}
      {{\sqrt{3}}}  ,
  \frac{1}{{\sqrt{5}}},
  -{\sqrt{\frac{3}{10}}} \right \}
\label{ierundaE6}
\end{equation}
is the highest weight of the  $\mathbf{20}$-dimensional
representation of $\mathrm{SU(1,5)}$. Finally in this basis
the highest root $\psi$
defined by eq.(\ref{psie6}) takes the expected form:
\begin{equation}
  \psi = \{ 0,0,0,0,0,0,\sqrt{2}\}
\label{expepsi7bis}
\end{equation}
\par
\paragraph{N=4}The case of $\mathcal{N}=4$ supergravity is the first where the scalar manifold is
not completely fixed, since we can choose the number $n_m$ of vector multiplets
that we can couple to the graviton multiplet. In any case, once $n_m$ is
fixed the scalar manifold is also fixed and we have:
\begin{equation}
  \mathcal{M}_{N=4,D=4} = \frac{\mathrm{SL(2,\mathbb{R})_0}}{\mathrm{O(2)}} \, \otimes\,
  \frac{\mathrm{SO(6,n_m)}}{\mathrm{SO(6) \times SO(n_m)}}
\label{n6manif}
\end{equation}
The total number of vectors $n_{\mathrm{v}} = 6+n_m$ is also fixed and the
symplectic representation $\mathbf{W}$ of the duality algebra
\begin{equation}
  \mathbb{U}_{D=4} =\mathrm{SL(2,\mathbb{R})_0} \, \times \,
  \mathrm{SO(6,n_m)}
\label{UD4}
\end{equation}
to which the vectors are assigned and which determines the embedding:
\begin{equation}
  \mathrm{SL(2,\mathbb{R})_0} \, \times \, \mathrm{SO(6) \times
  SO(n_m)}\, \mapsto \, \mathrm{Sp( 12+2\,n_m,\mathbb{R})}
\label{N4sympembed}
\end{equation}
is also fixed, namely $\mathbf{W} = \left( \mathbf{2_0,6+n_m} \right)$,
$\mathbf{2_0}$ being the fundamental representation of $\mathrm{SL(2,\mathbb{R})_0}$
and $\mathbf{6+n_m}$ the fundamental vector representation of
$\mathrm{SO(6,n_m)}$.
\vskip 0.2cm
\begin{figure}
\centering
\begin{picture}(100,40)
      \put (-180,35){$D_{\ell=4+k+1}$}\put (-75,56){\line (1,-1){20}}
\put (-109,59){$\alpha_\ell$}
      \put
(-79,59){\circle {10}} \put (-75,14){\line (1,1){20}}
\put (-109,10){$\alpha_{\ell-1}$}
\put
(-79,10){\circle {10}} \put (-45,35){\line (1,0){20}} \put
(-50,35){\circle {10}} \put (-53,20){$\alpha_{\ell-2}$}\put (-20,35){\circle {10}}
\put(-23,20){$\alpha_{\ell-3}$} \put (-9,35){$\dots$} \put
(10,35){$\dots$}  \put (15,35){$\dots$} \put (40,35){\circle {10}} \put (37,20){$\alpha_3$}
 \put (45,35){\line (1,0){20}}
 \put (70,35){\circle {10}}
 \put (70,35){\circle {9}}
 \put (70,35){\circle {8}}
 \put (70,35){\circle {7}}
 \put (70,35){\circle {6}}
 \put (70,35){\circle {5}}
 \put (70,35){\circle {4}}
 \put (70,35){\circle {3}}
 \put (70,35){\circle {2}}
 \put (70,35){\circle {1}}
 \put (67,20){$\alpha_{2}$} \put
(75,35){\line (1,0){20}} \put (100,35){\circle {10}} \put
(97,20){$\alpha_{1}$}
\end{picture}
$$\begin{array}{l}\psi = \alpha_1 +2\alpha_2 +2\alpha_3+\dots
+2\alpha_{\ell-2}+\alpha_{\ell-1} +\alpha_{\ell}
  \\(\psi \, , \,\alpha_2) = 1 \quad; \quad  (\psi \, , \, \alpha_i ) = 0 \quad i \ne 2 \
\end{array}$$
\vskip 1cm \caption{The Dynkin diagram of $D_{4+k+1}$.
The algebra $D_{4+k+1}$ is that of the group $\mathrm{SO(8,2k+2)}$
corresponding to the $\sigma$-model reduction of $\mathcal{N}=4$ supergravity coupled to
$n_m=2k$ vector multiplets. The only simple root which has non
vanishing grading with respect to the highest one $\psi$ is $\alpha_2$.
Removing it (black circle) we are left with the algebra $D_{4+k-1}
\oplus A_1$ which is indeed the duality algebra in $D=4$, namely
$\mathrm{SO(6,2k)} \oplus \mathrm{SL(2,\mathbb{R})_0}$.
 The black root $\alpha_2$  is the highest weight of the symplectic representation of the vector fields,
 namely the $\mathbf{W}=(\mathbf{2_0},\mathbf{6+2k})$. }
\label{except5}
\end{figure}
\vskip 0.2cm
The $D=3$ algebra  is,
$\mathbb{U}_{D=3}=\mathrm{SO(8,n_m+2)}$.
Correspondingly the form taken by the general decomposition
(\ref{gendecompo}) is the following one:
\begin{equation}
 \mbox{adj}(\mathrm{SO(8,n_m+2)}) =\mbox{adj}(\mathrm{SL(2,\mathbb{R})_0} )\, \oplus \,
 \mbox{adj}(\mathrm{SO(6,n_m)})
 \oplus\mbox{adj}(\mathrm{SL(2,\mathbb{R})_E})\oplus
(\mathbf{2_E},\mathbf{2_0,6+n_m})
\label{decompN4}
\end{equation}
where $\mathbf{2_{E,0}}$ are the fundamental representations
respectively of $\mathrm{SL(2,\mathbb{R})_E}$ and of
$\mathrm{SL(2,\mathbb{R})_0}$.
\par
\par
In order to give a Dynkin Weyl description of these algebras, we are
forced to distinguish the case of an odd and even number of vector
multiplets. In the first case both $\mathbb{U}_{D=3}$ and $\mathbb{U}_{D=4}$ are
non simply laced algebras of the $B$--type, while in the second case
they are both simply laced algebras of the $D$-type
\begin{equation}
  n_m = \cases{2k \quad \quad \quad\rightarrow \quad \mathbb{U}_{D=4} \simeq D_{k+3} \cr
  2k+1 \quad \, \rightarrow \quad \mathbb{U}_{D=4} \simeq B_{k+3} \cr}
\label{twocases}
\end{equation}
Just for simplicity and for shortness we choose to discuss only
the even case $n_m=2k$ which is described by fig.\ref{except5}.
\par
In this case we consider the $\mathbb{U}_{D=3} =\mathrm{SO(8,2k+2)}$
Lie algebra whose Dynkin diagram is that of $D_{5+k}$.
Naming $\epsilon_i$ the unit vectors in an Euclidean
$\ell$--dimensional space where $\ell=5+k$, a well adapted basis of
simple roots for the considered algebra is the following one:
\begin{eqnarray}
\alpha_1 & = & \sqrt{2} \, \epsilon _1 \nonumber\\
\alpha_2 & = & -\frac{1}{\sqrt{2}}\, \epsilon _1 -\epsilon _2 +\frac{1}{\sqrt{2}}\, \epsilon
_\ell\nonumber\\
\alpha_3 & = & \epsilon _2 -\epsilon _3 \nonumber\\
\alpha_4 & = & \epsilon _3 -\epsilon _4 \nonumber\\
\dots & = & \dots \nonumber\\
\alpha_{\ell-1} & = &\epsilon _{\ell-2} - \epsilon
_{\ell-1}\nonumber\\
\alpha_{\ell} & = &\epsilon _{\ell-2} + \epsilon
_{\ell-1}\nonumber\\
\label{welladapn4}
\end{eqnarray}
which is quite different from the usual presentation but yields the
correct Cartan matrix. In this basis the highest root of the algebra:
\begin{equation}
  \psi =  \alpha_1 +2\alpha_2 +2\alpha_3+\dots
+2\alpha_{\ell-2}+\alpha_{\ell-1} +\alpha_{\ell}
\label{highestn4}
\end{equation}
takes the desired form:
\begin{equation}
  \psi=\sqrt{2} \, \epsilon_\ell
\label{wellpsi}
\end{equation}
In the same basis the $\alpha_W =\alpha_2$ root has also the expect
form:
\begin{equation}
  \alpha_W = \left( \mathbf{w},\frac{1}{\sqrt{2}} \right)
\label{wellalphan4}
\end{equation}
where:
\begin{equation}
  \mathbf{w} =  -\frac{1}{\sqrt{2}}\, \epsilon _1 -\epsilon _2
\label{wellwn4}
\end{equation}
is the weight of the symplectic  representation $\mathbf{W}=\left( \mathbf{2_0} , \mathbf{6+2k}\right)
$. Indeed $-\frac{1}{\sqrt{2}}\, \epsilon _1$ is the fundamental
weight for the Lie algebra $\mathrm{SL(2,\mathbb{R})_0}$, whose root
is $\alpha_1 = \sqrt{2} \, \epsilon _1$, while $-\epsilon _2$ is the
highest weight for the vector representation of the algebra
$\mathrm{SO(6,2k)}$, whose roots are $\alpha_3, \alpha_4, \dots,
\alpha_\ell$.
\par
Next we briefly comment on the Tits Satake projection. The algebra
$\mathrm{SO(8,n_m+2)}$ is maximally split only for $n_m=5,6,7$. The case $n_m=6$,
from the superstring view point, corresponds to the case of
Neveu-Schwarz vector multiplets in a toroidal compactification.
For a different number of vector multiplets, in particular for
$n_m>7$ the study of extremal black holes involves considering the
Tits Satake projection, which just yields the universal algebra
\begin{equation}
  {\mathbb{U}}^{TS}_{N=4,D=3}={\so(8,9)}
\label{TSN8D3}
\end{equation}
\section{Tits Satake decompositions of  the $\mathbf{W}$ representations}
\label{titsu1}
As we stressed in the introduction, one of our main goals  is to compare the classification of extremal black holes by means of \textit{charge orbits} with their classification by means of $\mathrm{H}^\star$ \textit{orbits}. Charge orbits means orbits of the $\mathrm{U_{D=4}}$ group in the $\mathbf{W}$-representation.
\par
For this reason, in the present section we consider the decomposition of the $\mathbf{W}$-representations with respect to Tits-Satake subalgebras and Paint groups for all the non-exotic models. The relevant $\mathbf{W}$-representations are listed in table \ref{hstarrigruppine}.
In table \ref{hstarrigruppi} we listed the $\mathbf{W}$-representations for the exotic models.
\par
In \cite{titsusataku} the \textit{paint algebra} was defined as the algebra of external automorphism of the solvable Lie algebra $\Solv_\mathcal{M}$
generating the non-compact symmetric space: $\mathcal{M}\, = \, \mathrm{U/H}$, namely
\begin{equation}
  \mathbb{G}_{\mathrm{paint}} \, = \, \mathrm{Aut}_{\mathrm{Ext}} \, \left[
\Solv_\mathcal{M}\right]. \label{pittureFuori}
\end{equation}
where:
\begin{equation}
  \mathrm{Aut}_{\mathrm{Ext}} \, \left[ \Solv_\mathcal{M}\right] \,
  \equiv \, \frac{\mathrm{Aut} \, \left[
  \Solv_\mathcal{M}\right]}{\Solv_\mathcal{M}},
\label{outerauto}
\end{equation}
Given the paint algebra $\mathbb{G}_{\mathrm{paint}} \, \subset \, \mathbb{U}$ and the Tits Satake subalgebra $\mathbb{G}_{\mathrm{TS}}\, \subset \, \mathbb{U}$, whose construction we have briefly recalled above, following \cite{titsusataku}
one introduces the \textit{sub Tits Satake} and \textit{sub paint} algebras as the centralizers of the paint algebra and of the Tits Satake algebra, respectively. In other words we have:
\begin{equation}\label{subTs}
    \mathfrak{s} \, \in \, \mathbb{G}_{\mathrm{subTS}} \, \subset \, \mathbb{G}_{\mathrm{TS}}\, \subset \, \mathbb{U} \quad \Leftrightarrow \, \quad \left[ \mathfrak{s} \, , \, \mathbb{G}_{\mathrm{paint}}\right] \, = \, 0
\end{equation}
and
\begin{equation}\label{subpaint}
    \mathfrak{t} \, \in \, \mathbb{G}_{\mathrm{subpaint}} \, \subset \, \mathbb{G}_{\mathrm{paint}}\, \subset \, \mathbb{U} \quad \Leftrightarrow \, \quad \left[ \mathfrak{t} \, , \, \mathbb{G}_{\mathrm{TS}}\right] \, = \, 0
\end{equation}
A very important property of the paint and subpaint algebras is that they are conserved in the dimensional reduction, namely they are the same for $\mathbb{U}_{D=4}$ and $\mathbb{U}_{D=3}$.
\par
In the next lines we analyze the decomposition of the $\mathbf{W}$-representations with respect to these subalgebras for each Tits Satake universality class of non maximally split models. In the case of maximally split models there is no paint algebra and there is nothing with respect to which to decompose.
\subsection{Universality class $\sym(6,\mathbb{R})\Rightarrow \mathfrak{f}_{4(4)}$}
In this case the sub Tits Satake Lie algebra is
\begin{equation}\label{cuffio}
    \mathbb{G}_{\mathrm{subTS}} \, = \, \slal(2,\mathbb{R}) \oplus \slal(2,\mathbb{R}) \oplus \slal(2,\mathbb{R}) \subset \sym(6,\mathbb{R}) \, = \, \mathbb{G}_{\mathrm{TS}}
\end{equation}
and the $\mathbf{W}$-representation of the maximally split model decomposes as follows:
\begin{equation}\label{scompo14}
    \mathbf{14}^\prime  \, \stackrel{\mathbb{G}_{\mathrm{subTS}}}{\Longrightarrow} \, (\mathbf{2},\mathbf{1},\mathbf{1}) \oplus (\mathbf{1},\mathbf{2},\mathbf{1}) \oplus (\mathbf{1},\mathbf{1},\mathbf{2}) \oplus (\mathbf{2},\mathbf{2},\mathbf{2})
\end{equation}
This decomposition combines in the following way with the paint group representations in the various models belonging to the same universality class.
\subsubsection{$\su(3,3)$ model}
For this case the paint algebra is
\begin{equation}\label{furadino}
     \mathbb{G}_{\mathrm{paint}} \, = \, \so(2) \oplus \so(2)
\end{equation}
and the $\mathbf{W}$-representation is the $\mathbf{20}$ dimensional of $\su(3,3)$ corresponding to an antisymmetric tensor with a reality condition of the form:
\begin{equation}\label{realata}
    t_{\alpha\beta\gamma}^\star = \frac{1}{3!} \, \epsilon_{\alpha\beta\gamma\delta\eta\theta} \, t_{\delta\eta\theta}
\end{equation}
The decomposition of this representation with respect to the Lie algebra $ \mathbb{G}_{\mathrm{paint}}\oplus {\mathbb{G}_{\mathrm{subTS}}}$ is the following one:
\begin{equation}\label{ruppatoA}
    \mathbf{20}  \, \stackrel{\mathbb{G}_{\mathrm{paint}} \oplus \mathbb{G}_{\mathrm{subTS}}}{\Longrightarrow} \, (2,q_1|\mathbf{2},\mathbf{1},\mathbf{1}) \oplus (2,q_2|\mathbf{1},\mathbf{2},\mathbf{1}) \oplus (2,q_3|\mathbf{1},\mathbf{1},\mathbf{2}) \oplus (1,0|\mathbf{2},\mathbf{2},\mathbf{2})
\end{equation}
where $(2,q)$ means a doublet of $\so(2)\oplus\so(2)$ with a
certain grading $q$ with respect to the generators, while $(1,0)$
means the singlet that has $0$ grading with respect to both
generators. The subpaint algebra in this case is $
\mathbb{G}_{\mathrm{subpaint}}\, = \,0$ and the decomposition of
the same $\mathbf{W}$-representation with respect to
$\mathbb{G}_{\mathrm{subpaint}} \oplus \mathbb{G}_{\mathrm{TS}}$
is:
\begin{equation}\label{ruppatoB}
    \mathbf{20}  \, \stackrel{\mathbb{G}_{\mathrm{subpaint}} \oplus \mathbb{G}_{\mathrm{TS}}}{\Longrightarrow} \, \mathbf{6} \, \oplus \, \mathbf{14}
\end{equation}
This follows from the decomposition of the $\mathbf{6}$ of $\sym(6,\mathbf{R})$ with respect to the sub Tits Satake algebra (\ref{cuffio}):
\begin{equation}\label{seifracco}
    \mathbf{6} \,  \stackrel{\mathbb{G}_{\mathrm{subTS}}}{\Longrightarrow}  \, (\mathbf{2},\mathbf{1},\mathbf{1}) \oplus (\mathbf{1},\mathbf{2},\mathbf{1}) \oplus (\mathbf{1},\mathbf{1},\mathbf{2})
\end{equation}
\subsubsection{$\so^\star(12)$ model}
For this case the paint algebra is
\begin{equation}\label{furadinotwo}
     \mathbb{G}_{\mathrm{paint}} \, = \, \so(3) \oplus \so(3) \oplus \so(3)
\end{equation}
and the $\mathbf{W}$-representation is the $\mathbf{32}_s $
dimensional spinorial representation of $\so^\star(12)$. The
decomposition of this representation with respect to the Lie
algebra $ \mathbb{G}_{\mathrm{paint}}\oplus
{\mathbb{G}_{\mathrm{subTS}}}$ is the following one:
\begin{equation}\label{ruppatoAtwo}
    \mathbf{32}_s  \, \stackrel{\mathbb{G}_{\mathrm{paint}} \oplus \mathbb{G}_{\mathrm{subTS}}}{\Longrightarrow} \, (\underline{2},\underline{2},\underline{1}|\mathbf{2},\mathbf{1},\mathbf{1}) \oplus (\underline{2},\underline{1},\underline{2}|\mathbf{1},\mathbf{2},\mathbf{1}) \oplus (\underline{1},\underline{1},\underline{2}|\mathbf{1},\mathbf{1},\mathbf{2}) \oplus (\underline{1},\underline{1},\underline{1}|\mathbf{2},\mathbf{2},\mathbf{2})
\end{equation}
where $\underline{2}$ means the doublet spinor representation of
$\so(3)$. The subpaint algebra in this case is $
\mathbb{G}_{\mathrm{paint}}\, = \,\so(3)_{\mathrm{diag}}$ and the
decomposition of the same $\mathbf{W}$-representation with respect
to $\mathbb{G}_{\mathrm{subpaint}} \oplus
\mathbb{G}_{\mathrm{TS}}$ is:
\begin{equation}\label{ruppatoBtwo}
    \mathbf{32}_s  \, \stackrel{\mathbb{G}_{\mathrm{TS}}\oplus \mathbb{G}_{\mathrm{subpaint}}}{\Longrightarrow} \, (\mathbf{6}|\underline{3}) \, \oplus \, (\mathbf{14}^\prime|\underline{1})
\end{equation}
This follows from the decomposition of the product $\underline{2} \times \underline{2}$ of $\so(3)_{\mathrm{diag}}$ times the Tits Satake algebra (\ref{cuffio}):
\begin{equation}\label{seifraccotwo}
    \underline{2} \times \underline{2} \, =  \, \underline{3} \, \oplus \, \underline{1}
\end{equation}
\subsubsection{$\mathfrak{e}_{7(-25)}$ model}
For this case the paint algebra is
\begin{equation}\label{furadinothree}
     \mathbb{G}_{\mathrm{paint}} \, = \, \so(8)
\end{equation}
and the $\mathbf{W}$-representation is the fundamental $\mathbf{56} $ dimensional  representation of $\mathfrak{e}_{7(-25)}$
The decomposition of this representation with respect to the Lie algebra $ \mathbb{G}_{\mathrm{paint}}\oplus {\mathbb{G}_{\mathrm{subTS}}}$ is the following one:
\begin{equation}\label{ruppatoAthree}
    \mathbf{56}  \, \stackrel{\mathbb{G}_{\mathrm{paint}} \oplus \mathbb{G}_{\mathrm{subTS}}}{\Longrightarrow} \, (\mathbf{8}_v|\mathbf{2},\mathbf{1},\mathbf{1}) \oplus (\mathbf{8}_s|\mathbf{1},\mathbf{2},\mathbf{1}) \oplus (\mathbf{8}_c|\mathbf{1},\mathbf{1},\mathbf{2}) \oplus (\mathbf{1}|\mathbf{2},\mathbf{2},\mathbf{2})
\end{equation}
where $\mathbf{8}_{v,s,c}$ are the three inequivalent eight-dimensional representations of $\so(8)$, the vector, the spinor and the conjugate spinor. The subpaint algebra in this case is $ \mathbb{G}_{\mathrm{paint}}\, = \,\mathfrak{g}_{2(-14)}$ with respect to which all three $8$-dimensional representations of $\so(8)$ branch as follows:
\begin{equation}\label{seifraccothree}
    \mathbf{8}_{v,s,c} \, \stackrel{\mathfrak{g}_{2(-14)}}{\Longrightarrow}  \, \mathbf{7 }\, \oplus \, \mathbf{1}
\end{equation}
In view of this the decomposition of the same $\mathbf{W}$-representation with respect to $\mathbb{G}_{\mathrm{subpaint}} \oplus \mathbb{G}_{\mathrm{TS}}$ is:
\begin{equation}\label{ruppatoBthree}
    \mathbf{56}  \, \stackrel{\mathbb{G}_{\mathrm{TS}} \oplus \mathbb{G}_{\mathrm{subpaint}}}{\Longrightarrow} \, (\mathbf{6}|\mathbf{7}) \, \oplus \, (\mathbf{14}^\prime|\mathbf{1})
\end{equation}
\subsection{Universality class $\slal(2,\mathbb{R}) \oplus \so(2,3) \Rightarrow \so(4,5)$}
This case corresponds to one of the possible infinite families of $\mathcal{N}=2$ theories with a symmetric homogeneous special K\"ahler manifold and a number of vector multiplets larger than three ($n=3+p$).  The other infinite family corresponds instead to  one of the three exotic models.
\par
The generic element of this infinite class corresponds to the following algebras:
\begin{eqnarray}
  \mathbb{U}_{\mathrm{D=4}} &=& \slal(2,\mathbb{R}) \oplus \so(2,2+p)\nonumber\\
  \mathbb{U}_{\mathrm{D=3}} &=& \so(4,4+p) \label{feldane}
\end{eqnarray}
In this case the sub Tits Satake algebra is:
\begin{equation}\label{casilino1}
    \mathbb{G}_{\mathrm{subTS}} \, = \, \slal(2,\mathbb{R}) \oplus \slal(2,\mathbb{R}) \oplus \slal(2,\mathbb{R}) \, \simeq \, \slal(2,\mathbb{R}) \oplus \so(2,2) \, \, \subset \, \slal(2,\mathbb{R}) \oplus \so(2,3) \, = \, \mathbb{G}_{\mathrm{TS}}
\end{equation}
an the paint and subpaint algebras are as follows:
\begin{eqnarray}
  \mathbb{G}_{\mathrm{paint}} &=& \so(p) \nonumber\\
  \mathbb{G}_{\mathrm{subpaint}} &=& \so(p-1) \label{felanina1}
\end{eqnarray}
The symplectic $\mathbf{W}$ representation of $\mathbb{U}_{\mathrm{D=4}}$ is the tensor product of the fundamental representation of $\slal(2)$ with the fundamental vector representation of $\so(2,2+p)$, namely
\begin{equation}\label{erast1}
    \mathbf{W} \, = \, \left( \mathbf{2 | 4}+p\right) \quad ; \quad \mbox{dim} \,\mathbf{W} \, = \, 8+2p
\end{equation}
The decomposition of this representation with respect to $\mathbb{G}_{\mathrm{subTS}} \oplus \mathbb{G}_{\mathrm{subpaint}}$ is the following one:
\begin{equation}\label{decompusOne}
    \mathbf{W} \, \stackrel{\mathbb{G}_{\mathrm{subTS}} \oplus \mathbb{G}_{\mathbf{subpaint}}}{\Longrightarrow} \,  \left(\mathbf{2,2,2|1}\right) \oplus \left(\mathbf{2,1,1|1}\right) \oplus \left(\mathbf{2,1,1}|p-1\right)
\end{equation}
where $\mathbf{2,2,2}$ denotes the tensor product of the three fundamental representations of $\slal(2,\mathbb{R})^3$. Similarly $\mathbf{2,1,1}$ denotes the doublet of the first $\slal(2,\mathbb{R})$ tensored with the singlets of the following two $\slal(2,\mathbb{R})$ algebras. The representations appearing in (\ref{decompusOne}) can be grouped in order to reconstruct full representations either of the complete Tits Satake or of the complete paint algebras. In this way one obtains:
\begin{eqnarray}
   \mathbf{W} & \stackrel{\mathbb{G}_{\mathrm{subTS}} \oplus \mathbb{G}_{\mathbf{paint}}}{\Longrightarrow} &  \left(\mathbf{2,2,2|1}\right) \oplus  \left(\mathbf{2,1,1}|p+1\right)\nonumber\\
  \mathbf{W} & \stackrel{\mathbb{G}_{\mathrm{TS}} \oplus \mathbb{G}_{\mathbf{subpaint}}}{\Longrightarrow} &  \left(\mathbf{2,5|1}\right) \oplus  \left(\mathbf{2,1}|p-1\right)
\end{eqnarray}
\subsection{Universality class $\slal(2,\mathbb{R}) \oplus \so(6,7) \Rightarrow \so(8,9)$}
This case, which corresponds to an $\mathcal{N}=4$ theory with a number of vector multiplets larger than six ($n=6+p$) presents a very strong similarity with the previous $\mathcal{N}=2$ case.
\par
The generic element of this infinite class corresponds to the following algebras:
\begin{eqnarray}
  \mathbb{U}_{\mathrm{D=4}} &=& \slal(2,\mathbb{R}) \oplus \so(6,6+p)\nonumber\\
  \mathbb{U}_{\mathrm{D=3}} &=& \so(8,8+p) \label{feldane2}
\end{eqnarray}
In this case the sub Tits Satake algebra is:
\begin{equation}\label{casilino2}
    \mathbb{G}_{\mathrm{subTS}} \, = \, \slal(2,\mathbb{R}) \oplus \so(6,6) \, \, \subset \, \slal(2,\mathbb{R}) \oplus \so(6,7) \, = \, \mathbb{G}_{\mathrm{TS}}
\end{equation}
an the paint and subpaint algebras are the same as in the previous $\mathcal{N}=2$ case, namely:
\begin{eqnarray}
  \mathbb{G}_{\mathrm{paint}} &=& \so(p) \nonumber\\
  \mathbb{G}_{\mathrm{subpaint}} &=& \so(p-1) \label{felanina2}
\end{eqnarray}
The symplectic $\mathbf{W}$ representation of $\mathbb{U}_{\mathrm{D=4}}$ is the tensor product of the fundamental representation of $\slal(2)$ with the fundamental vector representation of $\so(6,6+p)$, namely
\begin{equation}\label{erast2}
    \mathbf{W} \, = \, \left( \mathbf{2 | 12}+p\right) \quad ; \quad \mbox{dim} \,\mathbf{W} \, = \, 24+2p
\end{equation}
The decomposition of this representation with respect to $\mathbb{G}_{\mathrm{subTS}} \oplus \mathbb{G}_{\mathrm{subpaint}}$ is the following one:
\begin{equation}\label{decompusTwo}
    \mathbf{W} \, \stackrel{\mathbb{G}_{\mathrm{subTS}} \oplus \mathbb{G}_{\mathbf{subpaint}}}{\Longrightarrow} \,  \left(\mathbf{2,12|1}\right) \oplus \left(\mathbf{2,1|1}\right) \oplus \left(\mathbf{2,1}|p\right)
\end{equation}
Just as above the three representations appearing in (\ref{decompusTwo}) can be grouped in order to obtain either representation of
the complete Tits Satake or of the complete paint algebras. This yields
\begin{eqnarray}
   \mathbf{W} & \stackrel{\mathbb{G}_{\mathrm{subTS}} \oplus \mathbb{G}_{\mathbf{paint}}}{\Longrightarrow} &  \left(\mathbf{2,12|1}\right) \oplus  \left(\mathbf{2,1}|p+1\right)\nonumber\\
  \mathbf{W} & \stackrel{\mathbb{G}_{\mathrm{TS}} \oplus \mathbb{G}_{\mathbf{subpaint}}}{\Longrightarrow} &  \left(\mathbf{2,13|1}\right) \oplus  \left(\mathbf{2,1}|p\right) \label{robetta1}
\end{eqnarray}
\subsection{The  universality classes $\slal(2,\mathbb{R}) \oplus \so(6,n) \Rightarrow \so(8,n+2)$ with $n\le 5$}
These classes correspond to the $\mathcal{N}=4$ theories with a number $n=1,2,3,4,5$ of vector multiplets.
In each case we have  the following algebras:
\begin{eqnarray}
  \mathbb{U}_{\mathrm{D=4}} &=& \slal(2,\mathbb{R}) \oplus \so(6,n)\nonumber\\
  \mathbb{U}_{\mathrm{D=3}} &=& \so(8,n+2) \label{feldane4}
\end{eqnarray}
In all these cases the Tits Satake  and sub Tits Satake algebras are:
\begin{eqnarray}
  \mathbb{G}_{\mathrm{TS}} &=& \slal(2,\mathbb{R})  \oplus \so(n+1,n) \nonumber\\
  \mathbb{G}_{\mathrm{subTS}} &=& \slal(2,\mathbb{R})  \oplus \so(n,n) \label{casilus}
\end{eqnarray}
and the paint and subpaint algebras are:
\begin{eqnarray}
  \mathbb{G}_{\mathrm{paint}} &=&  \so(6-n)\nonumber\\
  \mathbb{G}_{\mathrm{subpaint}} &=& \so(5-n) \label{feldane4}
\end{eqnarray}
The symplectic  $\mathbb{W}$ representation is the tensor product of the doublet representation of $\slal(2)$ with the fundamental representation of $\so(6,n) $, namely
\begin{equation}\label{wdoppiaN}
    \mathbf{W} \, = \, \left(\mathbf{2, 6+n}\right)
\end{equation}
and its decomposition with respect to the $\mathbb{G}_{\mathrm{subTS}} \oplus \mathbb{G}_{\mathrm{subpaint}}$  algebra is as follows
\begin{equation}\label{decompusThree}
    \mathbf{W} \, \stackrel{\mathbb{G}_{\mathrm{subTS}} \oplus \mathbb{G}_{\mathbf{subpaint}}}{\Longrightarrow} \,  \left(\mathbf{2,2n|1}\right) \oplus \left(\mathbf{2,1|1}\right) \oplus \left(\mathbf{2,1|5-n}\right)
\end{equation}
which, with the same procedure as above leads to:
\begin{eqnarray}
   \mathbf{W} & \stackrel{\mathbb{G}_{\mathrm{subTS}} \oplus \mathbb{G}_{\mathbf{paint}}}{\Longrightarrow} &  \left(\mathbf{2,2n|1}\right) \oplus  \left(\mathbf{2,1}|\mathbf{6-n}\right)\nonumber\\
  \mathbf{W} & \stackrel{\mathbb{G}_{\mathrm{TS}} \oplus \mathbb{G}_{\mathbf{subpaint}}}{\Longrightarrow} &  \left(\mathbf{2,2n+1|1}\right) \oplus  \left(\mathbf{2,1}|\mathbf{5-n}\right) \label{robetta2}
\end{eqnarray}
\subsection{$\mathbf{W}$-representations of the maximally split non exotic models}
In the previous subsections we have analysed the Tits-Satake decomposition of the $\mathbf{W}$-representation for all those models that are non maximally split. The remaining models are the maximally split ones for which there is no paint algebra and the Tits Satake projection is the identity map. For the reader's convenience we have extracted from table \ref{homomodels} the list of such models and  presented it in table \ref{maxsplittus}.
\begin{table}
\begin{center}
{\tiny
\begin{tabular}{|l|c|c||c|c||c|c||c||}
  \hline
  \null & TS & TS & coset &coset &  Paint & subP &  susy\\
  $\#$ & D=4 & D=3 & D=4 & D=3 &  Group & Group &  \\
  \hline
  \hline
1&\null & \null & \null & \null &\null & \null & \null \\
\null \null & $ \frac{E_{7(7)}}{\mathrm{SU(8)}}$ & $ \frac{\mathrm{E_{8(8)}}}{\mathrm{SO^\star(16)}}$  & $ \frac{\mathrm{E_{7(7)}}}{\mathrm{SU(8)}}$ & $ \frac{\mathrm{E_{8(8)}}}{\mathrm{SO^\star(16)}}$  & $1$ & $1$ & $\mathcal{N}=8$ \\
\null &\null & \null & \null & \null &\null& \null & \null \\
\hline
2 &\null & \null & \null & \null &\null & \null & \null \\
\null & $ \frac{\mathrm{SU(1,1)}}{\mathrm{U(1)}}$ & $ \frac{\mathrm{G_{2(2)}}}{\mathrm{SL(2,R)\times SL(2,R)}}$  & $ \frac{\mathrm{SU(1,1)}}{\mathrm{U(1)}}$ & $ \frac{\mathrm{G_{2(2)}}}{\mathrm{SL(2,R)\times SL(2,R)}}$  & $1$ & $1$ & $\mathcal{N}=2$ \\
\null &\null & \null & \null & \null &\null & \null & n=1 \\
\hline
\hline
3 &\null & \null & \null & \null &\null & \null & \null \\
\null & $ \frac{\mathrm{Sp(6,R)}}{\mathrm{SU(3)\times  U(1)}}$ & $ \frac{\mathrm{F_{4(4)}}}{\mathrm{Sp(6,R)\times SL(2,R)}}$  & $ \frac{\mathrm{Sp(6,R)}}{\mathrm{SU(3)\times  U(1)}}$ & $ \frac{\mathrm{F_{4(4)}}}{\mathrm{Sp(6,R)\times SL(2,R)}}$  & $1$ & $1$  & $\mathcal{N}=2$ \\
\null &\null & \null & \null & \null &\null & \null & $n=6$ \\
\cline{1-1} \cline{4-8}
\hline
11 & \null \null & \null & \null & \null &\null & \null & \null \\
\null & $ \frac{\mathrm{SL(2,\mathbb{R})}}{\mathrm{O(2)}}\times\frac{\mathrm{SO(6,5)}}{\mathrm{SO(6)\times SO(5)}}$ & $ \frac{\mathrm{SO(8,7)}}{\mathrm{SO(6,2)\times SO(2,5)}}$  & $ \frac{\mathrm{SL(2,\mathbb{R})}}{\mathrm{O(2)}}\times\frac{\mathrm{SO(6,5)}}{\mathrm{SO(6)\times SO(5)}}$ & $ \frac{\mathrm{SO(8,7)}}{\mathrm{SO(6,2)\times SO(2,5)}}$ & $\mathrm{1}$ & $\mathrm{1}$ & $\mathcal{N}=4$ \\
\null &\null & \null & \null & \null &\null & \null & n=5 \\
\hline
12 & \null \null & \null & \null & \null &\null & \null & \null \\
\null & $ \frac{\mathrm{SL(2,\mathbb{R})}}{\mathrm{O(2)}}\times\frac{\mathrm{SO(6,6)}}{\mathrm{SO(6)\times SO(6))}}$ & $ \frac{\mathrm{SO(8,8)}}{\mathrm{SO(6,2)\times SO(2,6)}}$  & $ \frac{\mathrm{SL(2,\mathbb{R})}}{\mathrm{O(2)}}\times\frac{\mathrm{SO(6,6)}}{\mathrm{SO(6)\times SO(6))}}$ & $ \frac{\mathrm{SO(8,8)}}{\mathrm{SO(6,2)\times SO(2,6)}}$ & $1$ & $1$ & $\mathcal{N}=4$ \\
\null &\null & \null & \null & \null &\null & \null & n=6 \\
\hline
13 & \null \null & \null & \null & \null &\null & \null & \null \\
\null & $ \frac{\mathrm{SL(2,\mathbb{R})}}{\mathrm{O(2)}}\times\frac{\mathrm{SO(6,7)}}{\mathrm{SO(6)\times SO(7))}}$ & $ \frac{\mathrm{SO(8,9)}}{\mathrm{SO(6,2)\times SO(2,7)}}$  & $ \frac{\mathrm{SL(2,\mathbb{R})}}{\mathrm{O(2)}}\times\frac{\mathrm{SO(6,7)}}{\mathrm{SO(6)\times SO(7))}}$ & $ \frac{\mathrm{SO(8,9)}}{\mathrm{SO(6,2)\times SO(2,7)}}$ & $1$ & $1$ & $\mathcal{N}=4$ \\
\null &\null & \null & \null & \null &\null & \null & n=7 \\
\hline
14 & \null \null & \null & \null & \null &\null & \null & \null \\
\null & $ \frac{\mathrm{SL(2,\mathbb{R})}}{\mathrm{O(2)}}\times\frac{\mathrm{SO(2,1)}}{\mathrm{SO(2)}}$ & $ \frac{\mathrm{SO(4,3)}}{\mathrm{SO(2,2)\times SO(2,1)}}$  & $ \frac{\mathrm{SL(2,\mathbb{R})}}{\mathrm{O(2)}}\times\frac{\mathrm{SO(2,1)}}{\mathrm{SO(2)}}$ & $ \frac{\mathrm{SO(4,3)}}{\mathrm{SO(2,2)\times SO(2,1)}}$ & $\mathrm{1}$ & $\mathrm{1}$ & $\mathcal{N}=2$ \\
\null &\null & \null & \null & \null &\null & \null & n=2 \\
\hline
15 & \null \null & \null & \null & \null &\null & \null & \null \\
\null & $ \frac{\mathrm{SL(2,\mathbb{R})}}{\mathrm{O(2)}}\times\frac{\mathrm{SO(2,2)}}{\mathrm{SO(2)\times SO(2)}}$ & $ \frac{\mathrm{SO(4,4)}}{\mathrm{SO(2,2)\times SO(2,2)}}$  & $ \frac{\mathrm{SL(2,\mathbb{R})}}{\mathrm{O(2)}}\times\frac{\mathrm{SO(2,2)}}{\mathrm{SO(2)\times SO(2)}}$ & $ \frac{\mathrm{SO(4,4)}}{\mathrm{SO(2,2)\times SO(2,2)}}$ & $\mathrm{1}$ & $\mathrm{1}$ & $\mathcal{N}=2$ \\
\null &\null & \null & \null & \null &\null & \null & n=3 \\
\hline
16 & \null \null & \null & \null & \null &\null & \null & \null \\
\null & $ \frac{\mathrm{SL(2,\mathbb{R})}}{\mathrm{O(2)}}\times\frac{\mathrm{SO(2,3)}}{\mathrm{SO(2)\times SO(3)}}$ & $ \frac{\mathrm{SO(4,5)}}{\mathrm{SO(2,2)\times SO(2,3)}}$  & $ \frac{\mathrm{SL(2,\mathbb{R})}}{\mathrm{O(2)}}\times\frac{\mathrm{SO(2,3)}}{\mathrm{SO(2)\times SO(3)}}$ & $ \frac{\mathrm{SO(4,5)}}{\mathrm{SO(2,2)\times SO(2,3)}}$ & $\mathrm{1}$ & $\mathrm{1}$ & $\mathcal{N}=2$ \\
\null &\null & \null & \null & \null &\null & \null & n=4 \\
\hline
\end{tabular}
}
\caption{The list  of \textit{non-exotic} homogenous symmetric scalar manifolds appearing in $D=4$ supergravity which are also maximally split. For these models the paint group is the identity group. \label{maxsplittus}}
\end{center}
\end{table}
As we see from the table we have essentially five type of models:
\begin{enumerate}
  \item The $\mathrm{E_{7(7)}}$ model corresponding to $\mathcal{N}=8$ supergravity where the $\mathbf{W}$-representation is the fundamental $\mathbf{56}$.
  \item The $\mathrm{SU(1,1)}$ non exotic model where the $\mathbf{W}$-representation is the $j=\ft 32$ of $\so(1,2)\sim \su(1,1)$
  \item The $\mathrm{Sp(6,\mathbb{R})}$ model where the $\mathbf{W}$-representation is the $\mathbf{14}^\prime$ (antisymmetric symplectic traceless three-tensor).
  \item The models $\slal(2,\mathbb{R}) \oplus \so(q,q)$ where the $\mathbf{W}$-representation is the $\left(\mathrm{2,2q}\right)$, namely the tensor product of the two fundamentals.
   \item The models $\slal(2,\mathbb{R}) \oplus \so(q,q+1)$ where the $\mathbf{W}$-representation is the $\left(\mathrm{2,2q+1}\right)$, namely the tensor product of the two fundamentals.
\end{enumerate}
Therefore, for the above maximally split models, the \textit{charge classification} of black holes  reduces  to the classification of $\mathrm{U_{D=4}}$ orbits in the mentioned $\mathbf{W}$-representations. Actually  such orbits are sufficient also for the non maximally split models. Indeed each of the above $5$-models correspond to one Tits Satake universality class
and, within each universality class, the only relevant part of the $\mathbf{W}$-representation is the subpaint group singlet which is universal for all members of the class. This is precisely what we verified in the previous subsections.
\par
For instance for all members of the universality class of $\mathrm{Sp(6,\mathbb{R})}$, the $\mathbf{W}$-representation splits as follows with respect to the subalgebra $\sym(6,\mathbb{R})\oplus \mathbb{G}_{\mathrm{subpaint}}$:
\begin{equation}\label{gribochky}
    \mathbf{W} \, \stackrel{\sym(6,\mathbb{R})\oplus \mathbb{G}_{\mathrm{subpaint}}}{\Longrightarrow} \, \left( \mathbf{6}\, | \, \mathcal{D}_{\mathrm{subpaint}}\right) \, + \, \left( \mathbf{14}^\prime \, | \, \mathbf{1}_{\mathrm{subpaint}}\right)
\end{equation}
where the representation $\mathcal{D}_{\mathrm{subpaint}}$ is the following one for the three non-maximally split members of the class:
\begin{equation}\label{reppisubpitturi}
    \mathcal{D}_{\mathrm{subpaint}} \, = \, \left\{ \begin{array}{ccccc}
                                                      \mathbf{1} & \mbox{of} & \mathbf{1} & \mbox{for the} & \su(3,3)-\mbox{model} \\
                                                      \mathbf{3} & \mbox{of} & \so(3) & \mbox{for the} & \so^\star(12)-\mbox{model} \\
                                                      \mathbf{7} & \mbox{of} & \mathfrak{g}_{2(-14)} & \mbox{for the} & \mathfrak{e}_{7(-25)}-\mbox{model}\\
                                                    \end{array}\right.
\end{equation}
Clearly the condition:
\begin{equation}\label{condosputta}
    \left( \mathbf{6}\, | \, \mathcal{D}_{\mathrm{subpaint}}\right)\, = \, 0
\end{equation}
imposed on a vector in the $\mathbf{W}$-representation breaks the group $\mathrm{U}_{D=4}$  to its Tits Satake subgroup. The key point is that each $\mathbf{W}$-orbit of the big group $\mathrm{U}_{D=4}$ crosses the locus (\ref{condosputta}) so that the classification  of $\mathrm{Sp(6,\mathbb{R})}$ orbits in the $\mathbf{14}^\prime$-representation exhausts the classification of $\mathbf{W}$-orbits for all members of the universality class.
\par
In order to prove that the gauge (\ref{condosputta}) is always reachable it suffices to show that the representation $\left( \mathbf{6}\, | \,\mathcal{D}_{\mathrm{subpaint}}\right )$ always appears at least once in the decomposition of the Lie algebra $\mathbb{U}_{D=4}$ with respect to the subalgebra $\sym(6,\mathbb{R})\oplus \mathbb{G}_{\mathrm{subpaint}}$. The corresponding parameters of the big group can be used to set to zero the projection of the $\mathbf{W}$-vector onto $\left( \mathbf{6}\, | \,\mathcal{D}_{\mathrm{subpaint}}\right )$.
\par
The required condition is easily verified since we have:
\begin{eqnarray}
  \underbrace{\mbox{adj}\, \su(3,3)}_{\mathbf{35}} &\stackrel{\sym(6,\mathbb{R})}{\Longrightarrow} \,& \underbrace{\mbox{adj}\, \sym(6,\mathbb{R})}_{\mathbf{21}} \, \oplus \, \mathbf{6} \, \oplus \, \mathbf{6} \, \oplus \, \mathbf{1} \, \oplus \, \mathbf{1}\nonumber \\
 \underbrace{\mbox{adj}\, \so^\star(12)}_{\mathbf{66}} &\stackrel{\sym(6,\mathbb{R})\oplus \so(3)}{\Longrightarrow} \,& \underbrace{\mbox{adj}\, \sym(6,\mathbb{R})}_{\mathbf{21}} \, \oplus \, \underbrace{\mbox{adj}\, \so(3)}_{\mathbf{3}}\,
  \oplus \, \left(\mathbf{6} , \mathbf{3}\right)\, \oplus \, \left(\mathbf{6} , \mathbf{3}\right)\oplus \, \left(\mathbf{1},\mathbf{3}\right)\,\oplus \, \left(\mathbf{1},\mathbf{3}\right)\,\nonumber \\
 \underbrace{\mbox{adj}\, \mathfrak{e}_{7(-25)}}_{\mathbf{133}} &\stackrel{\sym(6,\mathbb{R})\oplus \mathfrak{g}_{2(-14)}}{\Longrightarrow} \,& \underbrace{\mbox{adj}\, \sym(6,\mathbb{R})}_{\mathbf{21}} \, \oplus \, \underbrace{\mbox{adj}\, \mathfrak{g}_{2(-14)}}_{\mathbf{14}}\,
  \oplus \, \left(\mathbf{6} , \mathbf{7}\right)\, \oplus \, \left(\mathbf{6} , \mathbf{7}\right)\oplus \, \left(\mathbf{1},\mathbf{7}\right)\,\oplus \, \left(\mathbf{1},\mathbf{7}\right)\,\nonumber \\
\end{eqnarray}
The reader cannot avoid being impressed by the striking similarity of the above decompositions which encode the very essence of  Tits Satake universality. Indeed the representations of the common Tits Satake subalgebra appearing in the decomposition of the adjoint are the same for all members of the class. They are simply uniformly assigned to the fundamental representation of the subpaint algebra which is different in the three cases. The representation $\left( \mathbf{6}\, | \,\mathcal{D}_{\mathrm{subpaint}}\right )$ appears twice in these decompositions and can be used to reach the gauge (\ref{condosputta}) as we claimed above.
\par
For the models of type $\slal(2,\mathbb{R})\oplus \so(q,q+p)$ having $\slal(2,\mathbb{R})\oplus \so(q,q+1)$ as Tits Satake subalgebra and $\so(p-1)$ as subpaint algebra the decomposition of the $\mathbf{W}$-representation is the following one:
\begin{equation}\label{funghifritti}
    \mathbf{W} \, = \, \left(\mathbf{2,2q+p}\right) \, \stackrel{\slal(2,\mathbb{R})\oplus \so(q,q+1)\oplus \so(p-1)}{\Longrightarrow} \, \left(\mathbf{2,2q+1}|\mathbf{1}\right) \, \oplus \, \left(\mathbf{2,1}|\mathbf{p-1}\right)
\end{equation}
and the question is whether each $\slal(2,\mathbb{R})\oplus \so(q,q+p)$ orbit in the $\left(\mathbf{2,2q+p}\right)$ representation intersects the $\slal(2,\mathbb{R})\oplus \so(q,q+1)\oplus \so(p-1)$-invariant locus:
\begin{equation}\label{fittone}
    \left(\mathbf{2,1}|\mathbf{p-1}\right) \, = \,0
\end{equation}
The answer is yes since we always have enough parameters in the coset
\begin{equation}\label{ciabatta}
    \frac{\mathrm{SL(2,\mathbb{R})}\times \mathrm{SO(q,q+p)}}{\mathrm{SL(2,\mathbb{R})\times SO(q,q+1)\times SO(p-1})}
\end{equation}
to reach the desired gauge (\ref{fittone}). Indeed let us observe the decomposition:
\begin{equation}\label{fruttodimare}
    \mbox{adj}\, \left[\slal(2,\mathbb{R})\oplus \so(q,q+p)\right] \, = \, \mbox{adj}\, \left[\slal(2,\mathbb{R})\right] \oplus \mbox{adj}\, \left[\so(q,q+1)\right] \, \oplus \mbox{adj}\, \left[\so(p-1)\right] \, \oplus \, \left(\mathbf{1,2q+1 | p-1}\right)
\end{equation}
The $2q+1$ vectors of $\so(p-1)$ appearing in (\ref{fruttodimare}) are certainly sufficient to set to zero the $2$ vectors of $\so(p-1)$ appearing in $\mathbf{W}$.
\par
The conclusion therefore is that the classification of charge-orbits for all supergravity models can be performed by restriction to the Tits Satake sub-model. The same we show, in the next section,  to be true at the level of the classification based on $\mathrm{H}^\star$ orbits of the Lax operators, so that the final comparison of the two classifications can be performed by restriction to the Tits Satake subalgebras.
\section{Tits Satake reduction of  the $\mathbb{H}^\star$ subalgebra and of its representation $\mathbb{K}^\star$}
In the $\sigma$-model approach to black hole solutions one arrives at the new coset manifold (\ref{qcosetto}). The structure of the enlarged group $\mathrm{U}_{D=3}$ and of its Lie algebra $\mathbb{U}_{D=3}$ was discussed in eq.(\ref{gendecompo}). The subgroups $\mathbb{H}^\star$ are listed in table (\ref{hstarrigruppine}) for the non exotic models and in table (\ref{hstarrigruppi}) for the exotic ones. The coset generators fall into a representation of $\mathbb{H}^\star$ that we name $\mathbb{K}^\star$. The Lax operator $L_0$ which determines the spherically symmetric black hole solution up to boundary conditions of the scalar fields at infinity is just an element of  such a representation:
\begin{equation}\label{crullini}
    L_0 \, \in \, \mathbb{K}^\star
\end{equation}
so that the classification of spherical black holes is reduced to the classification of $\mathbb{H}^\star$ orbits in the $\mathbb{K}^\star$ representation. On the other hand, in part one of the paper we already saw how nilpotent orbits can be associated to multicenter solutions.
\par
In this paper we focus on non-exotic models that admit a regular Tits Satake projection and we postpone the analysis of the exotic ones to a future publication.
\par
A first general remark concerns the structure of  $\mathbb{H}^\star$ in all those models that correspond to $\mathcal{N}=2$ supersymmetry. In these cases the $\mathbb{H}^\star$ subalgebra is isomorphic to $\slal(2,\mathbb{R}) \oplus \mathbb{U}_{D=4}$ so that we have a decomposition of the $\mathbb{U}_{D=3}$ Lie algebra with respect to $\mathbb{H}^\star$ completely analogous to that in equation (\ref{gendecompo}), namely:
\begin{equation}
\mbox{adj}(\mathbb{U}_{D=3}) =
\underbrace{\mbox{adj}(\widehat{\mathbb{U}_{D=4}})\oplus\mbox{adj}(\slal(2,\mathbb{R})_{\mathrm{h}^\star})}_{\mathbb{H}^\star }\oplus
\underbrace{{(2_{\mathrm{h}^\star},\widehat{\mathbf{W}})}}_{\mathbb{K}^\star}
\label{KgendecompoN2}
\end{equation}
Hence the representation $\mathbb{K}^\star$ which contains the Lax operators has a structure analogous to the representation   which contains  the generators of $\mathbb{U}_{D=4}$ that originate from the vector fields, namely: $(2_{\mathrm{h}^\star},\widehat{\mathbf{W}}) $. This means that in all these models, by means of exactly the same argument as utilized above, we can always reach the gauge where the $\mathbb{K}^\star$ representation is localized on the image of the Tits Satake projection $\mathbb{K}^\star_{\mathrm{TS}}$. For instance, for the models in the $\mathfrak{f}_{4(4)}$ universality class we have:
\begin{equation}\label{frucchiotto}
    \mathbb{H}_{\mathrm{TS}}^\star \, = \, \slal(2,\mathbb{R})_{\mathrm{h}^\star} \, \oplus \, \widehat{\sym (6,\mathrm{R})}
\end{equation}
and:
\begin{eqnarray}\label{finkello}
    \mathbb{H}^\star & \stackrel{\mathbb{H}_{\mathrm{TS}}^\star \oplus \mathbb{G}_{\mathrm{subpaint}}}{\Longrightarrow} & \mbox{adj} \, \slal(2,\mathbb{R})_{\mathrm{h}^\star} \, \oplus \, \mbox{adj} \, \widehat{\sym (6,\mathrm{R})}\nonumber\\
    &&\, \oplus \, \left( \mathbf{6}\, | \, \mathcal{D}_{\mathrm{subpaint}}\right)\, \oplus \, \left( \mathbf{6}\, | \, \mathcal{D}_{\mathrm{subpaint}}\right)\nonumber\\
   && \oplus \, \left( \mathbf{1}\, | \, \mathcal{D}_{\mathrm{subpaint}}\right)\, \oplus \, \left( \mathbf{1}\, | \, \mathcal{D}_{\mathrm{subpaint}}\right) \nonumber\\
   \mathbb{K}^\star & \stackrel{\mathbb{H}_{\mathrm{TS}}^\star \oplus \mathbb{G}_{\mathrm{subpaint}}}{\Longrightarrow} &
    \, \left( 2_{\mathrm{h}^\star} \, , \, \mathbf{14}^\prime\, | \, \mathbf{1}_{\mathrm{subpaint}}\right) \oplus \, \left( 2_{\mathrm{h}^\star} \, , \, \mathbf{6}\, | \, \mathcal{D}_{\mathrm{subpaint}}\right)\,
\end{eqnarray}
and the two representations $\left( \mathbf{6}\, | \, \mathcal{D}_{\mathrm{subpaint}}\right)$ appearing in the adjoint representation of
$\mathbb{H}^\star $ can be utilized to get rid of $\left( 2_{\mathrm{h}^\star} \, , \, \mathbf{6}\, | \, \mathcal{D}_{\mathrm{subpaint}}\right)$ appearing in the decomposition of $\mathbb{K}^\star$.
\par
What is important to stress is that, although isomorphic $\mathbb{H}^\star$ and $\slal(2,\mathbb{R}) \oplus \mathbb{U}_{D=4} $ are different subalgebras of $\mathbb{U}_{D=3}$:
\begin{equation}\label{coincas}
 \mathbb{U}_{D=3} \, \supset\,   \slal(2,\mathbb{R})_{\mathrm{h}^\star} \, \ne \,  \slal(2,\mathbb{R})_{E} \, \subset \, \mathbb{U}_{D=3}  \quad ; \quad \mathbb{U}_{D=3} \, \supset\,   \widehat{\mathbb{U}_{D=4}} \, \ne \, {\mathbb{U}_{D=4}} \, \subset \, \mathbb{U}_{D=3}
\end{equation}
Morevover, while the decomposition (\ref{gendecompo}) is universal and holds true for all supergravity models, the structure (\ref{frucchiotto}) of the $\mathbb{H}^\star$ subalgebra is peculiar to the $\mathcal{N}=2$ models. In other cases the structure of $\mathbb{H}^\star$ is different.
\par
\begin{table}
\begin{center}
{
\scriptsize
\begin{tabular}{|c|c|c|c||c|c|c|}
  \hline
  \hline
  \hline
  $\#$ & $\mathbb{U}_{\mathrm{D=3}}$ &   $\mathbb{H}^\star$ & $\mathbb{K}^\star$ & $\mathbb{U}_{\mathrm{D=4}}$ & rep.$W$& $\mathbb{H}_c$\\
  \hline
  1 & $\mathfrak{e}_{8(8)}$ &   $\so^\star(16)$ & $\mathbf{128}_{\mathrm{s}}$ & $\mathfrak{e}_{7(7)}$ & $\mathbf{56}$& $\su(8)$\\
 \hline
  2 & $\mathfrak{g}_{2(2)}$ &   $\widehat{\slal\mathrm{(2,R)}}\oplus {\slal\mathrm{(2,R)_{h^\star}}} $ & $\left(\mathbf{4}_{3/2} \, , \, \mathbf{2}_{h^\star}\right)$ & $\slal\mathrm{(2,R)}$ & $\mathbf{4}_{3/2}$& $\so(2)$\\
  \hline
  3 & $\mathfrak{f}_{4(4)}$ &   $\widehat{\sym\mathrm{(6,R)}}\oplus {\slal\mathrm{(2,R)_{h^\star}}}$ & $\left(\widehat{\mathbf{14}}^\prime \, , \, \mathbf{2}_{h^\star}\right)$ & $\sym\mathrm{(6,R)}$ & $\mathbf{14}^\prime$& $\uu(3)$\\
  \hline
 4 & $\mathfrak{e}_{6(2)}$ &   $\widehat{\su\mathrm{(3,3)}}\oplus {\slal\mathrm{(2,R)_{h^\star}}}$ & $\left(\widehat{\mathbf{20}} \, , \, \mathbf{2}_{h^\star}\right)$ & $\su\mathrm{(3,3)}$ & $\mathbf{20}$& $\su(3)\oplus \su(3) $\\
 \null & \null & \null & \null & \null & \null & $\oplus \uu(1)$\\
  \hline
 5 & $\mathfrak{e}_{7(-5)}$ &   $\widehat{\so^\star\mathrm{(12)}}\oplus {\slal\mathrm{(2,R)_{h^\star}}}$ & $\left(\widehat{\mathbf{32}_{spin}} \, , \, \mathbf{2}_{h^\star}\right)$ & $\so^\star\mathrm{(12)}$ & $\mathbf{32}_{spin}$& $\uu(6)$\\
  \hline
6 & $\mathfrak{e}_{8(-24)}$ &   $\widehat{\mathfrak{e}_{7(-25)}}\oplus {\slal\mathrm{(2,R)_{h^\star}}}$ & $\left(\widehat{\mathbf{56}} \, , \, \mathbf{2}_{h^\star}\right)$ & $\mathfrak{e}_{7(-25)}$ & $\mathbf{56}$& $\uu(6)$\\
  \hline
7 & $\so\mathrm{(8,3)}$ &   $\so(6,2)\oplus \so(2,1)$ & $\left(\mathbf{8} \, , \, \mathbf{3}\right)$ & $\so(6,1) \oplus \slal\mathrm{(2,R)}$ & $\left(\mathbf{7},\mathbf{2}\right)$& $\so(6) \oplus\uu(1)$\\
  \hline
8 & $\so\mathrm{(8,4)}$ &   $\so(6,2)\oplus \so(2,2)$ & $\left(\mathbf{8} \, , \, \mathbf{4}\right)$ & $\so(6,2)\oplus \slal\mathrm{(2,R)}$ & $\left(\mathbf{8},\mathbf{2}\right)$ & $\so(6)\oplus \so(2)$ \\
  \null & \null & \null & \null & \null & \null & $\oplus \uu(1)$\\
  \hline
9 & $\so\mathrm{(8,5)}$ &   $\so(6,2)\oplus \so(2,3)$ & $\left(\mathbf{8} \, , \, \mathbf{5}\right)$ & $\so(6,3) \oplus \slal\mathrm{(2,R)}$ & $\left(\mathbf{9},\mathbf{2}\right)$& $\so(6)\oplus \so(3)$ \\
\null &\null &\null & \null & \null & \null & $\oplus \uu(1)$\\
  \hline
10 & $\so\mathrm{(8,6)}$ &   $\so(6,2)\oplus \so(2,4)$ & $\left(\mathbf{8} \, , \, \mathbf{6}\right)$ & $\so(6,4) \oplus \slal\mathrm{(2,R)}$ & $\left(\mathbf{10},\mathbf{2}\right)$& $\so(6)\oplus \so(4)$\\
\null &\null &\null & \null & \null & \null & $\oplus \uu(1)$\\
  \hline
11 & $\so\mathrm{(8,7)}$ &   $\so(6,2)\oplus \so(2,5)$ & $\left(\mathbf{8} \, , \, \mathbf{7}\right)$ & $\so(6,5) \oplus \slal\mathrm{(2,R)}$ & $\left(\mathbf{11},\mathbf{2}\right)$& $\so(6)\oplus \so(5)$\\
\null &\null &\null & \null & \null & \null & $\oplus \uu(1)$\\
  \hline
 12 & $\so\mathrm{(8,8)}$ &   $\so(6,2)\oplus \so(2,6)$ & $\left(\mathbf{8} \, , \, \mathbf{8}\right)$ & $\so(6,6) \oplus \slal\mathrm{(2,R)}$ & $\left(\mathbf{12},\mathbf{2}\right)$& $\so(6)\oplus \so(6)$\\
\null &\null &\null & \null & \null & \null & $\oplus \uu(1)$\\
  \hline
13 & $\so\mathrm{(8,8+p)}$ &   $\so(6,2)\oplus \so(2,6+p)$ & $\left(\mathbf{8} \, , \, \mathbf{8+p}\right)$ & $\so(6,6+p) \oplus \slal\mathrm{(2,R)}$ & $\left(\mathbf{12+p},\mathbf{2}\right)$& $\so(6)\oplus \so(6+p)$\\
\null &\null &\null & \null & \null & \null &  $\oplus \uu(1)$\\
  \hline
14 & $\so\mathrm{(4,3)}$ &   $\widehat{\slal\mathrm{(2,R)}}\oplus \widehat{\so(2,1)}$ & $\left(\widehat{\mathbf{2}}\, ,\, \widehat{\mathbf{3}} \, , \, \mathbf{2}_{h^\star}\right)$ & $\slal\mathrm{(2,R)}\oplus \so(2,1)$ & $\left(\mathbf{2}\, ,\, \mathbf{3}\right)$& $\so(2)\oplus \uu(1)$\\
\null & \null & $\quad\quad\quad\quad\oplus {\slal\mathrm{(2,R)_{h^\star}}}$ & \null & \null & \null & \null \\
  \hline
15 & $\so\mathrm{(4,4)}$ &   $\widehat{\slal\mathrm{(2,R)}}\oplus \widehat{\so(2,2)}$ & $\left(\widehat{\mathbf{2}}\, ,\, \widehat{\mathbf{4}} \, , \, \mathbf{2}_{h^\star}\right)$ & $\slal\mathrm{(2,R)}\oplus \so(2,2)$ & $\left(\mathbf{2}\, ,\, \mathbf{4}\right)$& $\so(2)\oplus\so(2) $\\
\null & \null & $\quad\quad\quad\quad\oplus {\slal\mathrm{(2,R)_{h^\star}}}$ & \null & \null & \null &$\oplus \uu(1)$ \\
  \hline
  16 & $\so\mathrm{(4,4+p)}$ &   $\widehat{\slal\mathrm{(2,R)}}\oplus \widehat{\so(2,2+p)}$ & $\left(\widehat{\mathbf{2}}\, ,\, \widehat{\mathbf{4+p}} \, , \, \mathbf{2}_{h^\star}\right)$ & $\slal\mathrm{(2,R)}\oplus \so(2,2)$ & $\left(\mathbf{2}\, ,\, \mathbf{4+p}\right)$& $\so(2)\oplus\so(2+p) $\\
  \null & \null & $\quad\quad\quad\quad\oplus {\slal\mathrm{(2,R)_{h^\star}}}$ & \null & \null & \null & $\oplus \uu(1)$ \\
  \hline
  \hline
\end{tabular}
}
\caption{Table of $\mathbb{H}^\star$ subalgebras of $\mathbb{U}_{D=3}$, $\mathbb{K}^\star$-representations and $\mathbf{W}$ representations of $\mathbb{U}_{D=4}$ for the supergravity models based on \textit{non-exotic} scalar symmetric spaces\label{hstarrigruppine}}
\end{center}
\end{table}
\begin{table}
\begin{center}
{
\small
\begin{tabular}{|c|c|c|c||c|c|c|}
  \hline
  \hline
  \hline
  $\#$ & $\mathbb{U}_{\mathrm{D=3}}$ &   $\mathbb{H}^\star$ & $\mathbb{K}^\star$ & $\mathbb{U}_{\mathrm{D=4}}$ & symp. rep. $W$& $\mathbb{H}_c$\\
  \hline
  $1_e$ & $\su\mathrm{(p+2,2)}$ &   $\widehat{\su\mathrm{(p+1,1)}}\oplus \widehat{\uu(1)} $ & $\left(\mathbf{p+2},\mathbf{2}_{h^\star}\right)$ & $\su\mathrm{(p+1,1)}\oplus\uu(1)$ & $\mathbf{p+2}$& $\su\mathrm{(p+1)}$\\
  \null & \null & $\quad\quad\quad\quad\oplus {\slal\mathrm{(2,R)_{h^\star}}}$ & \null & \null & \null & $\oplus \uu(1)$ \\
 \hline
  $2_e$ & $\su\mathrm{(p+2,4)}$ &   ${\su\mathrm{(p+1,2)}}\oplus {\uu(1)} $ & $\left(\mathbf{p+3},\mathbf{3}\right)$ & $\su\mathrm{(p+1,1)}\oplus\uu(1)$ & $\mathbf{p+4}$& $\su\mathrm{(p+1)}\oplus \su(3)$\\
  \null & \null & $\quad\quad\quad\quad\oplus {\su\mathrm{(1,2)}}$ & \null & \null & \null & $\oplus \uu(1)$ \\
 \hline
  $3_e$ & $\mathfrak{e}_{6(-14)}$ &   $\so^\star(10) \oplus \so(2)$ & $\left(\mathbf{16}_s,\mathbf{2}\right) $ & $\su(5,1)$ & $\mathbf{10}$& $\uu(5)$\\
  \hline
  \hline
\end{tabular}
}
\caption{ Table of $\mathbb{H}^\star$ subalgebras of $\mathbb{U}_{D=3}$, $\mathbb{K}^\star$-representations and $\mathbf{W}$ representations of $\mathbb{U}_{D=4}$ for the supergravity models based on \textit{exotic} scalar symmetric spaces\label{hstarrigruppi}}
\end{center}
\end{table}
The reduction to the Tits Satake projection however is universal and applies to all non maximally split cases.
\par
Indeed the remaining cases are of the form:
\begin{equation}\label{tartinarossa}
    \frac{\mathrm{U}_{D=3}}{\mathrm{H}^\star} \, = \, \frac{\mathrm{SO(2+q,q+2+p)}}{\mathrm{SO(q,2) \times SO(2,q+p)}}
\end{equation}
leading to
\begin{equation}\label{cavialino}
    \mathbb{K}^\star \, = \, \left( \mathbf{q+2,q+p+2}\right) \, \stackrel{\so(q,2) \oplus \so(2,q+1)\oplus \so(p-1)}{\Longrightarrow}\,
     \left( \mathbf{q+2,q+1,1}\right) \, \oplus \, \left( \mathbf{q+2,1,p-1}\right)
\end{equation}
where:
\begin{eqnarray}
  \so(q,2) \oplus \so(2,q+1) &=& \mathbb{H}^\star_{\mathrm{TS}} \\
  \so(p-1) &=& \mathbb{G}_{\mathrm{subpaint}}
\end{eqnarray}
Considering the coset:
\begin{equation}\label{critimanu}
    \frac{\mathrm{H}^\star}{\mathrm{H}^\star_{\mathrm{TS}}\times \mathrm{G}_{\mathrm{subpaint}}} \, = \, \frac{\mathrm{SO(2,q+p)}}{\mathrm{SO(q+1,2) \times SO(p-1)}}
\end{equation}
we see that its $(q+3)\times(p-1)$ parameters are arranged into the
\begin{equation}\label{diciaccia}
    \left (\mathbf{q+3 | p-1}\right)
\end{equation}
representation of $\so(q+1,2)\oplus \so(p-1)$ and can be used to put to zero the component $\left( \mathbf{q+2,1,p-1}\right)$ in the decomposition (\ref{cavialino}).
Note that the $\mathcal{N}=4$ cases with more than $6$ vector multiplets are covered by the above formulae by setting:
\begin{equation}\label{pq}
    q\, = \, 6 \quad ; \quad p \, > \, 1
\end{equation}
Similarly the $\mathcal{N}=2$ cases with more than $3$ vector multiplets are covered by the above formulae by setting:
\begin{equation}\label{pqbis}
    q\, = \, 2 \quad ; \quad p \, > \, 1
\end{equation}
Finally the $\mathcal{N}=4$ cases with less  than $6$ vector multiplets are covered by the above formulae by setting:
\begin{equation}\label{pqbis}
    q\, = \, n \quad ; \quad p \, = \, 6 \, - \, n \quad ; \quad n=1,2,3,4,5
\end{equation}
\section{The general structure of the $\mathbb{H}^\star\oplus \mathbb{K}^\star$ decomposition in the maximally split models}
\label{hstarrostrucco}
In the previous section we have shown that all $\mathrm{H}^\star$ orbits in the $\mathbb{K}^\star$ representation cross the locus
defined by:
\begin{equation}\label{proiezia}
    \Pi_{\mathrm{TS}} \left( \mathbb{K}^\star \right) \, = \, \mathbb{K}^\star
\end{equation}
where $\Pi_{\mathrm{TS}}$ is the Tits-Satake projection. In other words just as for the $\mathbf{W}$-representation of $\mathrm{U}_{D=4}$, it suffices to classify the orbits $\mathrm{H}^\star_{\mathrm{TS}}$ in the $\mathbb{K}^\star_{\mathrm{TS}}$ representation. In view of this result, in the present section we study the general structure of the $\mathbb{H}^\star \oplus \mathbb{K}^\star$ decomposition for maximally split algebras $\mathbb{U}_{D=3}$.
\par
A key point in our following discussion is provided by the structure of the root system of $\mathbb{U}_{D=3}$ as described in section
\ref{dynkino}.
The entire set of positive roots can be written as follows:
\begin{equation}
0 \, < \, \mathfrak{a} \, = \, \left \{
\begin{array}{rcl}
 \alpha &=& \left\{\overline{\alpha},0\right\} \\
  \mathfrak{w} &=&\left\{{\overline{\mathbf{w}}},\frac{1}{\sqrt{2}}\right\} \\
 \psi &=& \left\{0,{\sqrt{2}}\right\}\\
 \end{array}\right. \label{struttururutu}
\end{equation}
where $\overline{\alpha} >0$ denotes the set of all positive roots of $\mathbb{U}_{D=4}$, while $\overline{\mathbf{w}}$ denotes the complete set of weights (positive, negative and null) of the $\mathbf{W}$ representation of $\mathbb{U}_{D=4}$. The root $\psi$ is the highest root of the $\mathbb{U}_{D=3}$ root system and is also the root of the Ehlers subalgebra $\slal(2,\mathbb{R})_E$. Accordingly, a basis of the Cartan subalgebra of $\mathbb{U}_{D=3}$ is constructed as follows:
\begin{equation}\label{ciulifischio}
    \underbrace{\mathrm{CSA}}_{\mbox{of }\mathbb{U}_{D=3}} \, = \, \mbox {span of} \, \left \{\underbrace{\mathcal{H}_1\, ,\, \mathcal{H}_2\, , \, \dots \, , \, \mathcal{H}_r\,}_{\mbox{CSA generators of $\mathbb{U}_{D=4}$}} , \, \underbrace{\mathcal{H}_\psi }_{\mbox{CSA generator of $\slal(2,\mathbb{R})_E$}}\, \right\}
\end{equation}
\par
For all maximally split Lie algebras $\mathbb{U}$ of rank $r+1$, the maximal compact subalgebra $\mathbb{H}\subset \mathbb{U} $ is generated by:
\begin{equation}\label{calatrala}
    T^{\mathfrak{a}} \, = \, E^{\mathfrak{a}} \, - \, E^{-\mathfrak{a}}
\end{equation}
while the complementary orthogonal space $\mathbb{K}$ is generated by
\begin{eqnarray}
  K^{\mathfrak{a}} &=& \, E^{\mathfrak{a}} \, + \, E^{-\mathfrak{a}} \\
  K^{I}  &=& \mathcal{H}^I \quad ; \quad I=1,\dots, r+1
\end{eqnarray}
The splitting $\mathbb{H}^\star \oplus \mathbb{K}^\star$ is obtained by means of just one change of sign which, thanks to the structure
(\ref{struttururutu}) of the root system is consistent, namely still singles out a subalgebra.
\par
The generators of the $\mathbb{H}^\star$ subalgebra are as follows:
\begin{eqnarray}
  T^\alpha_\star &=& E^\alpha \, - \, E^{-\alpha}\nonumber \\
   T^{\mathfrak{w}}_\star &=& E^{\mathfrak{w}} \, + \, E^{-\mathfrak{w}} \nonumber \\
  T^{\psi}_\star &=& E^{\psi} \, - \, E^{-\psi} \label{hstargenera}
\end{eqnarray}
while the generators of the $\mathbb{K}^\star$ complementary subspace are as follows:
\begin{eqnarray}
  K^\alpha_\star &=& E^\alpha \, + \, E^{-\alpha}\nonumber \\
   K^{\mathfrak{w}}_\star &=& E^{\mathfrak{w}} \, - \, E^{-\mathfrak{w}}\nonumber \\
  K^{\psi}_\star &=& E^{\psi} \, + \, E^{-\psi} \nonumber \\
   K^{I}  &=& \mathcal{H}^I \quad ; \quad I=1,\dots, r+1 \label{kapstargenera}
\end{eqnarray}
From eq.(\ref{hstargenera}) we see that $\mathbb{H}^\star$ contains the maximal compact subalgebra of the original $\mathbb{U}_{D=4}$ and the maximal compact subalgebra $\so(2) \, \subset \, \slal(2,R)_E$ of the Ehlers group. Using this structure we can now compare the classification of $\mathbb{K}^\star$ orbits with the classification of $\mathbf{W}$-orbits.
\section{$\mathbb{K}^\star$ orbits versus $\mathbf{W}$-orbits}
\label{KWseczia} In the $\sigma$-model approach the complete black
hole spherically symmetric supergravity solution is obtained from
two data\footnote{See papers
\cite{Fre:2009dg,Chemissany:2010zp,noig22,titsblackholes} for
detailed explanations.}, namely the Lax operator $L_0$ evaluated
at spatial infinity (see eq.(\ref{crullini})) and the coset
representative $\mathbb{L}_0$ also evaluated at spatial infinity.
In terms of these data one defines  the matrix of conserved
Noether charges:
\begin{equation}\label{patroffio}
    Q^{Noether} \, = \, \mathbb{L}_0 \, L_0 \, \mathbb{L}_0 ^{-1} \, = \, \mathbb{L}(\tau) \, L(\tau) \, \mathbb{L} ^{-1} (\tau)
\end{equation}
from which the electromagnetic charges of the black hole, belonging to the $\mathbf{W}$-representation of $\mathrm{U}_{D=4}$, can be obtained by means of the following trace:
\begin{equation}\label{forcina}
    \mathcal{Q}^\mathbf{w} \, = \, \mbox{Tr} \, \left( Q^{Noether} \, \mathcal{T}^{\mathbf{w}} \right)
\end{equation}
where
\begin{equation}\label{crattolo}
    \mathcal{T}^{\mathbf{w}} \, \propto \, E^{\mathfrak{w}}
\end{equation}
are the generators of the solvable Lie algebra corresponding to the $\mathbf{W}$-representation.
\par
It is important to stress that, because of physical boundary conditions, the coset representative at spatial infinity $\mathbb{L}_0$ belongs to the subgroup $\mathrm{U}_{D=4} \subset \mathrm{U}_{D=3}$. Indeed it simply encodes the boundary values at infinity of the $D=4$ scalar fields:
\begin{equation}\label{consuelo}
\mathrm{U}_{D=3}\, \supset \,  \mathrm{U}_{D=4} \, \ni \,   \mathbb{L}_0 \, = \, \exp \left[ \phi_0^\alpha \, E^{\alpha} \, + \, \sum_{i=1}^r \phi_0^i \, \mathcal{H}_i \right ]
\end{equation}
Using this information in eq.(\ref{forcina}) we obtain
\begin{equation}
\mathcal{Q}^\mathbf{w} \, = \, \mbox{Tr} \, \left( L_0 \, \mathbb{L}_0^{-1}(\phi) \mathcal{T}^{\mathbf{w}}\mathbb{L}_0(\phi) \right) \, = \, R(\phi)^{\mathbf{w}}_{\phantom{\mathbf{w}}\mathbf{w}^\prime} \, \mathcal{Q}^{\mathbf{w}^\prime}
\end{equation}
where:
\begin{equation}\label{crittuco}
    \mathcal{Q}^{\mathbf{w}^\prime} \, = \, \mbox{Tr} \, \left( L_0 \,  \mathcal{T}^{\mathbf{w}^\prime} \right)
\end{equation}
are the electromagnetic charges obtained with no scalar field dressing at infinity and
\begin{equation}\label{cipolla}
    R(\phi)^{\mathbf{w}}_{\phantom{\mathbf{w}}\mathbf{w}^\prime} \, \in \, \mathrm{U}_{D=4}
\end{equation}
is the matrix representing the group element $ \mathbb{L}_0(\phi)$ in the $\mathbf{W}$-representation.
\par
This result has a very significant consequence. The scalar field dressing at infinity simply rotates the charge vector along the same $\mathbf{W}$-orbit and is therefore irrelevant.
\par
Hence we conclude that for each Lax operator, the $\mathbf{W}$-orbit of charges is completely determined and unique. The next question that was already tackled in the introduction is whether the charge-orbit $\mathbf{W}$ is the same for all Lax operators belonging to the same $\mathrm{H}^\star$-orbit. As already anticipated, the answer is no and it is quite easy to produce counter examples.
\par
Yet if we impose the condition that the Taub-NUT charge should be zero:
\begin{equation}\label{vanishiuTN}
    \mbox{Tr} \left ( L_0 \, L_-^E\right) \, = \, 0
\end{equation}
then for all Lax operators in the same $\mathrm{H}^\star$, satisfying the additional constraint (\ref{vanishiuTN}), the corresponding charges $Q^w \, = \, \mbox{Tr}\left (L_0 \, T^w\right)$ fall into the same $\mathbf{W}$-orbit.
\par
We were not able to prove this statement, but we assert it as a \textit{conjecture}, since we analyzed many cases and it was always true, no counter example being ever found.
\par
In the case of multicenter non spherically symmetric solutions our conjecture appears to be true as long as we impose the condition of vanishing of the Taub-NUT current:
\begin{equation}\label{carappolindi}
    j^{TN} \, = \, 0
\end{equation}
So doing, at every pole of the involved harmonic functions, we obtain a black hole that  always falls into the same $\mathbf{W}$-orbit.
\par
What happens instead when the Taub-NUT current is turned on cannot be predicted in general terms at the present status of our knowledge and more study is certainly in order.
\section{Conclusions and Perspectives}
\label{concludone}
In this paper we examined the systematics of  extremal stationary solutions of supergravity models whose scalar manifold is a symmetric coset space.
\par
We provided a comprehensive group theoretical analysis of all such theories, organizing them in universality classes according to their Tits Satake projection. This was instrumental to our double classification of stationary solutions according to orbits of the non-compact isotropy subgroup $\mathrm{H}^\star$ in the three-dimensional approach, and to orbits  of the symplectic $\mathbf{W}$-representation of the duality group $\mathrm{U_{D=4}}$ in the four-dimensional approach. In both cases we provided full evidence that the solutions can be gauge rotated to the Tits Satake subalgebra so that classifying nilpotent orbits for the finite list of maximally split coset manifolds $\mathrm{U_{TS}/H_{TS}^\star}$ suffices to classify all stationary supergravity black holes. In force of this, restricting our attention to the maximally split cases, we provided the general form of the $\mathbb{H}^\star$ subalgebra which can be considered one of the new results attained by our paper (sect.\ref{hstarrostrucco}).
\par
Within this general group-theoretical setup we analyzed  the
construction of multicenter solutions following the strategy laid
down in
\cite{Bossard:2009my,Bossard:2009at,Bossard:2009bw,Bossard:2009mz,Bossard:2011kz,Bossard:2012ge}
which utilizes the nilpotent subalgebra singled out by every
nilpotent orbit. We provided the missing link necessary to
transform such a strategy into a complete constructive algorithm:
such link is the general procedure described in subsection
\ref{zurlina} for the transformation of the coset representative
from the symmetric to the solvable gauge where the supergravity
fields can be read off. We consider this another relevant result
of our paper.
\par
Next we performed a complete survey of the stationary solutions associated with the nilpotent  orbits of the $S^3$-model, considering the structure of charge-orbits for the corresponding black holes. In this case the $D=4$ duality algebra is
$\mathbb{U}_{\mathrm{D=4}} \, = \, \slal(2,\mathbb{R})$ and the $\mathbf{W}$-representation encoding the charges $\mathcal{Q}$, is the $j=\ft 32$ of $\slal(2,\mathbb{R})$. We found  that there are the following orbits\footnote{As we already mentioned the authors express their gratitude to Alessio Marrani who attracted their attention, after the first appearance of the present paper in the arXive, to paper \cite{Borsten:2011ai} where the existence of a discrete stability subgroup for the charges of the regular BPS orbit in the $S^3$ model had already been found. }:
\begin{itemize}
  \item $\mathfrak{J}_4(\mathcal{Q}) \, = \, 0$ with stability subgroup $\Gamma\, \mathcal{Q} \, = \, \mathcal{Q}$ given by $\Gamma \, = \, \left(\begin{array}{cc}
                                              1 & 0 \\
                                              c & 1
                                            \end{array}\right)$, \textit{i.e.} by a parabolic subgroup of $\mathrm{SL(2,\mathbb{R})}$. This orbit corresponds to the \textit{very small black holes}.
 \item $\mathfrak{J}_4(\mathcal{Q}) \, = \, 0$ with stability subgroup $\Gamma\, \mathcal{Q} \, = \, \mathcal{Q}$ given by $\Gamma \, = \, \left(\begin{array}{cc}
                                              1 & 0 \\
                                              0 & 1
                                            \end{array}\right)$ \textit{i.e.} by the identity of $\mathrm{SL(2,\mathbb{R})}$. This orbit corresponds to the \textit{small black holes}.
 \item $\mathfrak{J}_4(\mathcal{Q}) \, > \, 0$ with stability subgroup $\Gamma\, \mathcal{Q} \, = \, \mathcal{Q}$ given by $\Gamma \, = \, \mathbb{Z}_3 \, \subset \, \mathrm{SL(2,\mathbb{R})}$ . This orbit corresponds to the regular \textit{BPS black holes}.
 \item $\mathfrak{J}_4(\mathcal{Q}) \, < \, 0$ with stability subgroup $\Gamma\, \mathcal{Q} \, = \, \mathcal{Q}$ given by $\Gamma \, = \, \left(\begin{array}{cc}
                                              1 & 0 \\
                                              0 & 1
                                            \end{array}\right)$ \textit{i.e.} by the identity of $\mathrm{SL(2,\mathbb{R})}$. This orbit corresponds to the regular \textit{non BPS black holes}..
 \item $\mathfrak{J}_4(\mathcal{Q}) \, > \, 0$ with stability subgroup $\Gamma\, \mathcal{Q} \, = \, \mathcal{Q}$, given by $\Gamma \, = \, \left(\begin{array}{cc}
                                              1 & 0 \\
                                              0 & 1
                                            \end{array}\right)$ \textit{i.e.} by the identity of $\mathrm{SL(2,\mathbb{R})}$ This orbit corresponds to the generically singular  \textit{ very large black holes}..
\end{itemize}
Up to our knowledge, this detailed structure of the $S^3$ model  was not discussed  in the literature so far and can be considered still another of our main results. The above pattern is quite inspiring and leads to the conjecture that also in other models the charges of BPS black holes might be characterized by their invariance under some suitable finite subgroup of the duality group.
\par
In our analysis of the multicenter solutions we came to the conclusion that, within each $\mathrm{H}^\star$ orbit, when the vanishing of the Taub-NUT current is imposed, all the black holes that are located at the various poles of the involved harmonic functions, emerge with charges always assigned to the same  $\mathbf{W}$-orbit. Hence at vanishing Taub-NUT current the main question that motivated our paper has been answered. At least in the $S^3$ model there is a rigid association between the $\mathrm{H}^\star$ orbit utilized to construct the solution and the $\mathbf{W}$-orbit of their charges.
\par
Confirming this rigid association for other Tits Satake universality classes is one of the issue that emerge from our paper and that we live for future publications. Other issues raised by our results that we plan to further investigate are:
\begin{enumerate}
  \item The appropriate interpretation of the solutions associated with higher nilpotency orbits.
  \item Fitting of extremal and non extremal rotating black holes into the scheme.
\end{enumerate}
\section{Acknowledgements}
\par
The authors would like to express their gratitude  to  Sergio Ferrara and Mario Trigiante for enlightening discussions during the development of this work. Furthermore it is also a pleasure to aknowledge some remarks from and useful communications with Alessio Marrani.
\par
A.S. Sorin acknowledges partial financial support from the RFBR Grants No.
11-02-01335-a and  11-02-12232-ofi-m-2011.
\newpage
\appendix
\begin{landscape}
\section{The complete form of the fields for the $\mathcal{O}_{3|11}$ orbit}
In this appendix we present the complete result for the scalar fields of the three-dimensional $\sigma$-model parameterized by four harmonic functions $\mathcal{H}_{1,2,3,4}$ in the case of the Large BPS orbit $\mathcal{O}_{11}^3$
\begin{eqnarray}
  \mathfrak{W} &=& \frac{2 \sqrt{3}}{\sqrt{4 \mathcal{H}_1^4+8 \mathcal{H}_2 \mathcal{H}_1^3+3 \left(\mathcal{H}_3^2-6 \mathcal{H}_4 \mathcal{H}_3-3 \mathcal{H}_4^2\right) \mathcal{H}_1^2+2
   \mathcal{H}_2 \left(-4 \mathcal{H}_2^2+3 \mathcal{H}_3^2+9 \mathcal{H}_4^2\right) \mathcal{H}_1-4 \mathcal{H}_2^4-12 \mathcal{H}_3^3 \mathcal{H}_4+3 \mathcal{H}_2^2
   \left(\mathcal{H}_3^2+6 \mathcal{H}_4 \mathcal{H}_3-3 \mathcal{H}_4^2\right)}}\nonumber\\
   \null & \null & \null \nonumber\\
  \mbox{Im} z &=&\frac{\sqrt{4 \mathcal{H}_1^4+8 \mathcal{H}_2 \mathcal{H}_1^3+3 \left(\mathcal{H}_3^2-6 \mathcal{H}_4 \mathcal{H}_3-3 \mathcal{H}_4^2\right) \mathcal{H}_1^2+2 \mathcal{H}_2
   \left(-4 \mathcal{H}_2^2+3 \mathcal{H}_3^2+9 \mathcal{H}_4^2\right) \mathcal{H}_1-4 \mathcal{H}_2^4-12 \mathcal{H}_3^3 \mathcal{H}_4+3 \mathcal{H}_2^2
   \left(\mathcal{H}_3^2+6 \mathcal{H}_4 \mathcal{H}_3-3 \mathcal{H}_4^2\right)}}{2 \sqrt{3} \left(\mathcal{H}_1^2-\mathcal{H}_2^2+\mathcal{H}_3^2\right)}\nonumber \\
   \null & \null & \null \nonumber\\
  \mbox{Re} z &=& \frac{\mathcal{H}_2 \left(\mathcal{H}_3-3 \mathcal{H}_4\right)+\mathcal{H}_1 \left(\mathcal{H}_3+3 \mathcal{H}_4\right)}{2 \sqrt{3}
   \left(\mathcal{H}_1^2-\mathcal{H}_2^2+\mathcal{H}_3^2\right)}\label{fullyfilli1}
\end{eqnarray}
\begin{eqnarray}
  Z^1&=& -\frac{3 \sqrt{2} \left(2 \mathcal{H}_1^3+2 \mathcal{H}_2 \mathcal{H}_1^2+\left(\mathcal{H}_3 \left(\mathcal{H}_3-3 \mathcal{H}_4\right)-2 \mathcal{H}_2^2\right)
   \mathcal{H}_1+\mathcal{H}_2 \left(\mathcal{H}_3 \left(\mathcal{H}_3+3 \mathcal{H}_4\right)-2 \mathcal{H}_2^2\right)\right)}{4 \mathcal{H}_1^4+8 \mathcal{H}_2
   \mathcal{H}_1^3+3 \left(\mathcal{H}_3^2-6 \mathcal{H}_4 \mathcal{H}_3-3 \mathcal{H}_4^2\right) \mathcal{H}_1^2+2 \mathcal{H}_2 \left(-4 \mathcal{H}_2^2+3
   \mathcal{H}_3^2+9 \mathcal{H}_4^2\right) \mathcal{H}_1-4 \mathcal{H}_2^4-12 \mathcal{H}_3^3 \mathcal{H}_4+3 \mathcal{H}_2^2 \left(\mathcal{H}_3^2+6 \mathcal{H}_4
   \mathcal{H}_3-3 \mathcal{H}_4^2\right)}\nonumber \\
 \null & \null & \null \nonumber\\
 Z^2 &=& \left( 4 \mathcal{H}_1^4+8 \mathcal{H}_2 \mathcal{H}_1^3+3 \left(\mathcal{H}_3^2-6 \left(\mathcal{H}_4+1\right) \mathcal{H}_3-3 \mathcal{H}_4
   \left(\mathcal{H}_4+2\right)\right) \mathcal{H}_1^2+2 \mathcal{H}_2 \left(3 \left(\mathcal{H}_3^2+3 \mathcal{H}_4 \left(\mathcal{H}_4+2\right)\right)-4
   \mathcal{H}_2^2\right) \mathcal{H}_1-4 \mathcal{H}_2^4 \right.\nonumber\\
  && \left. - 12 \mathcal{H}_3^3 \left(\mathcal{H}_4+1\right)+3 \mathcal{H}_2^2 \left(\mathcal{H}_3^2+6
   \left(\mathcal{H}_4+1\right) \mathcal{H}_3-3 \mathcal{H}_4 \left(\mathcal{H}_4+2\right)\right) \right)
   \times \, \left(
   \sqrt{2} \left(4 \mathcal{H}_1^4+8 \mathcal{H}_2 \mathcal{H}_1^3+3
   \left(\mathcal{H}_3^2-6 \mathcal{H}_4 \mathcal{H}_3
   -3 \mathcal{H}_4^2\right) \mathcal{H}_1^2 \right.\right. \nonumber\\
   &&\left.\left.+2 \mathcal{H}_2 \left(-4 \mathcal{H}_2^2+3 \mathcal{H}_3^2+9
   \mathcal{H}_4^2\right) \mathcal{H}_1-4 \mathcal{H}_2^4-12 \mathcal{H}_3^3 \mathcal{H}_4+3 \mathcal{H}_2^2 \left(\mathcal{H}_3^2+6 \mathcal{H}_4 \mathcal{H}_3-3
   \mathcal{H}_4^2\right)\right)\right)^{-1}\nonumber \\
\null & \null & \null \\
 Z^3 &=&-\left( \sqrt{\frac{3}{2}} \left(4 \mathcal{H}_1^4+8 \mathcal{H}_2 \mathcal{H}_1^3+\left(3 \mathcal{H}_3^2-2 \left(9 \mathcal{H}_4+1\right) \mathcal{H}_3+3 \left(2-3
   \mathcal{H}_4\right) \mathcal{H}_4\right) \mathcal{H}_1^2 \right.\right.\nonumber\\
   &&\left.\left.+2 \mathcal{H}_2 \left(-4 \mathcal{H}_2^2+3 \mathcal{H}_3^2+9 \mathcal{H}_4^2-2 \mathcal{H}_3\right)
   \mathcal{H}_1-4 \mathcal{H}_2^4-12 \left(\mathcal{H}_3-1\right) \mathcal{H}_3^2 \mathcal{H}_4+\mathcal{H}_2^2 \left(3 \mathcal{H}_3^2+2 \left(9 \mathcal{H}_4-1\right)
   \mathcal{H}_3-3 \mathcal{H}_4 \left(3 \mathcal{H}_4+2\right)\right)\right)\right) \times \nonumber\\
   &&\times \, \left( 4 \mathcal{H}_1^4+8 \mathcal{H}_2 \mathcal{H}_1^3+3 \left(\mathcal{H}_3^2-6 \mathcal{H}_4
   \mathcal{H}_3-3 \mathcal{H}_4^2\right) \mathcal{H}_1^2\right.\nonumber\\
   &&\left.+2 \mathcal{H}_2 \left(-4 \mathcal{H}_2^2+3 \mathcal{H}_3^2+9 \mathcal{H}_4^2\right) \mathcal{H}_1-4
   \mathcal{H}_2^4-12 \mathcal{H}_3^3 \mathcal{H}_4+3 \mathcal{H}_2^2 \left(\mathcal{H}_3^2+6 \mathcal{H}_4 \mathcal{H}_3-3 \mathcal{H}_4^2\right)\right)^{-1}\\
 \null & \null & \null \\
  Z^4 &=& -\frac{\sqrt{\frac{2}{3}} \left(2 \mathcal{H}_1^3+6 \mathcal{H}_2 \mathcal{H}_1^2+\left(6 \mathcal{H}_2^2-9 \mathcal{H}_4 \left(\mathcal{H}_3+\mathcal{H}_4\right)\right)
   \mathcal{H}_1+\mathcal{H}_2 \left(2 \mathcal{H}_2^2+9 \mathcal{H}_4 \left(\mathcal{H}_4-\mathcal{H}_3\right)\right)\right)}{4 \mathcal{H}_1^4+8 \mathcal{H}_2
   \mathcal{H}_1^3+3 \left(\mathcal{H}_3^2-6 \mathcal{H}_4 \mathcal{H}_3-3 \mathcal{H}_4^2\right) \mathcal{H}_1^2+2 \mathcal{H}_2 \left(-4 \mathcal{H}_2^2+3
   \mathcal{H}_3^2+9 \mathcal{H}_4^2\right) \mathcal{H}_1-4 \mathcal{H}_2^4-12 \mathcal{H}_3^3 \mathcal{H}_4+3 \mathcal{H}_2^2 \left(\mathcal{H}_3^2+6 \mathcal{H}_4
   \mathcal{H}_3-3 \mathcal{H}_4^2\right)}\nonumber\\
   \label{fullyfilli2}
\end{eqnarray}
\begin{eqnarray}
  a &=& \frac{\mathcal{N}}{\mathcal{D}}\nonumber\\
  \mathcal{N}&=& 240 \mathcal{H}_1^{11}+768 \mathcal{H}_2 \mathcal{H}_1^{10}-24 \left(2 \mathcal{H}_2^2-32 \mathcal{H}_3^2+27 \mathcal{H}_4^2+63 \mathcal{H}_3 \mathcal{H}_4\right)
   \mathcal{H}_1^9+4 \mathcal{H}_2 \left(8 \mathcal{H}_2^2 \left(3 \mathcal{H}_4-70\right) \right.\nonumber\\
   &&\left.-3 \left(\left(6 \mathcal{H}_4-187\right) \mathcal{H}_3^2+162 \mathcal{H}_4
   \mathcal{H}_3+3 \mathcal{H}_4^2 \left(6 \mathcal{H}_4+7\right)\right)\right) \mathcal{H}_1^8+3 \left(32 \left(\mathcal{H}_4-15\right) \mathcal{H}_2^4-8 \left(\left(3
   \mathcal{H}_4-10\right) \mathcal{H}_3^2 \right.\right.\nonumber\\
   &&\left.\left.-216 \mathcal{H}_4 \mathcal{H}_3+9 \left(\mathcal{H}_4-8\right) \mathcal{H}_4^2\right) \mathcal{H}_2^2+299 \mathcal{H}_3^4+81
   \mathcal{H}_4^4+486 \mathcal{H}_3 \mathcal{H}_4^3+108 \mathcal{H}_3^2 \mathcal{H}_4^2-1518 \mathcal{H}_3^3 \mathcal{H}_4\right) \mathcal{H}_1^7+\mathcal{H}_2 \left(-32
   \left(9 \mathcal{H}_4-68\right) \mathcal{H}_2^4\right.\nonumber\\
   &&\left.+144 \left(\left(2 \mathcal{H}_4-35\right) \mathcal{H}_3^2+\left(40-3 \mathcal{H}_4\right) \mathcal{H}_4
   \mathcal{H}_3+\mathcal{H}_4^2 \left(3 \mathcal{H}_4-5\right)\right) \mathcal{H}_2^2-162 \mathcal{H}_3^2 \mathcal{H}_4^2+\mathcal{H}_3^4 \left(2667-54
   \mathcal{H}_4\right)+324 \mathcal{H}_3^3 \left(\mathcal{H}_4-15\right) \mathcal{H}_4\right.\nonumber\\
   &&\left.+324 \mathcal{H}_3 \mathcal{H}_4^3 \left(3 \mathcal{H}_4-1\right)+81 \mathcal{H}_4^4
   \left(6 \mathcal{H}_4+7\right)\right) \mathcal{H}_1^6-\left(96 \left(2 \mathcal{H}_4-27\right) \mathcal{H}_2^6-16 \left(2 \mathcal{H}_3^3+9 \left(\mathcal{H}_4-31\right)
   \mathcal{H}_3^2+9 \mathcal{H}_4 \left(3 \mathcal{H}_4-41\right) \mathcal{H}_3\right.\right.\nonumber\\
   &&\left.\left.+27 \mathcal{H}_4^2 \left(2 \mathcal{H}_4-3\right)\right) \mathcal{H}_2^4+3 \left(8
   \mathcal{H}_3^5-435 \mathcal{H}_3^4+66 \mathcal{H}_4 \left(2 \mathcal{H}_4-55\right) \mathcal{H}_3^3+36 \mathcal{H}_4^2 \left(3 \mathcal{H}_4+17\right)
   \mathcal{H}_3^2+54 \mathcal{H}_4^3 \left(6 \mathcal{H}_4+17\right) \mathcal{H}_3+81 \mathcal{H}_4^4 \left(4 \mathcal{H}_4+7\right)\right) \mathcal{H}_2^2\right.\nonumber\\
   &&\left. -18
   \mathcal{H}_3^2 \left(25 \mathcal{H}_3^4-272 \mathcal{H}_4 \mathcal{H}_3^3+216 \mathcal{H}_4^2 \mathcal{H}_3^2+180 \mathcal{H}_4^3 \mathcal{H}_3+27
   \mathcal{H}_4^4\right)\right) \mathcal{H}_1^5+\mathcal{H}_2 \left(64 \left(3 \mathcal{H}_4-14\right) \mathcal{H}_2^6+8 \left(8 \mathcal{H}_3^3+\left(411-27
   \mathcal{H}_4\right) \mathcal{H}_3^2\right.\right.\nonumber\\
   &&\left.\left.+18 \mathcal{H}_4 \left(6 \mathcal{H}_4-47\right) \mathcal{H}_3-81 \left(\mathcal{H}_4-3\right) \mathcal{H}_4^2\right)
   \mathcal{H}_2^4+3 \left(-16 \mathcal{H}_3^5+\left(18 \mathcal{H}_4-1271\right) \mathcal{H}_3^4-8 \mathcal{H}_4 \left(45 \mathcal{H}_4-514\right) \mathcal{H}_3^3+18
   \mathcal{H}_4^2 \left(6 \mathcal{H}_4+17\right) \mathcal{H}_3^2\right.\right.\nonumber\\
   &&\left.\left.-216 \mathcal{H}_4^3 \left(3 \mathcal{H}_4-1\right) \mathcal{H}_3+27 \mathcal{H}_4^4 \left(6
   \mathcal{H}_4+23\right)\right) \mathcal{H}_2^2+18 \mathcal{H}_3^2 \left(77 \mathcal{H}_3^4+2 \mathcal{H}_4 \left(6 \mathcal{H}_4-149\right) \mathcal{H}_3^3-108
   \mathcal{H}_4^2 \mathcal{H}_3^2+6 \mathcal{H}_4^3 \left(6 \mathcal{H}_4-41\right) \mathcal{H}_3-81 \mathcal{H}_4^4\right)\right) \mathcal{H}_1^4\nonumber\\
   &&+3 \left(16 \left(4
   \mathcal{H}_4-39\right) \mathcal{H}_2^8-24 \left(\left(3 \mathcal{H}_4-70\right) \mathcal{H}_3^2+4 \mathcal{H}_4 \left(3 \mathcal{H}_4-10\right)
   \mathcal{H}_3+\mathcal{H}_4^2 \left(9 \mathcal{H}_4-20\right)\right) \mathcal{H}_2^6+\left(8 \mathcal{H}_3^5-\left(30 \mathcal{H}_4+1423\right) \mathcal{H}_3^4\right.\right.\nonumber\\
   &&\left.\left.+2
   \mathcal{H}_4 \left(144 \mathcal{H}_4-1361\right) \mathcal{H}_3^3+36 \mathcal{H}_4^2 \left(3 \mathcal{H}_4+17\right) \mathcal{H}_3^2+54 \mathcal{H}_4^3 \left(12
   \mathcal{H}_4-1\right) \mathcal{H}_3+27 \mathcal{H}_4^4 \left(6 \mathcal{H}_4-7\right)\right) \mathcal{H}_2^4-6 \mathcal{H}_3^2 \left(\mathcal{H}_3^5-2 \left(3
   \mathcal{H}_4+29\right) \mathcal{H}_3^4\right.\right.\nonumber\\
   &&\left.\left.+4 \mathcal{H}_4 \left(3 \mathcal{H}_4-103\right) \mathcal{H}_3^3-18 \left(\mathcal{H}_4-24\right) \mathcal{H}_4^2
   \mathcal{H}_3^2+3 \mathcal{H}_4^3 \left(9 \mathcal{H}_4+28\right) \mathcal{H}_3-54 \mathcal{H}_4^4\right) \mathcal{H}_2^2+3 \mathcal{H}_3^4 \left(9 \mathcal{H}_3^4-242
   \mathcal{H}_4 \mathcal{H}_3^3+468 \mathcal{H}_4^2 \mathcal{H}_3^2\right.\right.\nonumber\\
   &&\left.\left.+234 \mathcal{H}_4^3 \mathcal{H}_3+27 \mathcal{H}_4^4\right)\right) \mathcal{H}_1^3-\mathcal{H}_2
   \left(128 \mathcal{H}_2^8+16 \left(2 \mathcal{H}_3^3-9 \mathcal{H}_3^2+9 \mathcal{H}_4 \left(3 \mathcal{H}_4-28\right) \mathcal{H}_3-27 \left(\mathcal{H}_4-3\right)
   \mathcal{H}_4^2\right) \mathcal{H}_2^6\right.\nonumber\\
   &&\left.+\left(-48 \mathcal{H}_3^5-441 \mathcal{H}_3^4-12 \mathcal{H}_4 \left(63 \mathcal{H}_4-905\right) \mathcal{H}_3^3+162
   \mathcal{H}_4^2 \left(2 \mathcal{H}_4+3\right) \mathcal{H}_3^2-324 \mathcal{H}_4^3 \left(3 \mathcal{H}_4-1\right) \mathcal{H}_3+81 \mathcal{H}_4^4 \left(12
   \mathcal{H}_4+19\right)\right) \mathcal{H}_2^4\right.\nonumber\\
   &&\left.+18 \mathcal{H}_3^2 \left(\mathcal{H}_3^5+38 \mathcal{H}_3^4+2 \mathcal{H}_4 \left(9 \mathcal{H}_4-256\right)
   \mathcal{H}_3^3-216 \mathcal{H}_4^2 \mathcal{H}_3^2+3 \mathcal{H}_4^3 \left(15 \mathcal{H}_4-88\right) \mathcal{H}_3-54 \mathcal{H}_4^4\right) \mathcal{H}_2^2\right.\nonumber\\
   &&\left. -27
   \mathcal{H}_3^4 \left(9 \mathcal{H}_3^4-92 \mathcal{H}_4 \mathcal{H}_3^3-78 \mathcal{H}_4^2 \mathcal{H}_3^2-132 \mathcal{H}_4^3 \mathcal{H}_3-27
   \mathcal{H}_4^4\right)\right) \mathcal{H}_1^2+\left(\left(528-96 \mathcal{H}_4\right) \mathcal{H}_2^{10}-8 \left(4 \mathcal{H}_3^3-6 \left(3 \mathcal{H}_4-37\right)
   \mathcal{H}_3^2\right.\right.\nonumber\\
   &&\left.\left.+27 \left(3-2 \mathcal{H}_4\right) \mathcal{H}_4 \mathcal{H}_3+81 \mathcal{H}_4^2\right) \mathcal{H}_2^8+3 \left(16 \mathcal{H}_3^5+\left(30
   \mathcal{H}_4+737\right) \mathcal{H}_3^4+6 \left(123-34 \mathcal{H}_4\right) \mathcal{H}_4 \mathcal{H}_3^3-108 \mathcal{H}_4^2 \mathcal{H}_3^2-162 \mathcal{H}_4^3
   \left(2 \mathcal{H}_4-3\right) \mathcal{H}_3\right.\right.\nonumber\\
   &&\left.\left.+81 \mathcal{H}_4^4 \left(2 \mathcal{H}_4+3\right)\right) \mathcal{H}_2^6-6 \mathcal{H}_3^2 \left(3 \mathcal{H}_3^5+\left(34
   \mathcal{H}_4+201\right) \mathcal{H}_3^4-36 \left(\mathcal{H}_4-13\right) \mathcal{H}_4 \mathcal{H}_3^3+54 \left(\mathcal{H}_4-12\right) \mathcal{H}_4^2
   \mathcal{H}_3^2\right.\right.\nonumber\\
   &&\left.\left.-27 \mathcal{H}_4^3 \left(5 \mathcal{H}_4-16\right) \mathcal{H}_3+243 \mathcal{H}_4^4\right) \mathcal{H}_2^4+9 \mathcal{H}_3^4 \left(\left(8
   \mathcal{H}_4+27\right) \mathcal{H}_3^4+174 \mathcal{H}_4 \mathcal{H}_3^3+12 \mathcal{H}_4^2 \left(2 \mathcal{H}_4-39\right) \mathcal{H}_3^2+90 \mathcal{H}_4^3
   \mathcal{H}_3+81 \mathcal{H}_4^4\right) \mathcal{H}_2^2\right.\nonumber\\
   &&\left.-324 \mathcal{H}_3^7 \mathcal{H}_4 \left(\mathcal{H}_3^2-4 \mathcal{H}_4
   \mathcal{H}_3-\mathcal{H}_4^2\right)\right) \mathcal{H}_1+3 \mathcal{H}_2 \left(\mathcal{H}_2^2-\mathcal{H}_3^2\right)^2 \left(64 \mathcal{H}_2^6-12 \left(7
   \mathcal{H}_3^2+30 \mathcal{H}_4 \mathcal{H}_3-9 \mathcal{H}_4^2\right) \mathcal{H}_2^4+3 \left(9 \mathcal{H}_3^4+136 \mathcal{H}_4 \mathcal{H}_3^3\right.\right.\nonumber\\
   &&\left.\left.+90 \mathcal{H}_4^2
   \mathcal{H}_3^2-27 \mathcal{H}_4^4\right) \mathcal{H}_2^2-108 \mathcal{H}_3^3 \mathcal{H}_4 \left(\mathcal{H}_3+\mathcal{H}_4\right)^2\right)\nonumber\\
  \null & \null & \null  \\
  \mathcal{D} &=& 3 \sqrt{3}
   \left(\mathcal{H}_1^2-\mathcal{H}_2^2+\mathcal{H}_3^2\right)^2 \left(4 \mathcal{H}_1^4+8 \mathcal{H}_2 \mathcal{H}_1^3+3 \left(\mathcal{H}_3^2-6 \mathcal{H}_4
   \mathcal{H}_3-3 \mathcal{H}_4^2\right) \mathcal{H}_1^2+2 \mathcal{H}_2 \left(-4 \mathcal{H}_2^2+3 \mathcal{H}_3^2+9 \mathcal{H}_4^2\right) \mathcal{H}_1-4
   \mathcal{H}_2^4\right.\nonumber\\
   &&\left.-12 \mathcal{H}_3^3 \mathcal{H}_4+3 \mathcal{H}_2^2 \left(\mathcal{H}_3^2+6 \mathcal{H}_4 \mathcal{H}_3-3 \mathcal{H}_4^2\right)\right)^2  \label{fullyfilli3}
  \end{eqnarray}
\end{landscape}
\newpage

\end{document}